%% file: 0.main.tex
\documentclass[11pt]{article}
\usepackage[hmargin=1in,vmargin=1in]{geometry}
\usepackage[dvipsnames,usenames,table]{xcolor}
\usepackage{hyperref}
\hypersetup{colorlinks=true, urlcolor=Blue, citecolor=Green, linkcolor=BrickRed, breaklinks, unicode}
\usepackage{hchang}

\usepackage{url}
\usepackage{graphicx}
\usepackage{algpseudocode}
\usepackage{bbm}
\usepackage{float}
\usepackage{framed}
\usepackage{enumerate}
\usepackage{color}
\usepackage{thm-restate}
\usepackage{comment}
\usepackage{subcaption}
\usepackage{tikz}
\usepackage{cleveref}
\usepackage{subfiles}

\usepackage{mdframed}

\usepackage{enumitem}
\newlist{inlinelist}{enumerate*}{1}
\setlist[inlinelist]{label=(\roman*),font={\color{red!50!black}\bfseries}, itemjoin={{, }}, itemjoin*={{, }}}

\newtheorem{theorem}{Theorem}

\newtheorem{lemma}{Lemma}
\newtheorem{claim}{Claim}
\newtheorem{definition}{Definition}
\newtheorem{observation}{Observation}
\newtheorem{remark}{Remark}
\newtheorem{invariant}{Invariant}

\newcommand{\eps}{\varepsilon}

\begin{document}
\begin{titlepage}

\title{Dynamic Locality Sensitive Orderings in Doubling Metrics}
\author{
      An La \thanks{University of Massachusetts Amherst. Email: {\tt anla@umass.edu}.}\and
       Hung Le\thanks{University of Massachusetts Amherst. Email: {\tt hungle@cs.umass.edu}.}
}
\date{}

\maketitle
	
\thispagestyle{empty}

\begin{abstract}
In their pioneering work, Chan, Har-Peled, and Jones (SICOMP 2020)  introduced \emph{locality-sensitive ordering} (LSO), and constructed an LSO with a constant number of orderings for point sets in the $d$-dimensional Euclidean space. Furthermore, their LSO could be made dynamic effortlessly under point insertions and deletions, taking $O(\log(n))$ time per update by exploiting Euclidean geometry. Their LSO provides a powerful primitive to solve a host of geometric problems in Euclidean spaces in both dynamic and static settings.   Filtser and Le (STOC 2022) constructed the first LSO with a constant number of orderings in the more general setting of doubling metrics.  However, their algorithm is inherently static since it relies on several sophisticated constructions in intermediate steps, none of which is known to have a dynamic version.  Making their LSO dynamic would recover the full generality of LSO and provide a general tool to dynamize a vast number of static constructions in doubling metrics. 

In this work, we give a dynamic algorithm that has $O(\log n)$ time per update for constructing an LSO in doubling metrics under point insertions and deletions. To this end, we introduce a toolkit of several new data structures: \emph{a pairwise index tree} (PIT) which augments the standard net tree with the pairwise property, \emph{a pairwise tree cover} which in a certain sense is a tree counterpart of LSO, \emph{a net tree cover} for stabilizing the net tree, and \emph{a leaf tracker} for keeping track of a DFS ordering of leaves in a dynamic tree. A key technical problem that we solves in this work is stabilizing the dynamic net tree of Cole and Gottlieb (STOC 2006), a central dynamic data structure in doubling metrics, using  a dynamic \emph{net tree cover}. Specifically,  we show that every update to the dynamic net tree can be decomposed into a few very simple updates to trees in the net tree cover. As stability is the key to any dynamic algorithm, our technique could be useful for other problems in doubling metrics.

We obtain several algorithmic applications from our dynamic LSO, including dynamic fault-tolerant spanner, dynamic tree cover, dynamic nearest neighbor search with optimal search time, dynamic (bichromatic) closest pair of points,  all in doubling metrics. Most notably, we obtain the first dynamic algorithm for maintaining an $k$-fault tolerant spanner in doubling metrics with optimal sparsity in optimal $O(\log n )$ time per update.
\end{abstract}
\end{titlepage}

\tableofcontents

\pagebreak

\section{Introduction}
\subfile{1.intro.tex}

\section{Preliminaries}\label{sec:prelim}
\subfile{2.preliminaries.tex}

\section{Pairwise Tree Cover: Static Construction}\label{sec:collect_pit}
    \subfile{5.static.tex}

   \section{Pairwise Tree Cover: Dynamic Construction}\label{sec:dynamic_ptc}
    \subfile{6.dynamic.tex}

    \subfile{8.nettreecover_new.tex}
   \subsection{Dynamic Pairing}\label{sec:dynamic_pairing}
     \subfile{7.dynamic_pairing.tex}
    
    \section{Leaf Tracker}\label{sec:leaf_tracker}
    \subfile{9.leaf_tracker.tex}

    \section{Dynamic Net Tree}\label{sec:dynamic_nettree}
    \subfile{10.nettree.tex}

\section{Applications of LSO}\label{sec:dynamic-LSO}

Here we give the details of the applications of LSO mentioned in \Cref{subsec:app}.

\subfile{4.application.tex}

\paragraph*{Acknowledgements.~}  This work was supported by the NSF CAREER Award No. CCF-223728, an NSF Grant No. CCF-2121952, and a Google Research Scholar Award. Thank Aditya Kumar Roy Chowdhury for joining the early states of this work.

\bibliography{ref}

\end{document}

%% file: 1.intro.tex
Chan, Har-Peled and Jones~\cite{CHJ20} introduced locality-sensitive ordering (LSO) as a powerful tool for solving geometric problems. Roughly speaking, an LSO of a point set $S$ in a metric space $(X,d_X)$ is a collection of linear orderings of points in $S$ that has a \EMPH{locality property}, namely, any two points $x,y\in X$ are close in some ordering of the collection. Here being close means the points between $x$ and $y$ in the ordering are either close to $x$ or to $y$. 

\begin{definition}[$(\tau, \varepsilon)$-LSO]\label{def:lso} Let $S$ be a set of points in a metric space $(X,d_X)$. A collection of linear orderings, denoted by $\Sigma$ is a \emph{$(\tau,\eps)$-locality sensitive ordering} if:
\begin{itemize}
    \item \textnormal{[Size.]}  $\Sigma$ has at most $\tau$ ordering.
    \item \textnormal{[Covering.]} there is a bijection between points in each ordering and $S$.
    \item \textnormal{[Locality.]} for every $x,y\in S$, there exists an ordering $\sigma\in \Sigma$ such that any point $z$ between $x$ and $y$ on $\sigma$  is in distance $\eps d_X(x,y)$ either from $x$ or from $y$. That is, $\min\{d_X(z,x),d_X(z,y)\}\leq \eps d_X(x,y)$.  
\end{itemize}
\end{definition}

One could think of an LSO as an ``embedding'' of $S$ into a collection of lines so that geometric constructions for $S$ in the (complicated and high dimensional) metric space $(X,d_X)$ could be reduced to the $1$-dimensional line. Therefore, LSO allows significant simplification of the constructions of many complicated objects, such as fault-tolerant spanners~\cite{CHJ20} and reliable spanners~\cite{BHO19,BHO20,FL22}. 

Chan, Har-Peled and Jones~\cite{CHJ20} constructed a $(\tau,\eps)$-LSO $\Sigma$ for point sets in $\mathbb{R}^d$ where the number of orderings $\tau = 2^{O(d)}\eps^{-d}$. Thus, for a fixed $\eps$ and $d$, the number of orderings is a constant. Furthermore, their LSO could be easily made dynamic, since the construction is based on space partitioning. More specifically, for any given two points $p,q \in \mathbb{R}^d$, one could determine their relative positions in a given ordering $\sigma \in \Sigma$ --- determine whether $p\prec_\sigma q$ or $q\prec_\sigma p$--- by applying bitwise operations on their coordinates. Therefore, one could represent each ordering in the LSO as a binary search tree, which supports point updates in $O(\log n)$ time per operation. The main takeaway here is that Euclidean geometry allows simple dynamization of their LSO.

Using their dynamic LSO in $\mathbb{R}^d$, Chan, Har-Peled and Jones~\cite{CHJ20} obtained a host of dynamic algorithms for geometric problems, such as dynamic bichromatic closest pair of points,  dynamic spanners, dynamic vertex-fault-tolerant spanners, dynamic approximate nearest neighbors, dynamic approximate MST; these algorithms all have $O(\log n)$ time per update. For several of these problems, they were the first to achieve logarithmic update time. 

The existence of an LSO in Euclidean metrics naturally motivates the question of constructing an LSO for doubling metrics. While doubling metrics vastly generalize Euclidean metrics, many nice properties of Euclidean geometry are lost in doubling metrics. The technique of Chan, Har-Peled and Jones~\cite{CHJ20} relied extensively on Euclidean geometry, and it is unclear if their technique can be easily extended to doubling metrics. Nevertheless, they are able to construct a $(\tau,\eps)$-LSO for point sets in doubling metrics of dimension $\lambda$ where the number of orderings is $\tau = O(\log(n)/\eps)^{O(\lambda)}$, which depends on the number of points. An open problem left by their work is to reduce the number of orderings to\footnote{We use the notation $O_{\eps,\lambda}$ to hide the dependency on $\eps$ and $\lambda$.} $O_{\eps,\lambda}(1)$. This problem was recently solved by Filtser and Le~\cite{FL22}; the number of orderings in their construction is $\tau = \eps^{-O(\lambda)}$ orderings. Their LSO (and its variants) are powerful primitives to solve various problems in metric spaces~\cite{FL22,Filtser23}.

An arguably more important problem is to construct a \EMPH{dynamic LSO} with a small number of orderings in doubling metrics. As mentioned above, a dynamic LSO will give dynamic algorithms for a host of problems in doubling metrics, recovering the full power of LSO. If all possible points in the metric occurring during the course of the algorithm are given in advance, that is, the points under insertions/deletions belong to a specific set $P$ given at the beginning of the algorithm, Chan, Har-Peled, and Jones~\cite{CHJ20} gave a simple dynamic LSO by simulating their Euclidean counterpart. The key observation is that if $P$ is given, one can construct a net tree, which then can be used to ``partition the space'' in the same way that a quadtree partitions $\mathbb{R}^d$. However, knowing $P$ in advance is a very strong and artificial assumption for a dynamic data structure, as concurred by Chan, Har-Peled, and Jones~\cite{CHJ20}.

Another major problem of the dynamic LSO  for doubling metrics by Chan, Har-Peled, and Jones~\cite{CHJ20} is the size: the number of orderings is poly-logarithmic instead of a constant. On the other hand, for the LSO of constant size by Filtser and Le~\cite{FL22}, achieving a \EMPH{static construction} in  $O(n\log n)$ time remains an open problem; their LSO construction is rather complicated, relying on sophisticated objects, such as ultrametric covers and pairwise partition covers. Therefore, even if all the points are given in advance, it is not easy to dynamize the construction of Filtser and Le with a poly-logarithmic time per update. We note that it might be possible to construct the LSO by Filtser and Le~\cite{FL22} statically in time $O(n\log(\Delta))$ where $\Delta$ is the \emph{spread}\footnote{The spread is the ratio between the maximum pairwise distance over the minimum pairwise distance.} of the point set. However, $\Delta$ could be exponential in $n,$ and hence the worst case running time remains $\Omega(n^2)$. Removing the dependency on $\Delta$ is a central problem in designing algorithms, both static and dynamic, for doubling metrics~\cite{HM06,GC06,GL08,GL08B}.

In this work, we give the first data structure, as formally defined in \Cref{def:dynamic_lso} below, for maintaining a \EMPH{dynamic LSO with a constant number of orderings} in doubling metrics. Our data structure could handle point insertions/deletions to the LSO in $O(\log n)$ time per update.  

\begin{definition}[Dynamic LSO Data Structure]\label{def:dynamic_lso} $(\tau, \varepsilon)$-Dynamic LSO is a data structure maintaining a $(\tau, \varepsilon)$-LSO  $\Sigma$ for a dynamic set of points $S$ and supporting the following operations:
\begin{itemize}
    \item \textsc{Insert}$(q, \Sigma)$: insert a point $q$ to $\Sigma$.
    \item \textsc{Delete}$(q, \Sigma)$: remove $q$ from $\Sigma$.
    \item \textsc{GetPredecessor}$(q, i, \Sigma)$: return the predecessor of $q$ in $i^{th}$ ordering of $\Sigma$, return null if $q$ is the first point in the ordering.
    \item \textsc{GetSuccessor}$(q, i, \Sigma)$: return the successor of $q$ in $i^{th}$ ordering of $\Sigma$, return null if $q$ is the last point in the ordering.
\end{itemize}
\end{definition}

One important property of our dynamic LSO is stability. We say that a data structure for maintaining an LSO of a dynamic point set $S$ is \EMPH{stable} if for every dynamic ordering $\sigma$ in the LSO, when a point is inserted or deleted from $S$, the data structure does not change the relative ordering of existing points in the LSO. Intuitively, when a point is deleted from or inserted to $S$, a stable data structure simply deleting or inserting it, respectively, from each ordering in the LSO without mixing up the order of other points. A prior, it is unclear (even in the static setting) a stable LSO exists, and specifically, if there is a way to insert a new point to an existing LSO to get a new LSO that is also good for the new point.  In some applications of LSO (to be discussed in more detail in \Cref{subsec:app}) such as dynamic closest pair or approximate bichromatic closest pair, we consider adjacent pairs of points in all orderings. The stability of LSO allows us to keep track of these pairs in $O(1)$ time per ordering. On the other hand, without the stability, we have to update the set of adjacent pairs of points in all orderings, which could cost $\Omega(n)$ time. The stability is even more crucial in dynamic vertex-fault-tolerant (VFT) spanners, since in this application, we need to query $k$ nearest predecessors and $k$ nearest successors of every point in each ordering. In the following theorem, which is our main result, we construct a stable dynamic LSO.

\begin{theorem}\label{thrm:dynamic_lso}
    Given $\varepsilon \in (0,1)$, there is a data structure maintaining $(\varepsilon^{-O(\lambda)}, \varepsilon)$-LSO for a dynamic point set $S$ in doubling metrics of dimension $\lambda$ supporting \textsc{Insert}/\textsc{Delete} in $O\left( \varepsilon^{-O(\lambda)}\log(n)\right)$ time per operation and \textsc{GetPredecessor}/\textsc{GetSuccessor} in $O(1)$ time per operation. Furthermore, our LSO is stable. 
\end{theorem}

We emphasize that the running time of each operation \textsc{GetPredecessor}/\textsc{GetSuccessor} does not depend on $n,\eps$ or $\lambda$. As we will see, in some applications, such as fault-tolerant spanner, achieving $O(1)$ time per \textsc{GetPredecessor}/\textsc{GetSuccessor} operation as in \Cref{thrm:dynamic_lso} is important to get optimal running time, matching the best static algorithms.

\paragraph{Model assumptions.} Our dynamic algorithm makes the same two common assumptions either explicitly or implicitly used in prior works in doubling metrics~\cite{GC06,GL08,GL08B}. First, we have access to an exact distance oracle that, given any two points in the metric, computes their distance in $O(1)$ time. Second, after a point is deleted from the point set, the distance between the current point and the deleted point could still be computed in $O(1)$ time. These assumptions can be naturally realized in some special cases such as low dimensional Euclidean or $\ell_p$ spaces for a constant $p\geq 1$. 

Next, we give an overview of our technical ideas for maintaining a dynamic LSO of constant size. Then in \Cref{subsec:app}, we discuss the applications of our dynamic LSO; some of these applications were studied in Euclidean spaces by  Chan, Har-Peled and Jones~\cite{CHJ20}.

\subsection{Key Technical Ideas}\label{subsc:tech}

Our first step is to interpret the (only) existing construction of LSO in doubling metrics by Filtser and Le~\cite{FL22} in terms of trees since (dynamic) trees are the basic building block of many dynamic algorithms. Filtser and Le~\cite{FL22} constructed their LSO via a so-called \emph{pairwise partition cover}, which is a family of hierarchical partitions of the input metric space, and an \emph{ultrametric cover}, which a family of ultrametrics that have a certain distance covering property. Their overall construction is rather involved, and it is not clear even how to implement it statically in $O(n\log n)$ time for constants $\eps$ and $\lambda$. Here we introduce a new type of trees and tree covers called \EMPH{pairwise index tree} (PIT) and \EMPH{pairwise tree cover}, respectively. A pairwise tree cover consists of $O(\log(1/\eps))$ different PITs where each pair of points is ``covered'' by one of the PITs.  In a PIT, each internal node is labeled with one or two points in $S$, and each leaf is labeled by exactly one point (to form a bijection to $S$).  For a given pair of points $(x,y)$, loosely speaking, we would like to have a node $\eta_{xy}$ in some PIT $T$ labeled with both $x$ and $y$ such that (the points associated with) leaves of the subtree rooted at $\eta_{xy}$ of $T$  is either in the distance $\eps d_X(x,y)$ from $x$ or from $y$. (Both $x$ and $y$ will be associated with leaves in the subtree of $T$ rooted at $\eta_{xy}$.) If so, then visiting each PIT in the cover by depth-first search (\EMPH{DFS}) would give us a linear ordering of leaves, called \EMPH{DFS leaf ordering}, satisfying the locality property (in \Cref{def:lso}) for $x$ and $y$: every point between $x$ and $y$ in the DFS linear ordering will be children of $\eta_{xy}$ and hence within $\eps d_X(x,y)$. Thus, all the DFS leaf orderings from all the trees in the pairwise tree cover together would be an LSO. At a more technical level, having such node $\eta_{xy}$ for every pair $(x,y)$ would mean the total number of nodes would be $\Omega(n^2)$, rendering any hope for efficient dynamic maintenance. So we relax it slightly: $\eta_{xy}$ would be labelled by a pair $(x',y')$ such that $d_X(x,x'), d_X(y,y')\leq \eps d_X(x,y)$. Both $x$  and $y$ remain associated with leaves of the subtree rooted at $\eta_{xy}$. This relaxation allows many pairs to share the same node and hence could potentially be maintained efficiently. The formal definitions, therefore, are less intuitive than described here.

\begin{definition}[Pairwise index tree (PIT)]\label{def:pit} Let $\delta \geq 1$ and $\eps$ be parameters. A $(\delta, \varepsilon)$-pairwise index tree of $S$ is a rooted tree with  the following properties:
\begin{enumerate}
    \item \textnormal{[Pairwise labelling.]} Each node is labeled with one or two elements in $S$. A node at level $i$ is denoted as $(x, y, i)$ where $x, y \in S$, $x$ could be the same or different from $y$. 
    A leaf is labeled with exactly one point in $S$. 
    \item \textnormal{[Packing.]} For two nodes $(x, y, i)$ and $(u, v, i)$, the distance between any pair of points in $\{x, y, u, v\}$ is $\Omega(\frac{\delta}{\varepsilon^{i-1}})$.
    \item \textnormal{[Covering.]} 
    Label points in all children of $(x, y, i)$ are within the distance $O(\frac{\delta}{\varepsilon^{i-1}})$ from $x$ or $y$. 
    Let $C_i(x, y)$ be the union of all labels (or leaf labels) in the subtree rooted as $(x, y, i)$.   The diameter of $C_i(x, y)$ is $O(\delta/\varepsilon^i)$. We call  $C_i(x, y)$ \EMPH{the cluster of node $(x,y,i)$}.
\end{enumerate}
\end{definition}

The packing and covering properties in the definition of PIT are very similar to the packing/covering properties of a net tree, a standard tool for navigating doubling metrics. The key difference between a PIT and a net tree is that these properties applied to pairs of points. In the construction of pairwise tree cover, different PITs in the cover will be obtained by varying the parameter $\delta$ in \Cref{def:pit}. 

Given a PIT $T$, we say that a node at level $i$ of $T$, denoted by $(u,v,i)$, is \EMPH{$\eps$-close} to a pair $(x,y)$ if every point $p\in C_i(u,v)$, the cluster of $(u,v,i)$, has 
\begin{inlinelist}
    \item $x,y\in C_i(u,v)$
    \item $d_X(p,x)\leq \eps d_X(x,y)$ or $d_X(p,y)\leq \eps d_X(x,y)$ (that is, either $p$ is close to $x$ or close to $y$)
\end{inlinelist}.

\begin{definition}[Pairwise tree cover] \label{def:collect_pit} A $(\tau,\eps)$-pairwise tree cover of a point set $S$, denoted by $\mathcal{T}$, is a collection of $(\delta,\eps)$-PITs (for different values of $\delta$) such that:

\begin{itemize}
    \item \textnormal{[Size.]} $\mathcal{T}$ contains at most $\tau$ PITs.
    \item \textnormal{[Pairwise covering.]} For any pair of points $x, y \in S$ whose distance in $[\frac{\delta}{\varepsilon^i}, \frac{2\delta}{\varepsilon^{i}})$ for some $\delta\in \{1,2^{1},2^{2},\ldots, 2^{\lceil \log(1/\eps) \rceil}\}$, there exists a $(\delta,\eps)$-PIT $T\in \mathcal{T}$ such that a node at level $i$ of $T$ is $O(\eps)$-close to pair $(x,y)$.
\end{itemize}
\end{definition}

We remark that there could be more than one $(\delta,\eps)$-PITs in a pairwise tree cover $\mathcal{T}$ that shares the same value of parameter $\delta$; they are different in internal representations as they cover different sets of pairs. The key points are (a) there are only $O(\log(1/\eps))$ different values of $\delta$, and (b) as we will show later, for each $\delta$, there are only $\eps^{-O(\lambda)}$ different PITs sharing the same $\delta$.

\begin{figure}[!ht]
\vspace{0.5cm}
    \hspace*{-2.cm}
    \begin{tikzpicture}
	 \node at (0,0){\includegraphics[width=1.2\linewidth]{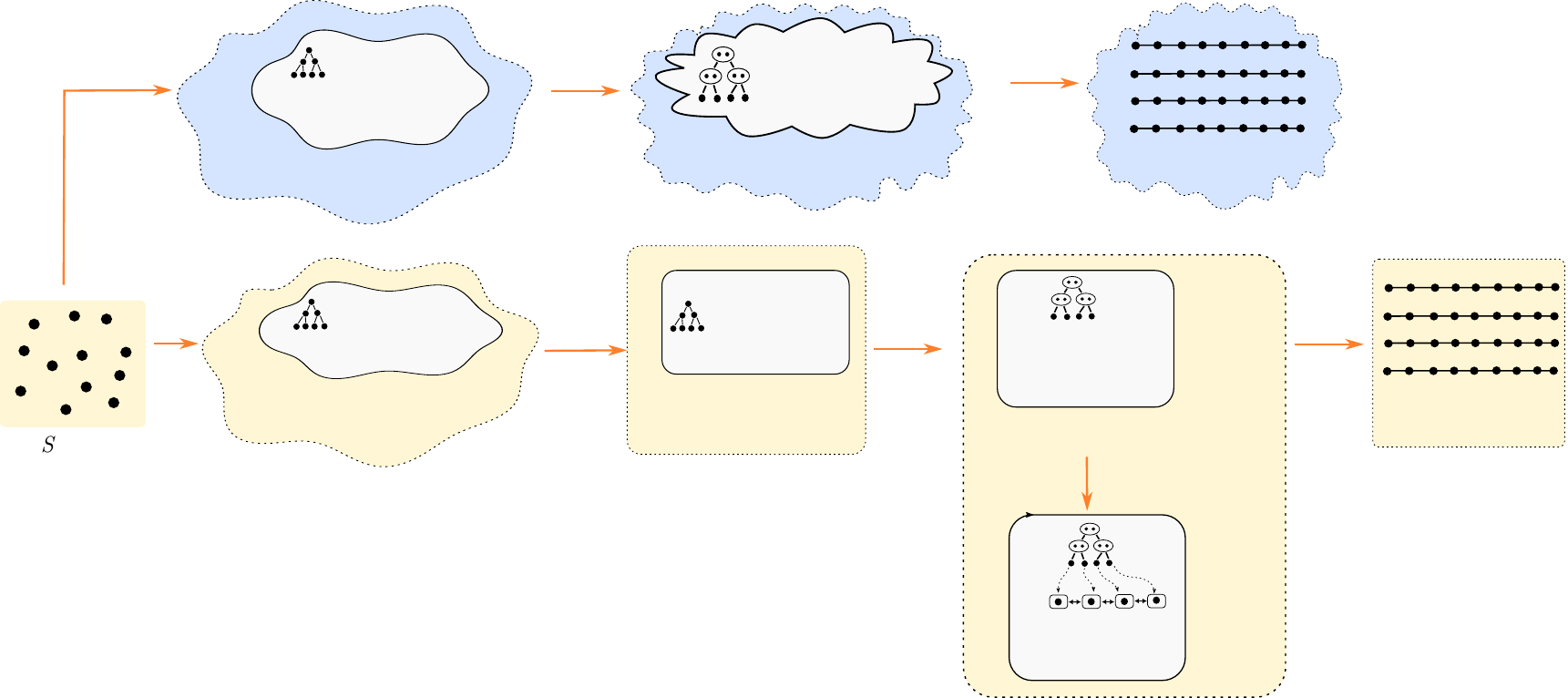}};	
            \node at (-5.1, 3.4){\footnotesize static};
			\node at (-5.3, 3.0){\footnotesize $(\delta,\eps)$-net tree};
             \node at (-5.7, 2.2){\footnotesize $\delta = 2^{0},2^{1},\ldots, 2^{\lceil \log(1/\eps) \rceil}$};
             
              \node at (0.0, 4.7){\footnotesize(\Cref{sec:collect_pit})};
              \node at (0.6, 3.5){\footnotesize static};
			\node at (0.5, 3.1){\footnotesize $(\delta,\eps)$-PITs};
             \node at (0.2, 2.4){\footnotesize $(\tau,\eps)$-pairwise tree cover};

            \node at (3.2, 3.7){\footnotesize DFS leaf};
            \node at (3.2, 3.0){\footnotesize ordering};

            \node at (5.2, 5.0){\footnotesize(\Cref{sec:LSO-from-PIT})};
             \node at (5.4, 2.3){\footnotesize static $(\tau,\eps)$-LSO};

            \node at (-4.8, 0.4){\footnotesize dynamic};
			\node at (-5.0, 0.0){\footnotesize $(\delta,\eps)$-net tree};
             \node at (-5.4, -0.7){\footnotesize $\delta = 2^{0},2^{1},\ldots, 2^{\lceil \log(1/\eps) \rceil}$};
             \node at (-5.4, -1.7){\footnotesize(\Cref{sec:dynamic_nettree})};

            \node at (-0.2, 0.7){\footnotesize stable};
            \node at (-0.2, 0.35){\footnotesize dynamic};
			\node at (-0.2, 0.0){\footnotesize net tree};
             \node at (-0.5, -0.6){\footnotesize stable dynamic};
             \node at (-0.5, -0.95){\footnotesize net tree cover};
             \node at (-0.5, -1.7){\footnotesize(\Cref{sec:nettreecover})};

              \node at (1.7, 0.3){\footnotesize dynamic};
            \node at (1.7, -0.35){\footnotesize pairing};
            
            \node at (3.8, -0.0){\footnotesize stable dynamic};
			\node at (3.7, -0.35){\footnotesize  $(\delta,\eps)$-PITs};
            \node at (3.7, -1.0){\footnotesize(\Cref{sec:dynamic_pairing})};
            
             \node at (5.4, -0.95){\footnotesize dynamic};
             \node at (5.4, -1.3){\footnotesize pairwise};
              \node at (5.4, -1.65){\footnotesize tree cover};

              \node at (6.8, 0.4){\footnotesize DFS leaf};
            \node at (6.8, -0.3){\footnotesize ordering};
 
             \node at (3.9, -3.7){\footnotesize leaf tracker};
             \node at (3.9, -4.7){\footnotesize(\Cref{sec:leaf_tracker})};
             
             \node at (8.7, -1.5){\footnotesize(\Cref{sec:LSO-from-PIT})};
             \node at (8.7, -0.9){\footnotesize dynamic LSO};
             
			\node at (-2.7, 0.6){\footnotesize  stabilize};
            \end{tikzpicture}    
            \vspace{0.5cm}

    \caption{Data structures highlighted in light blue are static while those highlighted in light yellow are dynamic. Data structures with rectangular shapes are stable; others are unstable.}
    \label{fig:organize}
\end{figure}

The pairwise tree cover is our attempt to combine the strengths of the LSO construction by Chan, Har-Peled and Jones~\cite{CHJ20}  for Euclidean spaces and the LSO construction by Filtser and Le~\cite{FL22} for doubling metrics. Specifically,  Chan, Har-Peled, and Jones~\cite{CHJ20} constructed a collection of shifted quadtrees, a well-studied space partitioning data structure in Euclidean spaces, and visited each quadtree by \EMPH{$Z$-order} to form an LSO. The geometrical nature of the quadtree makes it easy to dynamize their static LSO. An analogous but less powerful counterpart of quadtree in doubling metrics is the net tree. However, it is unclear if an analogous $Z$-order in doubling metrics exists. For this reason, Filtser and Le~\cite{FL22} developed a very different technique to construct an LSO in doubling metrics. First, they constructed a \emph{pairwise partition cover} that has a certain pairwise property. They then used the partition to construct an ultrametric covers, and each LSO is constructed from an ultrametric in the cover by induction.   Here we combine the strengths of both works in the PITs: we start with a net tree and augment it with the pairwise property by Filtser and Le~\cite{FL22}, which can be seen as a replacement for the $Z$-order. Given the pairwise tree cover, we simply apply a DFS leaf ordering to each tree to get an LSO. \Cref{fig:organize} (the top part) illustrates building blocks to construct a (static) LSO for a point set $S$.

In the dynamic setting, we could use the algorithm by Cole and  Gottlieb~\cite{GC06} to maintain a dynamic net tree.  As PITs are built on top of net trees, in principle, one could adapt their technique to maintain a dynamic PIT. In our (static) construction, we show that our PIT has a certain locality condition, and specifically, the neighborhood of a node in a PIT is a subset of the neighborhood of the corresponding node in a net tree $T$. This allows us to maintain a dynamic PIT from a dynamic net tree in a black-box manner. While  the high-level ideas are relatively simple, there are some conceptual difficulties in translating the static construction to the dynamic construction, mostly due to that a dynamic net tree has to be compressed and hence some nodes are not directly accessible.

\begin{figure}[!ht]
    \begin{tikzpicture}
	 \node at (0,0){\includegraphics[width=1.0\linewidth]{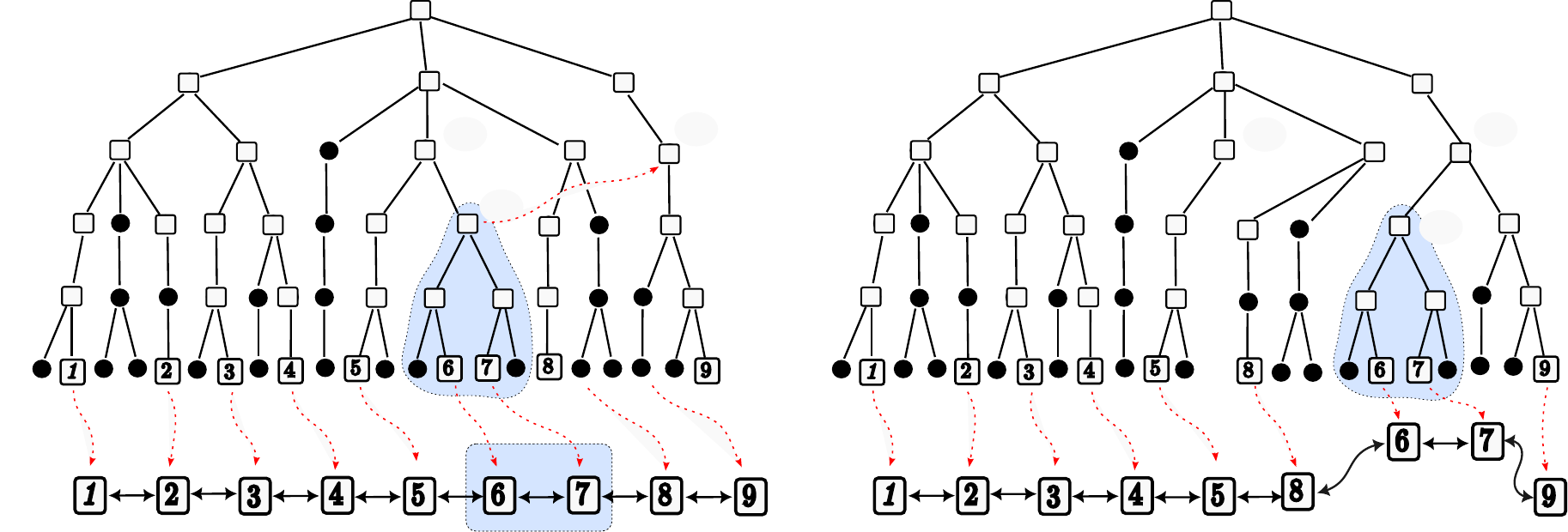}};	
            \node at (-2.9, 0.6){\footnotesize  $x$};
            \node at (-3.5, 1.4){\footnotesize  $u$};
            \node at (-1.0, 1.4){\footnotesize  $v$};
            
             \node at (6.8, 0.4){\footnotesize $x$};
              \node at (4.9, 1.4){\footnotesize $u$};
              \node at (7.5, 1.4){\footnotesize $v$};
            \end{tikzpicture}    

    \caption{A node $x$ in a PIT changes parent from $u$ to $v$ leads to changes in DFS leaf orderings of $v$ and its ancestors, and of $u$ and its ancestors. Rectangular nodes are active nodes, which either leaves corresponding to non-deleted points or internal nodes with at least one active descendant leaf.}
    \label{fig:unstable}
\end{figure}

The much more difficult task is to maintain a dynamic leaf ordering of PIT, and indeed, all technical ideas we  develop herein are to solve this task. In the static setting, one simply applies DFS to visit each PIT to get an ordering of the leaves.  In the dynamic setting, there are two major challenges (see \Cref{fig:unstable}):
\begin{itemize}
    \item \textbf{(C1):} In dynamic net trees (and hence dynamic PITs derived from dynamic net trees), \EMPH{nodes are only marked deleted} rather than being explicitly deleted from the trees, and hence some leaves become inactive\footnote{Here an inactive leaf refers to a leaf that is marked deleted. Later, in technical sections, sometimes it is convenient for us to insert a null leaf, which is a leaf associated with no point, into a tree. There, an inactive leaf refers to a leaf that is either null or marked deleted.} when their corresponding points are deleted from $S$. In the DFS leaf ordering of a PIT, we only keep track of an ordering of active leaves. Furthermore, adding a single active leaf to a net tree could activate $\Omega(n)$ ancestors of the leaf to become active. (A node in the tree is active if it has at least one active descendant leaf.)

    \item  \textbf{(C2):} Active descendant leaves of each internal node in a PIT induce a contiguous subsequence of the DFS leaf ordering of the PIT. When a node $x$ changes parents from $u$ to $v$ as in \Cref{fig:unstable}, if we can identify the leftmost active leaf, say $l_v$, of $v$ in its subsequence, then we could simply slice the DFS leaf subsequence of $x$ and stitch it to the left of $l_v$ in the DFS leaf ordering. (This means that the \emph{recourse} of a parent update in the DFS leaf ordering is small.) The difficulty is in identifying the leftmost (and also rightmost) leaf of $v$, and more generally, of an internal node. A natural idea is to a pointer from each internal node to the leftmost/rightmost descendant leaves, but \EMPH{a parent update of a single node in a PIT could change the pointers to the leftmost/rightmost descendant leaves of all of its (both old and new) ancestors}. There could be $\Omega(n)$ such ancestors (if the aspect ratio $\Delta$ is large).
\end{itemize}

 In the special case of incremental dynamic, a.k.a. insertion only, (C1) does not happen since there is no deletion,  it remains challenging to resolve (C2). A well-known technique  for maintaining a certain kind of DFS of a dynamic tree is the Euler tour technique~\cite{TV84,KH99}. However, the Euler tour maintains a DFS ordering of all edges in the tree (in both directions), while we only maintain a list of leave nodes, and hence we need to be able to query a leftmost leaf in the subtree of an internal node.  There seems to be no easy way to modify existing dynamic tree data structures, including the Euler tour technique, for this purpose under parent updates. This problem is significantly compounded by the presence of deleted leaves: even designing a data structure for querying an (arbitrary) active descendant leaf of a given internal node in a PIT becomes non-trivial.

We will take several steps to resolve both challenges, see the lower part of \Cref{fig:organize}. Our key idea is to stabilize the net tree using a net tree cover. The formal definition of a net tree cover is somewhat unintuitive and hence we defer it to \Cref{def:nettree_cover} in \Cref{sec:dynamic_ptc}; here we describe its high-level intuition. A dynamic net tree cover is a collection of $O_{\lambda}(1)$ net tree where updates to every tree are restricted to one of three types: inserting a new leaf, marking a leaf deleted, and subdividing an edge. Therefore, the only non-trivial parent update in the tree is by edge subdivision: inserting a new node $z$ in the middle of an edge $(x,y)$ between a parent $x$ and a child $y$, effectively changing the parent of $y$ from $x$ to $z$. This type of parent change does not alter the DFS leaf ordering of a node in the tree, which is the key to our dynamic data structure.  We say that every tree in the net tree cover is \EMPH{stable}; updates that are not one of the three types above are \EMPH{unstable}. As stable updates are too restrictive, it should not be surprising that the dynamic net tree by Cole and Gottlieb~\cite{GC06}, as well as many other dynamic tree data structures, are unstable.

Our basic idea is to ``decompose'' an (unstable) update to a net tree $T$ into $O_{\lambda}(1)$ stable updates to $\mathcal{J}$, the corresponding net tree cover. The main observation is that when a node $(p,i-1)$ at level $i-1$ of a net tree $T$ changes its parent at level $i$ from $(u,i)$ to $(v,i)$, then  both $d_X(p,u)$ and $d_X(p,v)$ are bounded by $O(\delta/\eps^i)$ and hence  small compared to the radius at level $i$, by the covering of the net tree. Suppose in an \EMPH{ideal situation} in which we have a version of $T$, denoted by $J$, where we only keep a $\frac{c\cdot\delta}{\eps^i}$-net at level $i$ of $J$ for a sufficiently big constant $c$, and that $u$ happens to be in the net, then both $p$ and $v$ will be children of $u$, and there is no need to change the parent of $(p,i-1)$ to $(v,i)$. (The bound $\frac{\delta}{\eps^i}$ is the packing/covering radius at level $i$ of a $(\delta,\eps)$-net tree; see \Cref{def:netttree}.) Of course, the ideal situation will not always happen, and therefore, we construct many, but $O_{\lambda}(1)$, different versions of $T$ in the net tree cover, and we could show that, loosely speaking, the ideal situation will happen at one version of $T$. Our idea is to realize some kind of shifting strategy, in the same way the shifted quadtree was used in Euclidean space by  Chan, Har-Peled, and Jones~\cite{CHJ20}. However, we do not have Euclidean geometry; instead, we use a standard coloring trick (e.g.,~\cite{BNF22,KLMS22}) to color net points. To implement all of these ideas, we have to handle two major difficulties: (1) the ideal situation only happens at one of the trees, and we have to handle non-ideal situations in other trees---the key to this is that we have more leeway in other trees, as the important pairs were taken care of in the ideal situation; (2) when new node arrives due to insertions of new points to $S$, one has to \emph{merge} it with other nodes, leading to unstable parent changes. (Roughly speaking, a node $(v,i)$ is merged to a node $(u,i)$ if $(v,i)$'s children become $(u,i)$'s children in a version $J$ of $T$.)  We resolve both problems by developing two rules on top of the net tree cover, namely \EMPH{merging by distances} and \EMPH{merging through time}: when a new node arrives, we look at its distance to existing nodes, and decide to merge based on both the distances and the time when the node arrives. All in all, we are able to show that updates to $\mathcal{J}$ are stable. As stability is the key to dynamic algorithms, we believe that this construction is of independent interest.

Given a stable dynamic net tree (encapsulated in a dynamic net tree cover), we develop a dynamic algorithm, called \EMPH{dynamic pairing}, to construct dynamic PITs. Here, we exploit the stability of the net tree cover to simplify and adapt our static construction  to the dynamic setting. 
As we noted earlier, our static construction has a certain locality condition. An important guarantee of our dynamic pairing algorithm is that (updates to) our dynamic PITs are stable, given that the input dynamic net tree is stable; this is important for the next step: keeping track of DFS ordering of active leaves in PITs. 

Finally, we develop a data structure, called \EMPH{leaf tracker}, to keep track of the DFS ordering   of active leaves in a PIT. Recall that the ordering is obtained by visiting each PIT by DFS, breaking ties by the insertion time.  We will store the DFS ordering using a \emph{doubly linked list} $\sigma$. We also build a skip list on top of $\sigma$ to  perform some kind of binary search. We design the keys to the skip lists to be what we call \emph{ancestral arrays}. Roughly speaking,  an \EMPH{ancestral array} of a node $u\in T$ is an \emph{array} $O(\log n)$ ``important'' ancestors stemming from a centroid decomposition of $T$ (see \Cref{def:centr-decom-anc-arry}). Though there is no linear order between the ancestral arrays to use them as keys in the traditional sense, we could use them to determine if a leaf $x$ is a descendant of a query node $u$ or not by \Cref{lm:leaf-descedant}, which turns out to be sufficient for binary search using skip lists. There are several subtleties in the implementation, which we will discuss in detail later in \Cref{sec:track_leaf}. Here the stability of the updates in PITs helps in two ways: (i) only leaves get inserted into a PIT and hence the ancestral arrays of a node do not change by much after an insertion; and (ii) a node could only change parent due to edge subdivision, but edge subdivision does not change the DFS leaf ordering $\sigma$. Therefore, we could rely on the data structure of Kopelowitz and Lewenstein~\cite{KL07} to maintain ancestral arrays (and their associated centroid decomposition) under stable updates in $O(\log n)$ time. 

Now, to keep track of the DFS leaf ordering, when a new node $q$ is inserted to a PIT as a leaf child of a node $u$, we will locate the current rightmost active descendant leaf, say $r_u$ of $u$ in $\sigma$---assume for now that $u$ has at least one such leaf--- and then insert $q$ after $r_u$ in $\sigma$. The basic idea is to first find an arbitrary active descendant leaf $x$ of $u$ and then start a binary search procedure to search for $r_u$ in the skip list using ancestral arrays as keys. In the case where the parent $u$ of $q$ does not have any active descendant leaf before inserting $q$, our idea is to search for the lowest ancestor, say $v$, of $u$ that has at least one active descendant leaf, find the leftmost/rightmost active descendant leaf of $v$, and insert $q$ next to the leftmost/rightmost leaf.  Finding an arbitrary active descendant leaf and the lowest active ancestor are rather non-trivial: in the former case,  $u$ might contain up to $\Omega(n)$ inactive descendant leaves, while in the latter case, we might end up checking a large number of ancestors of $u$. Here we develop a new data structure called \EMPH{active tracker} to support both operations. 

With all ideas together, we are able to develop a dynamic data structure for a pairwise tree cover, formally defined in \Cref{def:dynamic_collpit} below, which maintains a set of stable dynamic PITs and their corresponding DFS leaf orderings (using leaf trackers). One important corollary is that the DFS leaf ordering of every stable PIT is also stable: we say that a \EMPH{DFS leaf ordering is stable} if the insertion or deletion of a new active leaf does not change the relative DFS ordering of all existing active leaves. The stability of the DFS leaf ordering is because edge subdivision does not change DFS leaf ordering and inserting a new active leaf does not change the relative order of existing nodes.  See \Cref{fig:organize} for a graphical illustration of all ideas.  

\begin{definition} [Dynamic pairwise tree cover data structure]\label{def:dynamic_collpit} 
    A data structure for maintaining a $(\tau, \varepsilon)$-pairwise tree cover $\mathcal{T}$ and a \EMPH{stable DFS leaf ordering} of every PIT in $\mathcal{T}$, and supporting the following operations:
    \begin{itemize}
        \item \textsc{Insert}$(q,\mathcal{T})$: insert a new point $q$ to $\mathcal{T}$.
        \item \textsc{Delete}$(q, \mathcal{T})$: remove an existing point $q$ from $\mathcal{T}$.
        \item \textsc{GetPredecessor}$(q, i, \mathcal{T})$:  return the predecessor of $q$ in $\sigma_i$ where $\sigma_i$ is the DFS leaf ordering of the $i^{th}$ tree of $\mathcal{T}$.
        The result is null if $q$ is the first element in $\sigma_i$.
        \item \textsc{GetSuccessor}$(q, i, \mathcal{T})$: return successor of $q$ in $\sigma_i$ where $\sigma_i$ is the DFS leaf ordering of the $i^{th}$ tree of $\mathcal{T}$. 
        The result is null if $q$ is the last element in $\sigma_i$.
    \end{itemize}
\end{definition}

Since we maintain DFS leaf orderings by a doubly linked list, we can support querying the predecessor or successor of a point in $O(1)$ time. The following theorem, whose proof will be given in \Cref{sec:dynamic_ptc}, summarizes our main technical result.

\begin{restatable}{theorem}{DynamicCover}\label{thrm:dynamic_collection} Given $\varepsilon> 0$, there is a data structure maintaining $(\tau, \varepsilon)$-pairwise tree cover with $\tau = \eps^{-O(\lambda)}$ supporting \textsc{Insert}/\textsc{Delete} in  $O\left(\varepsilon^{-O(\lambda)}\log(n)\right)$  time per operation, and \textsc{GetPredecessor}/\textsc{GetSuccessor} in $O(1)$ time per operation.
\end{restatable}

As we mentioned above, each ordering in an LSO is basically a DFS ordering of (active) leaves in a PIT. Hence, once we can maintain PITs and their DFS leaf orderings, we could obtain an LSO as a corollary. As the DFS leaf orderings are stable, the LSO we obtain is also stable. 
The following theorem formalizes our results; the proof is rather simple, and will be given in the preliminaries section (\Cref{sec:LSO-from-PIT}).

\begin{restatable}{theorem}{CoverToLSO}\label{thrm:collection_to_lso} If there is a data structure for dynamic $(\tau, \epsilon)$-pairwise tree cover supporting \textsc{Insert}/\textsc{Delete} in $T_1(n,\eps)$ time per operation and \textsc{GetPredecessor}/\textsc{GetSuccessor} in $T_2(n,\eps)$ time per operation, then we can construct a data structure for $(\tau, O(\varepsilon))$-LSO supporting \textsc{Insert}/\textsc{Delete} in $O(T_1(n,O(\eps)))$ time per operation, and \textsc{GetPredecessor}/\textsc{GetSuccessor} in $O(T_2(n,O(\eps)))$ time per operation. Furthermore, the LSO is stable (due to the stability of the DFS leaf orderings in the pairwise tree cover).  
\end{restatable}

We observe that \Cref{thrm:dynamic_lso} follows directly from \Cref{thrm:collection_to_lso} and \Cref{thrm:dynamic_collection}.

\subsection{Applications}\label{subsec:app}

We now give examples of applications of our dynamic LSO in \Cref{thrm:dynamic_lso}. We note that the list of applications mentioned here is not meant to be exhaustive. We believe that LSO could find many more applications in handling dynamic point sets in doubling metrics. As we remarked earlier, the stability of our dynamic LSO is the key to applications. All but the dynamic tree cover application were shown for Euclidean spaces by  Chan, Har-Peled and Jones~\cite{CHJ20}.

\paragraph{Dynamic vertex-fault-tolerant spanners.~} Given a set of points $S$ in a doubling metric of dimension $\lambda$, we denote by $G_S$ the complete graph representing the submetric induced on $S$. A $t$-spanner of $S$ is a spanning subgraph $H$ of $G_S$ such that $d_H(x,y)\leq t\cdot d_{G_S}(x,y)$ for every $x,y\in S$. Given $k\in [1,n-2]$, we say that $H$ is an  \EMPH{$k$-fault-tolerant $t$-spanner}, or $(k,t)$-VFTS for short, if for every subset $F\subset S$  of size at most $k$, called a \EMPH{faulty set}, $H\setminus F$, the graph obtained by removing every vertex in $F$ from $H$, is a $(1+\eps)$ of $S\setminus F$. 

Observe that in a $(k,1+\eps)$-VFTS, every vertex must have a degree at least $k$, and therefore at least $\Omega(nk)$ edges. Levcopoulos, Narasimhan, and Smid~\cite{LNS98} introduced and constructed the first $k$-fault-tolerant $(1+\eps)$-spanner in Euclidean spaces of constant dimensions that has $O(k^2n)$ edges. There was then a long line of work, see e.g.,\cite{Lukovszki99,CZ04,CLN12a,CLNS13,Sol14,LST23}, aiming to improve the Euclidean construction by Levcopoulos, Narasimhan, and Smid as well as extend their result to doubling metrics.  Specifically, in doubling metrics, it is possible to achieve degree bound $O(k)$ and/or $O(nk)$ number of edges~\cite{Sol14,LST23}. Some constructions are \emph{simple} but could only achieve $O(nk)$ number of edges (without any reasonable bound on the degree)~\cite{CLN12a}, or the degree is $\Omega(k^2)$~\cite{CLNS13}; other constructions achieving optimal degree bound of $O(k)$ (for constant $\eps$ and $\lambda$) are sophisticated~\cite{Sol14}. 

A more ambitious goal is to construct an optimal $k$-fault-tolerant $(1+\eps)$-spanner efficiently, even in the static setting.  Solomon~\cite{Sol14} devised an $O(n(\log n + k))$-time algorithm to construct a $k$-fault-tolerant $(1+\eps)$-spanner with degree $O(k)$ (and diameter $O(\log k)$ and lightness $O(k^2\log n)$); the running time is of Solomon's algorithm is optimal in both $n$ and $k$. Solomon's result settled an important open problem raised in the book of Narasimhan and Smid (Problems 26 and 27 in ~\cite{NS07}). Recently, Le,  Solomon, and Than~\cite{LST23} designed a different algorithm with the same running time but achieving both optimal degree and lightness. Given the slow progress on static algorithms, it is understandable that the problem of maintaining a dynamic $(k,t)$-VFTS in doubling metrics under point updates remains wide open.  Even maintaining a dynamic and \EMPH{non-fault-tolerant} spanner, a much simpler problem, proved to be very challenging. Gottlieb and Roddity~\cite{GL08B} were the first to achieve  $O(\log n)$ time per update after several attempts~\cite{Roditty11,GL08,GL08B}. Their dynamic algorithm is much more complicated than its static counterpart~\cite{CGMZ16}. It is, therefore, unlikely that their technique could be extended to handle $(k,t)$-VFTS.

Given our dynamic LSO in \Cref{thrm:dynamic_lso} as a black box, following~\cite{CHJ20}, we obtain a dynamic algorithm for maintaining $(k,1+\eps)$-VFTS in $O_{\lambda, \varepsilon}(\log n + k)$ time per update in a very simple way: for each point $p\in S$, add edges to its $k+1$ predecessors and $k+1$ successors in each ordering of $\Sigma$. As $|\Sigma| = \eps^{-O(\lambda)}$, our dynamic  $(k,1+\eps)$-VFTS spanners achieve both \emph{optimal running time per update}, \emph{optimal degree} (and hence the number of edges), and optimal running time to query all neighbors of a vertex. 

\begin{restatable}{theorem}{VFTS}\label{thm:vfts} Given $\varepsilon \in (0,1)$, $k\in [1,n-2]$ and a dynamic point set $S$ in doubling metrics of dimension $\lambda$, there is a data structure $\mathcal{D}$ such that $\mathcal{D}$ (implicitly) maintains a $(k,1+\eps)$-VFTS $H$ of degree $k\cdot\eps^{-O(\lambda)}$ for $S$ in $O(\log n \eps^{-O(\lambda)})$ time per update, and $\mathcal{D}$ returns all neighbours of a  given vertex of $H$ in $k\varepsilon^{-O(\lambda)}$ time. The update time and query time are optimal for fixed $\eps, \lambda$. 
\end{restatable}

\paragraph{Dynamic tree covers.~} This result is an application of our technique rather than a direct application of LSO. Given a set of points $S$ in a doubling metric $(X,d_X)$, a \EMPH{tree cover} for $S$ is a collection of edge-weighted trees $\mathcal{T}$ such that for every tree $T\in \mathcal{T}$, $S\subseteq V(T)$ and $d_X(x,y)\geq d_T(x,y)$. The \EMPH{size} of the tree cover  $\mathcal{T}$, denoted by $|\mathcal{T}|$, is the number of trees in $\mathcal{T}$. The \EMPH{stretch} of  $\mathcal{T}$ is the smallest $t \geq 1$ such that $d_X(x,y)\leq t\cdot \min_{T\in \mathcal{T}}d_T(x,y)$. Tree covers have been studied extensively both in general metrics~\cite{TZ05} and special metrics, such as Euclidean~\cite{ADMSS95}, planar~\cite{BNF22,CCLMST23}, and doubling~\cite{BNF22,KLMS22}. Tree covers also have many algorithmic applications, such as spanners, routing, and distance oracles; see~\cite{KLMS22} for a thorough discussion.  In doubling metrics, a tree cover for $n$ points could be (statically) constructed in $O(n\log n)$ time~\cite{KLMS22}. However, there is no known dynamic construction of tree covers. Indeed, dynamically maintaining a tree cover is at least as hard as maintaining a dynamic spanner, which, as discussed above, is a difficult problem. Our technique for maintaining a dynamic \emph{pairwise tree cover} could be adapted directly to maintain a tree cover with $O_{\lambda, \varepsilon}(\log n)$ per point update.

\begin{restatable}{theorem}{Treecover}\label{thm:treecover} Given a dynamic point set $S$ in doubling metrics of dimension $\lambda$ and any $\eps\in (0,1)$, there is a data structure $\mathcal{D}_\mathcal{J}$ explicitly maintaining a tree cover $\mathcal{J}$ for $S$ such that $J$ has stretch of $1+\eps$ and size of $\eps^{-O(\lambda)}$, and the running time per update is  $O(\eps^{-O(\lambda)}\log(n))$.
\end{restatable}

\paragraph{Closest pair of points.~} Finding the closest pair in a point set is a very well-studied problem in computational geometry. In Euclidean spaces, there is a vast amount of literature on this problem. In the Euclidean spaces of constant dimension, the dynamic closest pair can be maintained in $O(\log n)$ time per update~\cite{Bespamyatnikh98,GRSS98,CHJ20}. In metrics of bounded doubling dimension, there are two fast static algorithms for finding the closest pair: one based on well-separated pair decomposition (WSPD)~\cite{HM06}, and the other is divide and conquer~\cite{SMM20}. Both algorithms are \EMPH{randomized} and have an expected running time of $O(n\log n)$. Using our dynamic LSO, we could maintain the closest pair in $O(\log n)$ time per update. Applying our dynamic algorithm to the static setting, we obtain a \EMPH{deterministic} algorithm for the closest pair in metrics of bounded doubling dimension in time $O_{\lambda, \varepsilon}(n\log n)$. 

\begin{restatable}{theorem}{CP}\label{thm:closest-pair} Given a dynamic point set  $S$ in doubling metrics of dimension $\lambda$, we construct a data structure for maintaining the closest pair in $S$ in $2^{O(\lambda)}\log(n)$ time per update. 
\end{restatable}

\paragraph{Bichromatic closest pair of points.~} This is another fundamental problem in computational geometry: given two point sets $R$ (red) and $B$ (blue) in a metric space, find the closest pair of points, one red and one blue, among all red-blue pairs of points. In Euclidean metrics, both static and dynamic versions of this problem have been studied extensively (see, e.g.~\cite{Eppstein95} and references therein). However, in doubling metrics, there is no known dynamic algorithm for this problem. Here we use our dynamic LSO to provide the first approximate dynamic algorithm. 

\begin{restatable}{theorem}{BCP}\label{thrm:bichromatic_pair}
    Given a parameter $\varepsilon \in (0, 1)$ and two dynamic point sets $R, B$ in doubling metric of dimension $\lambda$, there is a data structure such that it maintains $(1+\varepsilon)$-closest pair $(r, b)$ where $r \in R, b \in B$, and runs in $O(\varepsilon^{-O(\lambda)}\log(n))$ per update of $R$ or $B$, where $n = |R| + |B|$.
\end{restatable}

\paragraph{Approximate nearest neighbor search.~} One problem that motivated the early study of dynamic algorithms for point sets in doubling metrics is the approximate nearest neighbor: given a query point $p$, find a point $q$ such that  $d_X(p,q) \leq (1+\eps)\min_{x\in X\setminus p}d_X(p,x)$. We would like to design a dynamic data structure that could support fast update time and query time.  The pioneering work of  Krauthgamer and Lee~\cite{KL04} proposed the first dynamic solution for this problem with $O_{\eps,\lambda}(\log \Delta\log\log(\Delta))$ update time and $O(\log \Delta + \eps^{-O(\lambda)})$ query time, which is optimal.  Cole and Gottlieb~\cite{GC06} then removed the dependency on the spread $\Delta$ and improved the update time to $O_{\eps,\lambda}(\log n)$ and the query time to  $O(\log(n) + \eps^{-O(\lambda)})$.  Ideally, we would like the query time to be much faster than the update time; for example, in database applications, querying nearest neighbors is done much more frequently than deleting/inserting points.  Our dynamic LSO in \Cref{thrm:dynamic_lso} gives a simple solution for this problem. For a given query point $p$, the idea is to first insert $p$ to the current LSO of the point set, return the closest neighbor in the orderings of $p$, and then delete $p$ from the LSO. As the number of orderings is $O_{\eps,\lambda}(1)$, and $p$ has at most 2 neighbors per ordering,  the query time is $O_{\eps,\lambda}(1)$, plus the time to insert and delete $p$ from the LSO, which is  $\eps^{-O(\lambda)}\log(n)$. We note that $\log(n)$ query time is optimal for a constant $\eps, \lambda$ for any data structure with linear space~\cite{AMNSW98}. 

\begin{restatable}{theorem}{ANN}\label{thm:ann} Given a dynamic point set  $S$ in doubling metrics of dimension $\lambda$, we can construct a $(1+\varepsilon)$-nearest neighbor data structure for supporting point deletions/insertions in $O(\eps^{-O(\lambda)}\log(n))$ time per update, and $\eps^{-O(\lambda)}\log(n)$ query time.
\end{restatable}

%% file: 2.preliminaries.tex
\subsection{Basic Notation}

Given a metric space $(X, d_X)$, let $\Delta$ be the ratio between the maximum and the minimum distance in the space.
A \emph{ball} of $p$ radius $r$ is a set of all points in distance $r$ from $p$: $B(p, r) = \{q \in X: d(p, q) \leq r\}$. 
$(X, d_X)$ has doubling dimension $\lambda$ if any ball with radius $2r$ can be covered by at most $2^\lambda$ balls of radius $r$. 
The \EMPH{packing property} of a doubling metric states that any set of points with maximum distance $R$ and minimum distance $r$ has at most $\left(\frac{4R}{r}\right)^\lambda$ points.

$Y$ is a \EMPH{$r$-net} of point set $S$ if $Y$ is a subset of $S$ such that: \begin{inlinelist}
    \item for all $x, y \in Y$ and $x\neq y$, $d_X(x, y) > r$ (this property is called \EMPH{packing}) 
    \item for every point $x \in S$, there exists a point $y \in Y$ such that $d_X(x, y) \leq r$ (this property is called \EMPH{covering})
\end{inlinelist}.
\EMPH{Net tree} is a hierarchical tree where: each node has a label where the set of leaf labels is a bijection into $S$, and the set of points at level $i$, denoted as $Y_i$, is $r^i$-net of $Y_{i-1}$.
We denote a node by a pair $(t, i)$, where $t$ is a point in $S$ and $i$ is the level of the node.
Sometimes we simply use $t$ instead of $(t, i)$ when the level is clear from the context.
The distance between two nodes in the net tree means the distance between two points labeled these nodes.

For a dynamic point set $S$, Cole and Gottlieb~\cite{GC06} showed how to construct a net tree with relaxed packing and covering properties: \begin{inlinelist} \item for $x, y, \in Y_i$ and $x \neq y$, $d_X(x, y) > \alpha \frac{1}{\varepsilon^i}$ \item for $x\in Y_{i-1}$, there exists $y \in Y_i$ such that $d_X(x, y) \leq \phi \frac{1}{\varepsilon^i}$ \end{inlinelist}, where $\alpha$ and $\phi$ are some constants.
In this work, we use the notion of \EMPH{$(\delta, \varepsilon)$-net tree} to mention the net tree with relaxed packing and covering properties, and $Y_i$ is the $\frac{\delta}{\varepsilon^i}$-net of $Y_{i-1}$ for any level $i$.

\begin{definition}[$(\delta, \varepsilon)$-net tree]\label{def:netttree}
The $(\delta, \varepsilon)$-net tree is a net tree with packing and covering properties as follows:
    \begin{itemize}
        \item \textnormal{[Packing.]} two nodes $(x, i)$, $(y, i)$ have $d_X(x, y) > \Omega(\frac{\delta}{\varepsilon^i})$.
        \item \textnormal{[Covering.]} if $(x, i)$ is the parent of $(y, i-1)$, then $d_X(x, y) \leq O(\frac{\delta}{\varepsilon^i})$. 
    \end{itemize}        
\end{definition}

\subsection{Dynamic LSO from Dynamic Pairwise Tree Cover}\label{sec:LSO-from-PIT}

We now show how to construct  LSO  from a pairwise tree cover; the proof addresses both static and dynamic settings. 

\CoverToLSO*

\begin{proof}
We show how to construct $(\tau, O(\varepsilon))$-LSO from a $(\tau, \varepsilon)$-pairwise tree cover $\mathcal{T}$, then we obtain a $(\tau, \varepsilon)$-LSO by scaling $\varepsilon$ with a constant factor.

First, we show a static construction.
\begin{quote}
Given a $(\tau, \varepsilon)$-pairwise tree cover $\mathcal{T} = \{T_1, T_2, \ldots, T_\tau\}$, let $\Sigma$ be the set of $\{\sigma_1, \sigma_2, \ldots \sigma_\tau\}$, where $\sigma_i$ is the DFS leaf ordering of $T_i \in \mathcal{T}$.
\end{quote}
By the covering of the pairwise tree cover, for any pair $x, y$ with $d_X(x, y) \in [\frac{\delta}{\varepsilon^i}, \frac{2\delta}{\varepsilon^i}]$, there is a tree $T_j$ such that a node at level $i$ of $T_j$ is $O(\varepsilon)$-close to $(x, y)$. 
Let that node be $(u, v, i)$.
Recall that $C_i(u, v)$ is a set of all leaf labels under the subtree rooted at $(u, v, i)$ and $x, y \in C_i(u, v)$.
By DFS, all points in $C_i(u, v)$ are written consecutively in $\sigma_j$. 
This implies that any point $p$ between $x$ and $y$  in $\sigma_j$ has $d_X(p, x) \leq O(\varepsilon) d_X(x, y)$ or $d_X(p, y) \leq O(\varepsilon) d_X(x, y)$.
Therefore, $\Sigma$ is a $(\tau, O(\varepsilon))$-LSO.

Here is the dynamic maintenance for $\Sigma$: 
\begin{quote}
Suppose that we are given a data structure $\mathcal{D}_\mathcal{T}$ maintaining the $(\tau, \varepsilon)$-pairwise tree cover $\mathcal{T}$ under insertions and deletions.
Our data structure $\mathcal{D}_\Sigma$ maintaining $(
\tau, O(\varepsilon))$-LSO $\Sigma$ dynamically invokes operations of $\mathcal{D}_T$ directly.
To update a point $p$, \textsc{Insert$(p, \Sigma)$} calls \textsc{Insert$(p, \mathcal{T})$},  and \textsc{Delete$(p, \Sigma)$} calls \textsc{Delete$(p, \mathcal{T})$}.
To access orderings, 
we use operations of getting the predecessor or the successor of $\mathcal{D}_\Sigma$, where \textsc{GetPredecessor}$(p, i, \Sigma)$ and \textsc{GetSuccessor}$(p, i, \Sigma)$ of $\mathcal{D}_\Sigma$ return the result of \textsc{GetPredecessor}$(p, i, \mathcal{T})$ and \textsc{GetPredecessor}$(p, i, \mathcal{T})$ respectively.    
\end{quote}

The running time follows directly from the construction. The stability of the LSO follows from that of DFS leaf ordering of $\mathcal{T}$, implying the theorem.
\end{proof}

%% file: 5.static.tex
In this section, we show the static construction for a collection of $(\delta, \varepsilon)$-PITs as claimed in~\Cref{thrm:delta_eps_pits}. Then we construct an $(\varepsilon^{O(-\lambda)}, \varepsilon)$-pairwise tree cover by simply constructing PITs for each $\delta\in \{1,2^{1},2^{2},\ldots, 2^{\lceil \lg(1/\eps) \rceil}\}$.

\begin{restatable}{theorem}{StaticPITs}\label{thrm:delta_eps_pits}
    Given a $(\delta, \varepsilon)$-net tree $T$ of a point set $S$, we can construct from $T$ a collection of $(\delta, \varepsilon)$-PITs $\mathcal{T}$ with  $\varepsilon^{-O(\lambda)}$ trees such that 
    for any pair of points $(x, y)$ whose distance in $[\frac{\delta}{\varepsilon^i}, \frac{2\delta}{\varepsilon^i})$, there exist a PIT  $T'$ in the collection and a node at level $i$ of $T'$ that is $O(\varepsilon)$-close to $(x, y)$. 
\end{restatable}

We call a node in a PIT $T'$ as a \EMPH{pairwise node}, to distinguish with a node in net tree $T$. 
We simply refer to a pairwise node as a node when the tree in the context is a PIT. A pairwise node at level $i$ can be labeled by a single point of the form $(p,p,i)$ or two different points of the form $(x,y,i)$. In the former case, we say that the node has a \EMPH{single-label} and in the latter case, \EMPH{double-label}. For a given point $p \in S$, we define \EMPH{the node (pairwise node) at level $i$ of point $p$} to be the ancestor at level $i$ of the leaf $(p, 0)$ ($(p, p, 0)$, resp.) in the net tree $T$ (PIT  $T'$, resp.).  If $(u, v, i)$ is the pairwise node at level $i$ in $T'$ of $p$, and $p \in Y_{i-1}$, we also say \EMPH{$(u, v, i)$ is the pairwise node in $T'$ of $(p, i-1)$} (in $T$), for some $u, v \in Y_{i-1}$; it could be that $p\not\in  \{u,v\}$, and hence it is \EMPH{not} always the case that the corresponding pairwise node of $(p,i)$ is labeled with the same point $p$.  Observe that level $i$ of a PIT corresponds to level $i-1$ in the corresponding net tree; they are off by one level.

First, we describe intuitively an $O(\varepsilon)$-close node for a pair of points $x_0, y_0$ with $d_X(x_0, y_0) \in [\frac{\delta}{\varepsilon^{i}}, \frac{2\delta}{\varepsilon^{i}})$. We denote by $Y_i$ the set of net points at level $i$; these are points associated with nodes at level $i$ of $T$. Recall that in $(\delta, \varepsilon)$-net tree, given a level $i$, any point $p_0$ in $S$ has a node $(p, i)$ such that $p_0 \in B(p, \frac{2\delta}{\varepsilon^i})$, we say the ball of $(p, i)$ \EMPH{covers $p_0$}.
Let $(x, i-1)$ and $(y, i-1)$ be nodes at level $i-1$ whose balls cover $x_0$ and $y_0$, respectively.  Observe that for every $p \in B(x, \frac{2\delta}{\varepsilon^{i-1}})$, $d_X(x_0, p) \leq \frac{4\delta}{\varepsilon^{i-1}}$ by triangle inequality.
Similarly, for every $p \in B(y, \frac{2\delta}{\varepsilon^{i-1}})$, $d_X(y_0, p) \leq \frac{4\delta}{\varepsilon^{i-1}}$.
Therefore, if we have a pairwise node in a PIT such that its cluster is the union of $B(x, \frac{2\delta}{\varepsilon^{i-1}})$ and $B(y, \frac{2\delta}{\varepsilon^{i-1}})$, this node is $O(\varepsilon)$-close to $(x_0, y_0)$.
To see this, suppose that we have a node $(x, y, i)$ in the PIT and its cluster $C_i(x, y) = B(x, \frac{2\delta}{\varepsilon^{i-1}}) \cup B(y, \frac{2\delta}{\varepsilon^{i-1}})$, then it satisfies: \begin{inlinelist}
\item $x_0, y_0 \in C_i(x, y)$
\item for any $p \in C_i(x, y)$, $d_X(p, x_0)$ or $d_X(p, y_0)$ is at most $\frac{4\delta}{\varepsilon^{i-1}}$, which is $4\varepsilon \frac{\delta}{\varepsilon^{i}}= O(\varepsilon) d_X(x_0, y_0)$.
\end{inlinelist} 

Now, we sketch our main idea for the static construction.
Observe by the triangle inequality, $||d_X(x, y) - d_X(x_0, y_0)|| \leq d_X(x, x_0) + d_X(y, y_0)$, which means $d_X(x, y) \in \left[\frac{\delta}{\varepsilon^{i}} - \frac{4\delta}{\varepsilon^{i-1}}, \frac{2\delta}{\varepsilon^{i}} + \frac{4\delta}{\varepsilon^{i-1}}\right)$.
To have $O(\varepsilon)$-close nodes for all pairs with the distance in $\left[\frac{\delta}{\varepsilon^i}, \frac{2\delta}{\varepsilon^i}\right)$, we consider all pairs $u,v$ of $Y_{i-1}$, for each pair if $d_X(u, v)  \in \left[\frac{\delta}{\varepsilon^{i}} - \frac{4\delta}{\varepsilon^{i-1}}, \frac{2\delta}{\varepsilon^{i}} + \frac{4\delta}{\varepsilon^{i-1}}\right)$, then we create a pairwise node $(u, v, i)$.
Next, we arrange these nodes into PITs such that: 
\begin{inlinelist}
\item for each PIT, every point in $Y_{i-1}$ belongs to the label of at most one level-$i$ pairwise node
\item clusters of nodes at level $i$ are the union of some clusters of nodes at level $i-1$
\item these level-$i$ clusters are disjoint. 
\end{inlinelist}
Then, in each PIT, for any point $p \in Y_{i-1}$ that does not belong to a cluster of any pairwise node at level $i$, we create a single-label node $(p, p, i)$ in that PIT. Finally, we create edges connecting pairwise nodes at level $i$ and level $i-1$: if the cluster of a node $u$ at level $i-1$ is a subset of the cluster of a node $v$ at level $i$, $v$ becomes the parent of $u$.

Assigning points to nodes requires careful attention. To guarantee that all clusters are disjoint as specified by the condition (iii) of arranging nodes, we maintain a property that for each PIT, given any two points $u, v \in Y_{i-1}$ which are in (the same or different) double-label nodes at level $i$, $d_X(u, v) > \frac{8\delta}{\varepsilon^{i-1}}$. 
Here, we reuse the red-blue matching algorithm of Filtser and Le~\cite{Filtser23} to determine which pairwise nodes could be placed in the same PIT while guaranteeing the three conditions above.

\begin{figure}[!ht]
    \centering        
    \includegraphics[width=1\linewidth]{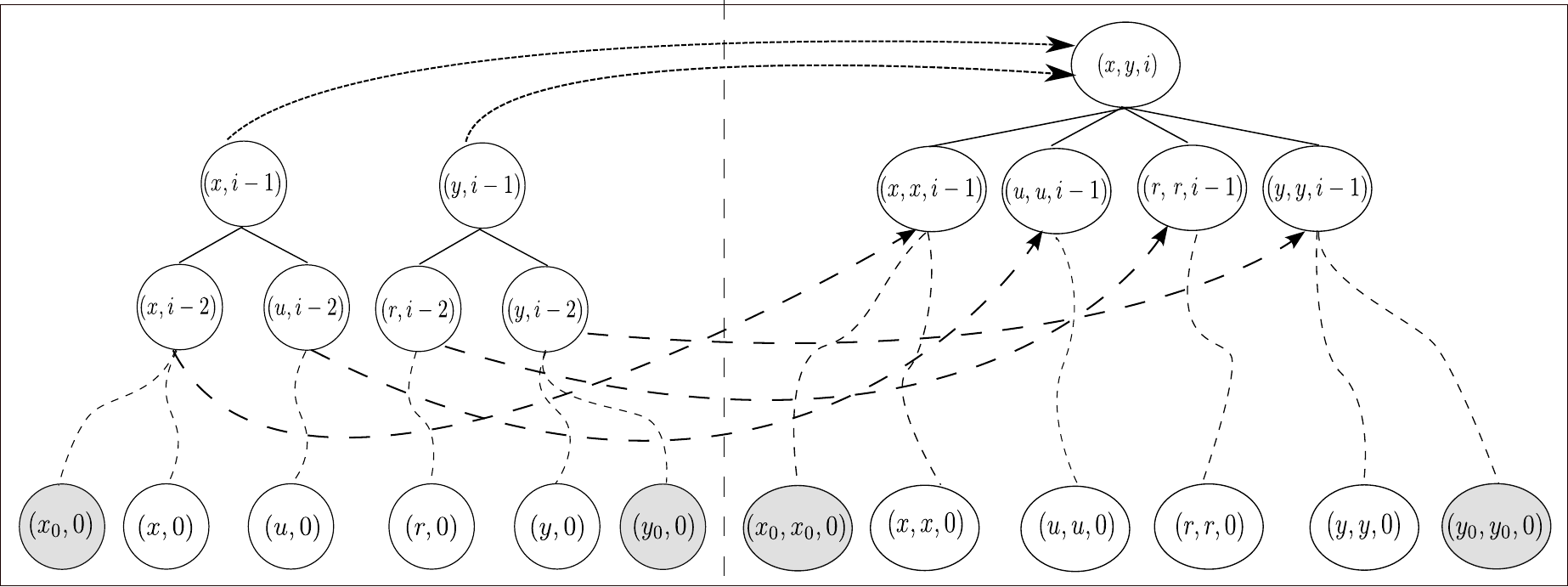}
    \caption{Illustrating a net tree $T$ (left), a PIT $T'$ derived from $T$ (right). We create $(x, y, i)$ by pairing up $(x, i-1)$ and $(y, i-1)$ in $T$, and this nodes is $O(\varepsilon)$-close to the pair $(x_0, y_0)$. The dashed arrows show corresponding single-label pairwise nodes, and the dot arrows show corresponding double-label pairwise nodes. } 
    \label{fig:net_pit}
\end{figure}

In the end, we obtain a collection of PITs $\mathcal{T}$, and the levels of PIT in  $\mathcal{T}$ are off from the levels of the net tree by $1$. At the leaf level, pairwise nodes are single-label, and these labels are exactly the points in $S$.  For a level $i > 0$, we construct pairwise nodes from the nodes of $T$  at level $i-1$, and they could be single-label or double-label depending on how we pair up points in $Y_{i-1}$. Some points in  $Y_{i-1}$ might not appear in the labels of pairwise nodes at level $i$ of a PIT. 
See~\Cref{fig:net_pit}.

Now, we show details of the static construction. We need the following lemma of Filtser and Le~\cite{FL22}:

\begin{lemma}[\cite{FL22}]\label{lmm:matching_alg}
    Consider a graph $G = (V, E_b \cup E_r)$ that consists of disjoint edge sets called blue and red respectively. Let $\delta_r \geq 1$ ($\delta_b >1)$ be the maximal red (blue) degree. There exists a set $\mathcal{M}$ of $O(\delta_b\delta_r)$ maximal matchings such that: 1) their union covers all blue edges; 2) there is no red edges whose both endpoints are matched by any matching in $\mathcal{M}$.
\end{lemma}

The algorithm to construct $\mathcal{M}$ of Filtser and Le~\cite{FL22} works roughly as follows. Let $\mathcal{M}$ be the collection of maximal matchings, initially empty.
Let $B$ be the set of blue edges remaining uncovered in $\mathcal{M}$, initially $B = E_b$.
We repeat the following process until $B$ is empty: 1) greedily find $M$ such that it is a maximal matching of $B$ and there are no red edges whose endpoints are matched in $M$, 2) add $M$ to $\mathcal{M}$, 3) remove edges in $M$ out of $B$. 
We refer readers to the work of Filtser and Le~\cite{FL22} for the analysis of the properties $\mathcal{M}$. We call this algorithm \EMPH{red-blue} matching. 

\subsection{The Static Construction} \label{subsec:PIT-static-const}

Given parameters $\delta > 0$, $\varepsilon < \frac{1}{16}$, a  $(\delta, \varepsilon)$-net tree $T$, our construction proceeds as follows. Initially, the collection has $\varepsilon^{-O(\lambda)}$ trees, each tree has a level $0$ such that the set of leaf labels is a bijection with the point set $S$. We construct trees in $\mathcal{T}$ by visiting $T$ in bottom-up order.

\begin{figure}[!ht]
\centering
\includegraphics[width=0.8\linewidth]{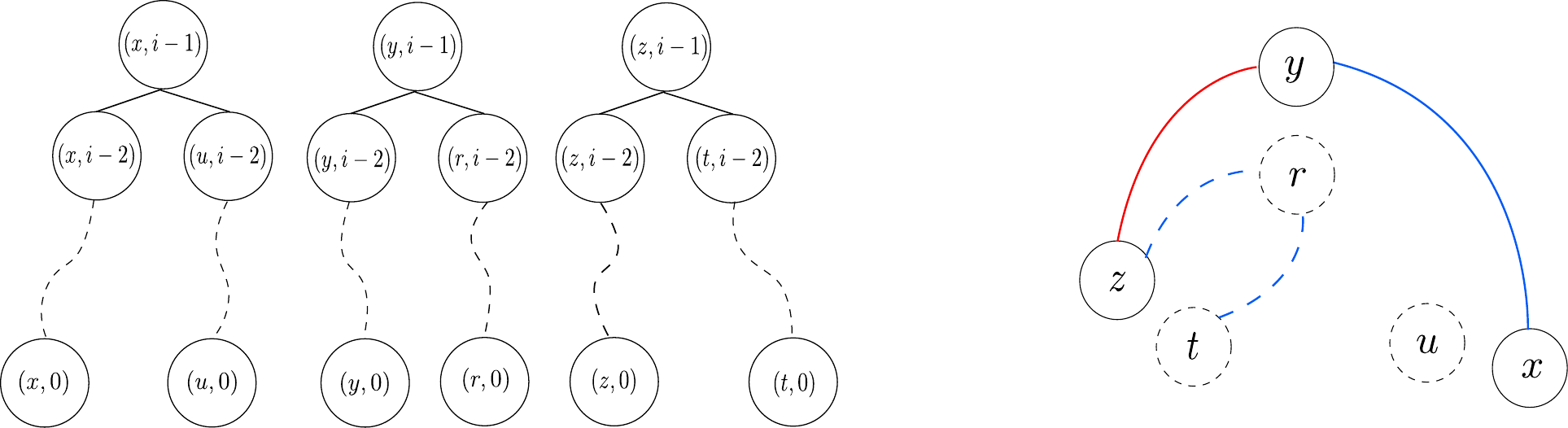}
\caption{Examples of red-blue graphs created from Step 1. The left figure is the net tree $T$, and the right figure includes $G_{i-1}$ and $G_i$. $G_i$ consists of solid edges and nodes; $G_{i-1}$ consists of dashed edges, and its vertex set includes dashed and solid nodes. The only blue edge in $G_i$ is $(x, y)$, while $G_{i-1}$ contains $(r, z)$ and $(r, t)$ as  blue edges.}
    \label{fig:matching_graph}
\end{figure}

\textbf{[Step 1 - Create matchings]}
To create pairwise nodes at level $i$ for all trees in the collection, let $Y_{i-1}$ be the set of all points at level $i-1$ of the net tree $T$. We define two important parameters: range 
$R_i = \left[(1-4\varepsilon)\frac{\delta}{\varepsilon^{i}}, (1+2\varepsilon)\frac{2\delta}{\varepsilon^{i}}\right)$, and threshold 
$s_i = \frac{10\delta}{\varepsilon^{i-1}}$.  Let $G_i = (V_i, E_b \cup E_r)$ be the graph where the vertex set is $V_i = Y_{i-1}$, $E_b = \{(u, v) \in V_i \times V_i: d_X(u, v) \in R_i\}$, $E_r = \{(u, v) \in V_i \times V_i: d_X(u, v) < s_i\}$. This graph consists of blue edges ($E_b$) and red edges ($E_r$), where blue edges contain pairs of points in $Y_{i-1}$ whose distances are in $R_i$, and red edges contain pairs whose distances are less than $s_i$. Since $\varepsilon < \frac{1}{16}$, we have $\frac{10\delta}{\varepsilon^{i-1}} < (1-4\varepsilon)\frac{\delta}{\varepsilon^i}$, which means an edge could be only red or blue. Applying the red-blue matching algorithm in~\Cref{lmm:matching_alg},  we obtain a set of matchings $\mathcal{M}_i$.  See~\Cref{fig:matching_graph}.

\begin{figure}[!ht]
    \begin{subfigure}{\textwidth}
    \centering     \includegraphics[width=\linewidth]{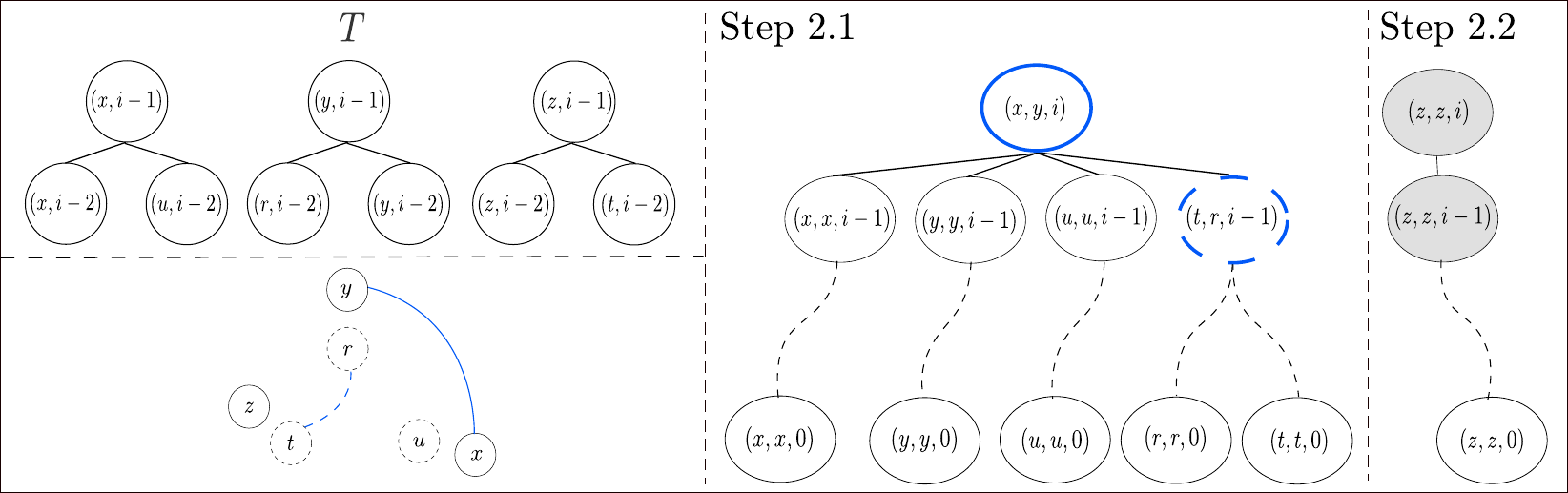}
    \caption{}
    \end{subfigure}
    \begin{subfigure}{\textwidth}
    \centering        \includegraphics[width=\linewidth]{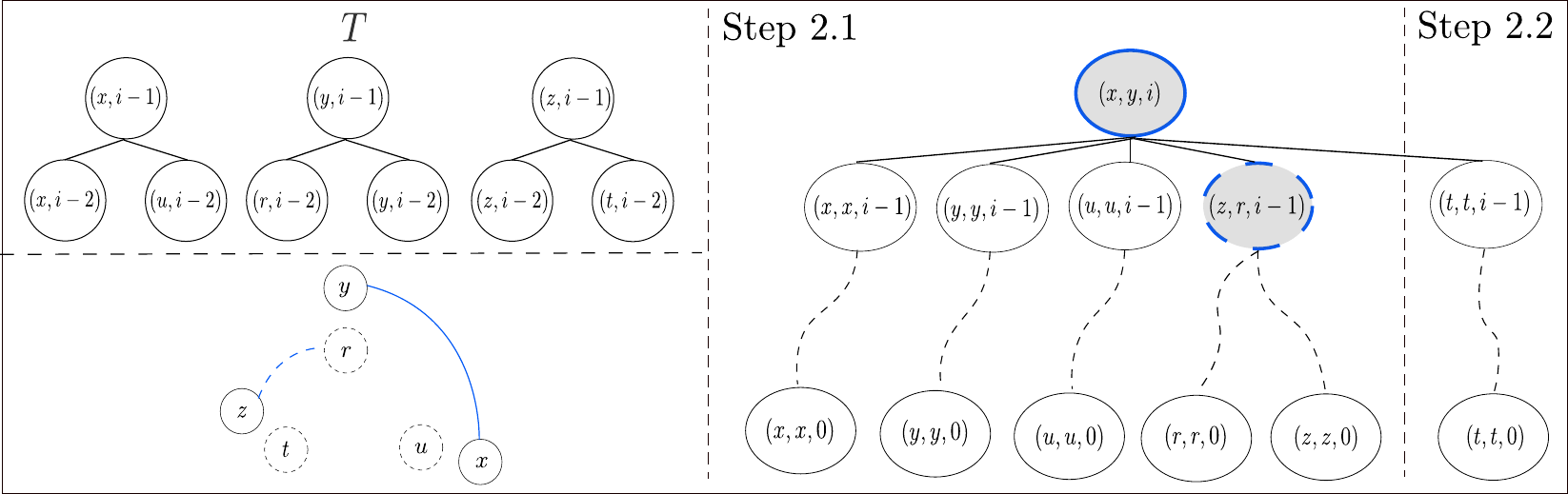}
    \caption{}
    \end{subfigure}   
    \caption{Illustration for step 2 - creating pairwise nodes. Depending on matched edges in $G_i$ and pairwise nodes at level $i-1$, we have different structures of PITs, as shown in figures (a) and (b). 
    Blue lines are matched edges, and  bold blue nodes are pairwise nodes created from these edges.
    The filled nodes are pairwise nodes of unmatched points, and specifically, $(x, y, i)$ is the pairwise node of unmatched point $(z, i-1)$.
    }
    \label{fig:pit_cons}
\end{figure}

\textbf{[Step 2 - Create pairwise nodes at level $i$]} 
For each matching in $\mathcal{M}_i$, we will create pairwise nodes at level $i$ for the corresponding PIT in $\mathcal{T}$, then find children for these nodes as described below.
Let $M_j^i$ be the $j^{th}$ matching of $\mathcal{M}_i$, and $T_j'$ be the $j^{th}$ PIT of $\mathcal{T}$.
Suppose by induction that we have already added pairwise nodes at level $i-1$ for all PITs in $\mathcal{T}$ from $T$ nodes at level $i-1$.
At this point, each PIT is a forest.
We add pairwise nodes at level $i$ to $T'_j$ from $M_j^i$ as follows:

\begin{itemize}
    \item \textbf{[Step 2.1 - Matched nodes]}  For each edge $(x, y) \in M_j^i$, we create a node $(x, y, i)$ in $T_j'$, and then assign $(x, y, i)$ as the \EMPH{corresponding pairwise node} of $(x, i-1)$ and of $(y, i-1)$. 
If $i \geq 2$, for $u \in Y_{i-2}$ such that $d_X(u, x) \leq \frac{3\delta}{\varepsilon^{i-1}}$ or $d_X(u, y) \leq \frac{3\delta}{\varepsilon^{i-1}}$ ($u$ might be $x$ or $y$), let $(u_1, u_2, i-1)$ be the corresponding pairwise node at level $i-1$ of $(u, i-2)$. We then set $(u_1, u_2, i-1)$ to be a child of $(x, y, i)$.
In~\Cref{lmm:pit_tree} below, we show that $(u_1, u_2, i-1)$ will not be set as a child of another pairwise node $(x', y', i)$ created from another matched edge $(x', y') \in M_j^i$ using the fact that no red edge has both endpoints matched by $M_{j}^i$. See~\Cref{fig:pit_cons} for an illustration. 

Now, children of $(x, y, i)$ include the corresponding pairwise nodes of $(x, i-1)$'s children, $(y, i-1)$'s children since if $(v, i-2)$ is a child of $(x, i-1)$ or $(y, i-1)$ then $d_X(v, \{x, y\}) \leq \frac{\delta}{\varepsilon^{i-1}}$ by the covering property of $T$. Children of $(x, y, i)$ might also contain corresponding pairwise nodes of children of some unmatched nodes. 

\item \textbf{[Step 2.2 - Unmatched nodes]} After going through all edges in $M_j^i$, we consider unmatched net point $z$ in $Y_{i-1}$.  

\begin{itemize}
    \item\textbf{[2.2.1]} We create the corresponding pairwise node in $T_j'$ for $(z, i-1)$ by considering the corresponding pairwise node of $(z, i-2)$:
    \begin{itemize}
        \item[(a)] If $i = 1$, $(z, i-2)$ does not exist, then we create $(z, z, i)$ as the pairwise node of $(z, i-1)$ in $T_j'$.
        \item[(b)] For $i > 1$, let $(z_1, z_2, i-1)$ be the corresponding pairwise node of $(z, i-2)$ in $T_j'$. If $(z_1, z_2, i-1)$ does not have a parent, then we create $(z, z, i)$ as the corresponding pairwise node of $(z, i-1)$ in $T_j'$.
        \item[(c)] If both (a) and (b) do not hold, meaning that $i\not=i$ and $(z_1, z_2, i-1)$ has a parent in $T_j'$, denoted by $(z_1', z_2', i)$, then we assign $(z_1', z_2', i)$ as the corresponding pairwise node of $(z, i-1)$ in $T_j'$.
    \end{itemize} 
    
    \item\textbf{[2.2.2]} For any child of $(t, i-2)$ of $(z, i-1)$ (this case only happens when $i > 1$), if the corresponding pairwise node of  $(t, i-2)$ in $T_j'$, denoted by $(t_1, t_2, i-1)$, has not been assigned a parent, then we make $(t_1, t_2, i-1)$ a child of the corresponding pairwise node of  $(z, i-1)$. Otherwise, we leave $(t_1, t_2, i-1)$ as it is.

\end{itemize}
\end{itemize}

\subsection{The Analysis}

In this section, we will analyze the properties of the PITs constructed in the previous section, and specifically, the packing and covering properties as defined in \Cref{def:pit} and the pairwise covering property of the pairwise tree cover as defined in \Cref{def:collect_pit}. When $(p,i-1)$ is unmatched by a matching and its level is clear from context, we will refer to $p$ as an \EMPH{unmatched point}. First, we observe that:

\begin{observation}\label{obs:label_distance}
    For any pairwise node $(u_1, u_2, i)$, $d_X(u_1, u_2) \leq (1+2\varepsilon)\frac{2\delta}{\varepsilon^i}$.
\end{observation}
\begin{proof}
    If $u_1 = u_2$, then $d_X(u_1, u_2) = 0$. If $u_1 \neq u_2$, $(u_1, u_2)$ must be a blue edge in $G_i$, thus $d_X(u_1, u_2) \leq (1+2\varepsilon)\frac{2\delta}{\varepsilon^i}$.
\end{proof}

\begin{observation}\label{obs:step_222c}
    In step 2.2.1 case (c), $(z_1, z_2, i-1)$ already has a parent, which is found in step 2.1.
\end{observation}
\begin{proof}
    We consider all matched nodes first in step 2.1, then unmatched nodes later in step 2.2. 
    In each step, we find parents for some pairwise nodes at level $i-1$ in $T'_j$. In case (c) of step 2.2.1, the corresponding pairwise node $(z_1, z_2, i-1)$ of $(z, i-2)$ already has a parent, which must be found by step 2.1. Thus the observation follows.
\end{proof}

In the next two lemmas, we will show basic facts about the PITs.

\begin{lemma}\label{lmm:cover_netpoint}
For any point $p \in Y_{i-1}$, let $(u_1, u_2, i)$
be the corresponding pairwise node of $p$ at level $i$ in $T_j'$, 
then $d_X(p, \{u_1, u_2\}) = \min\{d_X(p, u_1), d_X(p, u_2)\} \leq \frac{6\delta}{\varepsilon^{i-1}}$.
\end{lemma}

\begin{proof}
    If the corresponding pairwise node at level $i$ of $(p, i-1)$ is labeled by $p$, which means $p \in \{u_1, u_2\}$, then $d_X(p, \{u_1, u_2\}) = 0$.
    We remain to consider the case $p \not\in \{u_1, u_2\}$. 
    This occurs in step 2.2.1 case (c), when $p$ is unmatched by $M_j^i$ and the corresponding pairwise node of $(p, i-2)$, denoted by $(t_1, t_2, i-1)$, has a node $(u_1, u_2, i)$ as the parent for $i \geq 2$. Note that $t_1$ and $t_2$ could be the same or different points. 
    By~\Cref{obs:step_222c}, $(t_1, t_2, i-1)$ become a child of $(u_1, u_2, i)$ in step 2.1. Thus there exists $t \in Y_{i-2}$ such that $d_X(t,\{u_1, u_2\}) \leq \frac{3\delta}{\varepsilon^{i-1}}$ and $(t_1, t_2, i-1)$ is the corresponding pairwise node in $T_j'$ of $(t, i-2)$.
    Now, we prove that $d_X(p, \{u_1, u_2\}) \leq \frac{6\delta}{\varepsilon^{i-1}}$. By induction, $d_X(p, \{t_1, t_2\})$ and $d_X(t, \{t_1, t_2\})$ are at most $\frac{6\delta}{\varepsilon^{i-2}}$, since $(t_1, t_2, i-1)$ is the corresponding pairwise node in $T_j'$ of both $(p, i-2)$ and $(t, i-2)$.
    We have $d_X(t_1, t_2) \leq  (1+2\varepsilon)\frac{2\delta}{\varepsilon^{i-1}}$ by~\Cref{obs:label_distance}.
    Thus:
    \begin{equation*}
        \begin{aligned}
            d_X(p,t) &\leq d_X(p, \{t_1, t_2\}) + d_X(t_1, t_2) + d_X(\{t_1, t_2\}, t) \\
            &\leq \frac{6\delta}{\varepsilon^{i-2}} + (1+2\varepsilon)\frac{2\delta}{\varepsilon^{i-1}} + \frac{6\delta}{\varepsilon^{i-2}} \\
            &= \frac{2\delta}{\varepsilon^{i-1}} + \frac{16\delta}{\varepsilon^{i-2}} < \frac{3\delta}{\varepsilon^{i-1}} \textrm{\quad (since $\varepsilon < \frac{1}{16}$)}
        \end{aligned}
    \end{equation*}

    Finally, by triangle inequality, we have: 
    \begin{equation*}
    \begin{aligned}
    d_X(p, \{u_1, u_2\}) &\leq d_X(p,t) + d_X(t, \{u_1, u_2\}) \\
    &\leq \frac{3\delta}{\varepsilon^{i-1}} + \frac{3\delta}{\varepsilon^{i-1}} {\quad (d_X(t, \{u_1, u_2\}) \leq \frac{3\delta}{\varepsilon^{i-1}})}\\
    &= \frac{6\delta}{\varepsilon^{i-1}}        
    \end{aligned}
    \end{equation*}
    
    This completes the proof.
\end{proof}

\begin{lemma}\label{lmm:pit_tree}
    Consider step 2.1 of the construction, where we create pairwise nodes at level $i$ for $T_j'$ from edges matched in $M_j^i$. Let $(u_1, u_2, i-1)$ be a pairwise node at level $i-1$ of $T_j'$. There exists at most one matched blue edge $(x, y) \in M_j^i$ such that $(u_1, u_2, i-1)$ can be assigned as a child of the pairwise node $(x, y, i)$ created from the blue edge $(x, y)$.
\end{lemma}

\begin{proof}
    For contradiction, suppose there exists an edge $(x', y') \in M_j^i$ such that
    $(u_1, u_2, i-1)$ can be assigned as a child of both $(x, y, i)$ and $(x', y', i)$.
    By step 2.1, $(u_1, u_2, i-1)$ can be a child of $(x, y, i)$ if there exists $u \in Y_{i-2}$ such that $d_X(u, \{x, y\}) \leq \frac{3\delta}{\varepsilon^{i-1}}$, and $(u_1, u_2, i-1)$ is the corresponding pairwise node of $(u, i-2)$ in $T'_j$.
    Similarly, there exists $u' \in Y_{i-2}$ such that $d_X(u', \{x', y'\}) \leq \frac{3\delta}{\varepsilon^{i-1}}$, and $(u_1, u_2, i-1)$ is the corresponding pairwise node of $(u', i-2)$ in $T'_j$.
    First, we bound $d_X(u, u')$.
    By~\Cref{lmm:cover_netpoint}, 
    $d_X(u, \{u_1, u_2\}) \leq \frac{6\delta}{\varepsilon^{i-2}}$ and $d_X(u', \{u_1, u_2\})\leq \frac{6\delta}{\varepsilon^{i-2}}$.
    By~\Cref{obs:label_distance}, $d_X(u_1, u_2) \leq (1+2\varepsilon)\frac{2\delta}{\varepsilon^{i-1}}$.
    We obtain:
    \begin{equation}
        \begin{aligned}\label{equ:lmm3}
            d_X(u, u') &\leq d_X(u, \{u_1, u_2\}) + d_X(u_1, u_2) + d_X(\{u_1, u_2\}, u') \\
            &\leq \frac{6\delta}{\varepsilon^{i-2}} + (1+2\varepsilon)\frac{2\delta}{\varepsilon^{i-1}} + \frac{6\delta}{\varepsilon^{i-2}} \\
            &= \frac{2\delta}{\varepsilon^{i-1}} + \frac{16\delta}{\varepsilon^{i-2}}
        \end{aligned}
    \end{equation}
    Since $d_X(u, \{x, y\}) \leq \frac{3\delta}{\varepsilon^{i-1}}$ and $d_X(u', \{x, y\}) \leq \frac{3\delta}{\varepsilon^{i-1}}$, and by triangle inequality, we have:
    \begin{equation*}
        \begin{aligned}
            d_X(\{x, y\}, \{x', y'\}) &\leq d_X(\{x, y\}, u) + d_X(u, u') + d_X(u', \{x', y'\})\\
            &\leq \frac{3\delta}{\varepsilon^{i-1}} + \frac{2\delta}{\varepsilon^{i-1}} + \frac{16\delta}{\varepsilon^{i-2}} + \frac{3\delta}{\varepsilon^{i-1}} \textrm{\quad (By~\Cref{equ:lmm3})}\\
            &= \frac{8\delta}{\varepsilon^{i-1}} + \frac{16\delta}{\varepsilon^{i-2}} <\frac{10\delta}{\varepsilon^{i-1}}
            \textrm{ \quad (since $\varepsilon < \frac{1}{16})$}
        \end{aligned}
    \end{equation*}
    It follows that there is a red edge connecting a point in $\{x, y\}$ and a point in $\{x', y'\}$.
    Since $x$ and $y$ are matched, $(x', y')$ does not exist in $M_j^i$ by the red-blue matching algorithm, contradicting the assumption that $(x', y') \in M_j^i$.
\end{proof}

We are now ready to show the packing and covering of a PIT using \Cref{lmm:cover_netpoint} and \Cref{lmm:pit_tree} above.

\begin{lemma}\label{lmm:pit_packcover}
    Each tree $T'_j$ in the collection $\mathcal{T}'$ satisfies  packing and covering properties of PIT as defined in~\Cref{def:pit}: \textnormal{[packing]} for any two nodes $(x, y, i)$ and $(u, v, i)$, the distance between any pair of points in $\{x, y,u,v\}$ is $\Omega(\frac{\delta}{\varepsilon^{i-1}})$; \textnormal{[covering]} (i) label points $p$ in children of $(x, y, i)$ 
    has $d_X(p, \{x, y\}) \leq \frac{6\delta}{\varepsilon^{i-1}}$, and (ii) diameter of the cluster of a node at level $i$ is bounded by  $\frac{6\delta}{\varepsilon^{i-1}}$.  
\end{lemma}

\begin{proof}
Since $x, y, u, v \in Y_{i-1}$, by the packing property of net tree $T$, the distance between any pair of points in $\{x, y, u, v\}$ is $\Omega(\frac{\delta}{\varepsilon^{i-1}})$. This implies that $T_j'$ has the packing property of PITs.

To prove (i) of the covering, 
we consider a pairwise node $(u_1, u_2, i-1)$ of $T'_j$ for $u_1, u_2 \in Y_{i-2}$ and its parent $(x, y, i)$  for $x, y \in Y_{i-1}$.

If $(u_1, u_2, i-1)$ becomes a child of $(x, y, i)$ in step 2.1, then $(x, y) \in M_j^i$ and there exists $u\in Y_{i-2}$ such that $(u_1, u_2, i-1)$ is the pairwise node of $(u, i-2)$, and $d_X(u, \{x, y\}) \leq \frac{3\delta}{\varepsilon^{i-1}}$.
Now we bound $d_X(u_1, \{x, y\})$ and $d_X(u_2, \{x, y\})$ by $d_X(u, \{u_1, u_2\})$. 
By~\Cref{lmm:cover_netpoint}, $d_X(u, \{u_1, u_2\}) \leq \frac{6\delta}{\varepsilon^{i-2}}$.
By~\Cref{obs:label_distance}, $d_X(u_1, u_2) \leq (1+2\varepsilon)\frac{2\delta}{\varepsilon^{i-1}}$.
Thus:
\begin{equation*}
    \begin{aligned}
        d_X(u_1, \{x, y\}) &\leq d_X(u_1, u) + d_X(u, \{x, y\}) \\
        &\leq d_X(u_1, u_2) + d_X(\{u_1, u_2\}, u) + d_X(u, \{x, y\}) \\
        &\leq (1+2\varepsilon)\frac{2\delta}{\varepsilon^{i-1}} + \frac{6\delta}{\varepsilon^{i-2}} + \frac{3\delta}{\varepsilon^{i-1}}\\
        &\leq \frac{6\delta}{\varepsilon^{i-1}} \textrm{\quad (since $\varepsilon < \frac{1}{16}$)}
    \end{aligned}
\end{equation*}
By the same argument, we get $d_X(u_2, \{x, y\}) \leq \frac{6\delta}{\varepsilon^{i-1}}$.

Now we consider when $(u_1, u_2, i-1)$ is assigned as a child of $(x, y, i)$ in step 2.2.
Observe that $(u_1, u_2, i-1)$ must be the corresponding pairwise node of $(t, i-2)$ where $(t, i-2)$ is a child of an unmatched node $(z, i-1)$.
If $t = z$, we must create $(z, z, i)$ as the corresponding pairwise node of $(z, i-1)$ in step 2.2.1, thus $x = y = z$ and $d_X(t, \{x, y\}) = 0$.
If $t \neq z$, the corresponding pairwise node of $(z, i-1)$ has two cases:
\begin{itemize}
    \item If $z = x = y$, which means we create $(z, z, i)$, then $d_X(t, \{x, y\}) = d_X(t, z) \leq \frac{\delta}{\varepsilon^{i-2}}$ by the covering property of $T$.
    \item If $z \neq x$ and $z \neq y$, then the corresponding pairwise node of $(z, i-2)$, say $(z_1, z_2, i-1)$, must be assigned as a child of $(x, y, i)$ in step 2.1. This means $d_X(z, \{x, y\}) \leq \frac{3\delta}{\varepsilon^{i-1}}$. We have $d_X(t, z) \leq \frac{\delta}{\varepsilon^{i-2}}$ by the covering property of $T$. Therefore, $d_X(t, \{x, y\}) \leq d_X(t, z) + d_X(z, \{x, y\}) \leq \frac{3\delta}{\varepsilon^{i-1}} + \frac{\delta}{\varepsilon^{i-2}}$.
\end{itemize}

In any case, we obtain $d_X(t, \{x, y\}) \leq \frac{3\delta}{\varepsilon^{i-1}} + \frac{\delta}{\varepsilon^{i-2}}$.
By~\Cref{lmm:cover_netpoint}, $d_X(t, \{u_1, u_2\}) \leq \frac{6\delta}{\varepsilon^{i-2}}$. By~\Cref{obs:label_distance}, $d_X(u_1, u_2) \leq (1+2\varepsilon)\frac{2\delta}{\varepsilon^{i-1}}$.
Thus:
\begin{equation*}
    \begin{aligned}
        d_X(u_1, \{x, y\}) &\leq d_X(u_1, t) + d_X(t, \{x, y\}) \\
        &\leq  d_X(u_1, u_2) + d_X(t, \{u_1, u_2\}) +  d_X(t, \{x, y\}) \\
        &\leq (1+2\varepsilon)\frac{2\delta}{\varepsilon^{i-1}} + \frac{6\delta}{\varepsilon^{i-2}} + \frac{3\delta}{\varepsilon^{i-1}} + \frac{\delta}{\varepsilon^{i-2}}  \\
        &\leq \frac{6\delta}{\varepsilon^{i-1}} \textrm{\quad (since $\varepsilon < \frac{1}{16}$)}
    \end{aligned}
\end{equation*}

By the same argument, $d_X(u_2, \{x, y\}) \leq \frac{6\delta}{\varepsilon^{i-1}}$, giving item (i) of the covering property. 

Finally, we bound the cluster-diameter.
Recall that $C_{i}(x, y)$ is the set of leaves in the subtree rooted at $(x, y, i)$.
We denote by $diam(C_i(x, y))$ the diameter of $C_i(x, y)$.
To bound $diam(C_i(x, y))$, we consider $d_X(x, y)$, the distance from $\{x, y\}$ to labels of children of $(x, y, i)$, and the cluster-diameter of children nodes.
Let $(u, v, i-1)$ be a child of $(x, y, i)$.
We have $d_X(x, y) \leq (1+2\varepsilon)\frac{2\delta}{\varepsilon^i}$ by \Cref{obs:label_distance}, and $d_X(u, \{x, y\})$ and $d_X(v, \{x, y\})$ are at most $\frac{6\delta}{\varepsilon^{i-1}}$  by item (i) of the covering property of $T_j'$. By induction, suppose that $diam(C_{i-1}(u, v)) \leq \frac{6\delta}{\varepsilon^{i-1}}$.
By triangle inequality, we obtain:
\begin{equation*}\label{equ:pit_diam}
    \begin{aligned}
        diam(C_i(x, y)) &\leq d_X(x, y) + 2\cdot \max_{\textrm{a child $(u, v, i-1)$ of $(x, y, i)$}}\left\{d_X(\{u, v\}, \{x, y\}) + diam(C_{i-1}(u, v))\right\}
        \\
        &\leq (1+2\varepsilon)\frac{2\delta}{\varepsilon^i} + 2(\frac{6\delta}{\varepsilon^{i-1}} + \frac{6\delta}{\varepsilon^{i-1}}) \\
        &= \frac{2\delta}{\varepsilon^i} + \frac{28\delta}{\varepsilon^{i-1}} 
        < \frac{6\delta}{\varepsilon^i} \textrm{\quad (since $\varepsilon < \frac{1}{16})$}
    \end{aligned}
\end{equation*}
This completes the proof.
\end{proof}

Now we prove our main theorem of this section.

\begin{proof}[Proof of~\Cref{thrm:delta_eps_pits}]
Let $\mathcal{T}$ be the collection of PITs obtained by running the static construction in the previous algorithm to every $(\delta, \varepsilon)$-net tree $T$ with $\delta\in \{1,2^{1},2^{2},\ldots, 2^{\lceil \lg(1/\eps) \rceil}\}$. We have shown in \Cref{lmm:pit_packcover} that every PIT in $\mathcal{T}$ satisfies the packing and covering property. 

To bound the number of trees in $\mathcal{T}$, recall that for each level $i-1$ of the net tree $T$, we create a graph $G_i$ and run the matching algorithm. By~\Cref{lmm:matching_alg}, the algorithm returns $\mathcal{M}_i$ with $|\mathcal{M}_i| = \varepsilon^{-O(\lambda)}$. We create pairwise nodes of $T'_j$ by the mathching $j^{th}$ of $\mathcal{M}_i$, thus $\mathcal{T}'$ has $\varepsilon^{-O(\lambda)}$ PITs.

It remains to show the pairwise covering property of $\mathcal{T}$ as defined in \Cref{def:collect_pit}. Consider two points $x_0, y_0$ with $d_X(x_0, y_0)\in \left[\frac{\delta}{\varepsilon^i}, \frac{2\delta}{\varepsilon^{i}}\right)$ for an integer $i$.
By the covering property of $(\delta, \varepsilon)$-net tree,
the distance of a node at level $i-1$ to its children is at most $\frac{\delta}{\varepsilon^{i-1}}$.
This implies the distance of a node at level $i-1$ in $T$ to its descendants is at most $\frac{2\delta}{\varepsilon^{i-1}}$.
Therefore, there are two nodes $(x, i-1), (y, i-1)$ in $T$ such that $d_X(x, x_0) \leq \frac{2\delta}{\varepsilon^{i-1}}$ and $d_X(y, y_0) \leq \frac{2\delta}{\varepsilon^{i-1}}$. 
By triangle inequality, 
$||d_X(x, y) - d_X(x_0, y_0)|| \leq d_X(x, x_0) + d_X(y, y_0)$.
It follows that  $d_X(x, y) \in 
\left[(1-4\varepsilon)\frac{\delta}{\varepsilon^{i}},(1+2\varepsilon)\frac{2\delta}{\varepsilon^{i}}\right)$. 
By step 1 and step 2, there must be a PIT $T' \in \mathcal{T}$ such that $T'$ has a node $(x, y, i)$.
Furthermore, $C_i(x, y)$ contains $B(x, \frac{2\delta}{\varepsilon^{i-1}})$ and $B(y, \frac{2\delta}{\varepsilon^{i-1}})$. 
To see this, for any point $t$ where $d_X(t, \{x, y\}) \leq \frac{2\delta}{\varepsilon^{i-1}}$, let $(u_1, u_2, i-1)$ be the pairwise node of $T'$ at level $i-1$ such that its cluster, $C_{i-1}(u_1, u_2)$, contains $t$.
By~\Cref{lmm:pit_packcover}, the diameter of $C_{i-1}(u_1, u_2)$ is at most $\frac{6\delta}{\varepsilon^{i-2}}$, we have $d_X(t, u_1)$ and $d_X(t, u_2)$ are at most $\frac{6\delta}{\varepsilon^{i-2}}$. 
Therefore, $d_X(u_1, \{x, y\})$ and $d_X(u_2, \{x, y\})$ are at most $\frac{6\delta}{\varepsilon^{i-2}} + \frac{2\delta}{\varepsilon^{i-1}} < \frac{3\delta}{\varepsilon^{i-1}}$. By step 2.1, $(u_1, u_2, i-1)$ is a child of $(x, y, i)$, thus $C_i(x, y)$ contains $C_{i-1}(u_1, u_2)$, which contains $t$ whose $d_X(t, \{x, y\}) \leq \frac{2\delta}{\varepsilon^{i-1}}$.

We now prove that $(x, y, i)$ is $O(\varepsilon)$-close to the pair $(x_0, y_0)$.
Consider a point $p\in C_i(x, y)$, we bound $d_X(p, \{x_0, y_0\})$  by $d_X(p, \{x, y\})$ as follows.
Observe that $p$ is in a subtree rooted at a child $(u, v, i-1)$ of $(x, y, i)$ for some $u, v \in Y_{i-2}$.
By~\Cref{lmm:pit_packcover}, the diameter $C_{i-1}(u, v)$ is bounded by $\frac{6\delta}{\varepsilon^{i-1}}$, thus $d_X(p, u)\leq \frac{6\delta}{\varepsilon^{i-1}}$ and $d_X(p, v)\leq \frac{6\delta}{\varepsilon^{i-1}}$.
By item (i) in the covering property of $T'$ (~\Cref{lmm:pit_packcover}), we have $d_X(u, \{x, y\})$ and $d_X(v, \{x, y\})$ are at most $\frac{6\delta}{\varepsilon^{i-1}}$. 
Therefore:
\begin{equation}
    \begin{aligned}
        d_X(p, \{x, y\}) &\leq \min\{d_X(p, u) + d_X(u, \{x, y\}), d_X(p, v) + d_X(v, \{x, y\})\} \\
        &\leq \frac{6\delta}{\varepsilon^{i-1}} + \frac{6\delta}{\varepsilon^{i-1}} =  \frac{12\delta}{\varepsilon^{i-1}} 
    \end{aligned}
\end{equation}
It follows that:
\begin{equation}
    \begin{aligned}
    d_X(p, \{x_0, y_0\}) &\leq \min\{d_X(p, x) + d_X(x, x_0), d_X(p, y) + d_X(y, y_0)\} \\
    &\leq d_X(p, \{x, y\}) + \frac{2\delta}{\varepsilon^{i-1}} \quad \text{(since $d_X(x, x_0)\leq\frac{2\delta}{\varepsilon^{i-1}}$ and $d_X(y, y_0) \leq \frac{2\delta}{\varepsilon^{i-1}}$)}\\
    &\leq \frac{14\delta}{\varepsilon^{i-1}}
    \end{aligned}
\end{equation}
Since $d_X(x_0, y_0) \geq \frac{\delta}{\varepsilon^i}$, we obtain $d_X(p, \{x_0, y_0\})\leq 14\varepsilon d_X(x_0, y_0)$ as claimed.
\end{proof}

\begin{remark}\label{rm:hierarchy-PIT} The PITs constructed in \Cref{subsec:PIT-static-const} may not have the hierarchical property in the sense that  the points labeling a node $u$ at level $i$ may not be a subset of points labeling the children of $u$. We can enforce this hierarchical property by renaming the labels as follows. First, we claim that given a pairwise node $(x, y, i)$ of a PIT $T'$, children labels of $(x, y, i)$ can be partitioned into two disjoint sets $S_1$ and $S_2$ of diameter $\Theta(\frac{\delta}{\varepsilon^{i-1}})$, $S_1$ is close to $x$ and $S_2$ is close to $y$ (if $x = y$ then $S_2$ is empty). Observe that this claim follows by two items of the covering property of $T'$: for any child $(u_1, u_2, i-1)$ of $(x, y, i)$, $d_X(u_1, \{x, y\})$ and $d_X(u_2, \{x, y\})$ are at most $\frac{6\delta}{\varepsilon^{i-1}}$, and the diameter of $C_i(x, y)$ is at most $\frac{6\delta}{\varepsilon^{i}}$. 
By the packing property of $T$, any $u, v \in Y_i$ has $d_X(u, v) > \frac{\delta}{\varepsilon^i}$, thus $S_1 \cap Y_i$ (and in $S_2 \cap Y_i$) has at most one point. Therefore, there are at most two points in $Y_i$ that are also in children labels of $(x, y, i)$. Now, whenever we create a pairwise node $(x, y, i)$, we find the corresponding pairwise nodes $(x_1, x_2, i-1)$ and $(y_1, y_2, i-1)$ in $T'$ of $(x, i-2)$ and $(y, i-2)$.
If $x_1 = x_2$ and $x \neq x_1$, we rename $(x_1, x_2, i-1)$ to $(x, x, i-1)$.
If $x \neq x_1$ and $x \neq x_2$, we rename $(x_1, x_2, i-1)$ to $(x, x_2, i-1)$ if $d_X(x, x_1) \leq d_X(x, x_2)$, to $(x_1, x, i-1)$ otherwise. 
Similarly, if $y_1 = y_2$ and $y \neq y_1$, we rename $(y_1, y_2, i-1)$ to $(y, y, i-1)$. If $y \neq y_1$ and $y \neq y_2$, we rename $(y_1, y_2, i-1)$ to $(y, y_2, i-1)$ if $d_X(y, y_1) \leq d_X(y, y_2)$, to $(y_1, y, i-1)$ otherwise. 
\end{remark}

%% file: 6.dynamic.tex
In this section, we construct a data structure for maintaining a dynamic pairwise tree cover for a point set under updates as claimed in \Cref{thrm:dynamic_collection}, which we restate below.

\DynamicCover*

In \Cref{sec:collect_pit}, we outlined how a collection of PIT can be statically derived from a net tree. The static construction assumes the full net tree $T$ where the net points at every level are given explicitly. However, such a full net tree would have size $\Omega(n\log \Delta)$.  In dynamic construction, we cannot afford to maintain every level of $T$ explicitly. Instead, we need to maintain a \EMPH{compressed net tree}, for every level $i$, some nodes will be hidden (and hence can only be accessed indirectly) to guarantee that the total size is $O(n)$. 

\paragraph{Dynamic compressed net tree.}  Nodes at some level $i$ of the (uncompressed) net tree will be hidden via jumps: A \EMPH{jump} is an edge in $T$ connecting a node $(x, h)$ at level $h$ and a node $(x, l)$ at a lower level $l$ where $l < h-1$. The jump from $(x, h)$ down to $(x, l)$ effectively hides all level-$i$ nodes $(x,i)$ for every $l < i < h-1$; we call such a node $(x, i)$ a \EMPH{hidden node}. We call $(x, h)$ the \EMPH{top} of the jump and $(x, l)$ the \EMPH{bottom} of the jump. For a technical reason, we will maintain that every jump in $T$  starting from a node $(x, h)$ down to  $(x,l)$ will be \EMPH{$b$-isolated}: given a jump, for any node $(y, k)$ who is not a descendant of $(x, h)$ for $k < h$,  $d_X(x, y) > b\frac{\delta}{\varepsilon^k}$.  Furthermore, in a dynamic net tree, nodes are marked deleted rather than explicitly deleted; we will elaborate more details by the end of this section. (Herein, we use the term dynamic net tree to refer to the dynamic compressed net tree.) Note that the compressed net tree still has a degree bounded by $\eps^{-O(\lambda)}$, since the packing and covering properties still hold.

\begin{restatable}[$(\delta, \varepsilon)$-dynamic net tree]{definition}{DefDynamicNetTree}\label{def:dynamic_nettree} 
    $(\delta, \varepsilon)$-dynamic net tree is a data structure maintaining a $(\delta, \varepsilon)$-net tree $T$ under insertions and deletions.
    The data structure supports the following operations:
    \begin{itemize}
    \item \textsc{Insert}$(p, T)$:  Insert (possibly more than one) nodes at different levels associated with a new point $p$ to $T$, and return a list of $O_\lambda(1)$ new nodes or nodes whose parents in $T$ are updated due to inserting $p$.  There are three types of nodes in the list: \begin{enumerate}
    \item\EMPH{new-point node}: when a new point $p$ is added to $S$, up to three new-point nodes associated with $p$ might be created:  $(p, 0)$, $(p, i)$,  and $(p, i-1)$ for some level $i  > 0$.  Node $(p, i)$ is created as a new child of some node $(u, i+1)$ in $T$, and furthermore, it will be the top of the jump down to $(p, 0)$, making $(p,0)$ the only child of $(p,i)$. Once  $(p, 0)$ and $(p, i)$ are created, the algorithm might additionally create  $(p, i-1)$ as a node between $(p, 0)$ and $(p, i)$ (to split the jump from $(p, i)$  down to $(p,0)$) for maintaining the jump isolation property. 
    \item \EMPH{splitting-jump node}: which is new node $(q, i)$ added at the middle of the jump from $(q, l)$ down to $(q, h)$ for $l<i<h$. 
    \item \EMPH{promoting node}: which is a new node $(q, i)$ in $T$ created by applying  an operation called \textsc{Promote}$(q,i-1,T)$ to the  node $(q,i-1)$ at level $i-1$. As a result of this operation, the parent in $T$ of  $(q,i-1)$ was changed from some node $(u,i)$, with $u\not=q$, at level $i$ to $(q,i)$. Furthermore,  another node at level $i+1$ will be designated as the parent of $(q, i)$. We call $(q,i-1)$ a \EMPH{promoted} node, as the point $q$ was ``promoted'' to level $i$ from level $i-1$. 
    \end{enumerate}
    
    \item \textsc{Delete}$(p, T)$: mark the leaf of $p$
    as deleted and return the pointer to the leaf.
\end{itemize}
\end{restatable}

In \Cref{sec:dynamic_nettree}, we review and slightly simplify the dynamic net tree construction of Cole and Gotlieb~\cite{GC06}. Readers who are not familiar with the work of Cole and Gotlieb~\cite{GC06} are strongly encouraged to read \Cref{sec:dynamic_nettree} to have a complete understanding of how a dynamic net tree changes under updates. Our \emph{intuition} for constructing other data structures will be built on top of the dynamic net tree. However, our technical proofs presented here will only rely on the facts stated in the following theorem, whose proof will be given in \Cref{sec:dynamic_nettree}. The key properties are packing and covering; the jump isolation and close-containment properties are needed for technical purposes only.

\begin{restatable}{theorem}{ThmDynamicNetTree}\label{thm:nettree_ds}
    Given $b \geq 2$ a parameter of the jump isolation, $\varepsilon \leq \frac{1}{4b}$, there is a data structure maintaining a $(\delta,\varepsilon)$-net tree $T$ such that $T$ has the following properties:
    \begin{itemize}
    \item\textnormal{[Packing.]} Two nodes at the same level $(x, i)$ and $(y, i)$ have $d_X(x, y) > \frac{1}{4}\frac{\delta}{\varepsilon^i}$.
    \item\textnormal{[Covering.]} If $(x, i)$ is the parent of $(y, i')$ where $i' < i$, then $d_X(x, y) \leq \phi \frac{\delta}{\varepsilon^i}$, where $\phi = \frac{3}{4}$.
    \item\textnormal{[$b$-Jump isolation.]} Any jump is \EMPH{$b$-isolated}: given a jump starting from a node $(x, i)$, for any node $(y, k)$ who is not a descendant of $(x, i)$ for $k < i$,  $d_X(x, y) > b\frac{\delta}{\varepsilon^k}$. 
    \item\textnormal{[Close-containment.]} For any $(y, k)$ and any ancestor $(z, i)$ of $(y, k)$, $d_X(y, z) \leq  \frac{\delta}{\varepsilon^i} - \frac{\delta}{\varepsilon^k}$.
    This implies that every point $p$ in the subtree rooted at $(z, i)$ is contained in $B(z, \frac{\delta}{\varepsilon^i})$, i.e., $d_X(p, z) \leq \frac{\delta}{\varepsilon^i}$.
    \end{itemize}
    Furthermore, given access to a node $(x, i)$ in $T$ at level $i$, if $(x, i)$ is not the bottom node or a hidden node in a jump, then we can find all the nodes $(y, i)$ at level $i$ such that $d_X(x,y) = g\cdot\frac{\delta}{\varepsilon^i}$  for any parameter $g \geq 1$ in $O(g)^{\lambda}$ time. The data structure has space $O(n)$ and runs in $O_{\lambda}(\log n)$ time per update.
\end{restatable}

Since not all the nodes are explicitly accessible in a compressed net tree, the construction of a PIT from a compressed net tree is somewhat cumbersome. The key observation to keep in mind is the locality of our static construction in \Cref{sec:collect_pit}; specifically, the neighborhood of a node in a PIT is a subset of the neighborhood of the corresponding node in a net tree $T$. This locality alone allows one to maintain a dynamic PIT from a dynamic net tree $T$.

The much more difficult task is to maintain a dynamic leaf ordering of PIT due to two key challenges (C1) and (C2) outline in \Cref{subsc:tech}. We accomplish this task in several steps; see the block diagram in \Cref{fig:organize} for an overview. The first step we take is to stabilize the net tree using a net tree cover as defined in \Cref{def:nettree_cover} below. (The formal definition is somewhat involved; we will briefly describe the idea afterward.) In this definition, to distinguish nodes between different trees, we denote a \EMPH{node $(u, i)$ of $T$ by $(u, i, T)$}, and a node $(u, i)$ of $J$ by $(u, i, J)$.  When the context is clear about which tree is used, we simply denote a node by $(u, i)$.

\begin{restatable}[Net tree cover]{definition}{DefNetTreeCover}\label{def:nettree_cover}
 Let $T$ be a $(\delta, \varepsilon)$-net tree of a point set $S$ in a doubling metric with dimension $\lambda$, and $c\geq 4$ be a constant parameter.
 Given $\varepsilon \leq \frac{1}{20}$, a $(\delta, \varepsilon)$-net tree cover of $T$ is a collection of trees $\mathcal{J} = \{J_1, J_2, \ldots \}$ such that: 
\begin{itemize}
    \item \textnormal{[Size.]} $|\mathcal{J}| = O_\lambda(1)$ trees.

    \item \textnormal{[Net.]} For each tree $J \in \mathcal{J}$, points at level $i+1$ of $J$ is an $O(\frac{\delta
    }{\varepsilon^i})$-net of $S$, where the set of nodes at level $i+1$ of $J$ is a subset of nodes at level $i$ of $T$. 
    Specifically, there exists a \EMPH{surjective map $\psi_J$} that maps a node of $T$ to a node of $J$, $\psi_J(x, i) = (w, i+1, J)$, where $w$ can be $x$ or a different point, and 
    for any node $(w, i+1, J)$ in $J$ where $i\geq 0$, there exists $(w, i, T)$ in $T$ such that $\psi_J(w, i) = (w, i+1, J)$.

    \item \textnormal{[Partial isomorphism.]} 
    In every $J \in \mathcal{J}$, consider a node $(x, i+1, J)$ where $i \geq 0$. 
    If $(x, i, T)$ does not have a parent update except by splitting a jump,  then $(x, i+1, J)$ is a child of $\psi_J(u, i')$ where $(u, i', T)$ is the parent of $(x, i, T)$.
    We say that $J$ is \EMPH{partially isomorphic} to $T$.

    \item \textnormal{[Shifting.]} For every pair of node $(x, i)$ and $(y, i)$ in $T$ with $d_X(x, y) < \frac{c \cdot\delta}{\varepsilon^{i+1}}$, there exists a tree $J \in \mathcal{J}$ such that $\psi_J(x, i) = (x, i+1, J)$, $\psi_J(y, i) = (y, i+1, J)$ and they have the same parent. 
    
    \item \textnormal{[Pairwise covering.]} 
        For every pair of points $x_0, y_0 \in S$ such that $d_X(x_0, y_0) \in [\frac{\delta}{\varepsilon^i}, \frac{2\delta}{\varepsilon^{i}})$, 
        there exists a tree $J$ such that $(x_0, 0, J)$ and $(y_0, 0, J)$ have the same ancestor at level $i+1$. 
\end{itemize}
\end{restatable}

 The basic idea of net tree cover is to start from a net tree $T$, construct a constant number of trees in a set $\mathcal{J}$ where the net points in level in each tree $J\in \mathcal{J}$ is a subset of net points in the corresponding level of $T$. Therefore, each tree in $\mathcal{J}$ in some sense resembles $T$; this is formalized in the net and partial isomorphism properties in \Cref{def:nettree_cover}. As a $(\delta,\varepsilon)$-net tree ``takes care'' of distances in the ranges $[\frac{\delta}{\varepsilon^i}, \frac{2\delta}{\varepsilon^{i}})$, the cover  $\mathcal{J}$  also has to take care of these distances; this explains the covering property. The shifting property, on the other hand, captures the intuition that the cover $\mathcal{J}$ was constructed by the shifting technique similar to grid shifting in Euclidean spaces~\cite{HM85,CHJ20}. (We can fix the constant $c$ in the shifting property to be $4$, but this leads to a somewhat artificial-looking bound.)  While the definition of tree cover is more complicated and somewhat unnatural, we are able to show that updates to $T$ due to an insertion of a point to $S$ can be decomposed into two types of very simple updates to a tree in $\mathcal{J}$: leaf insertions or edge subdivisions. We say that these updates are \EMPH{stable}. For a technical reason, we need the dynamic net tree $T$ to have the $3c$-jump isolation property by simply setting $b=3c$ in \Cref{thm:nettree_ds}.

\begin{restatable}[Dynamic Net Tree Cover]{theorem}{ThrmNetTreeCover}\label{thm:nettreecover} Let $T$ be a dynamic $(\delta, \varepsilon)$-net tree for a dynamic point set $S$ such that every jump in $T$ is $3c$-isolated where $c$ is the constant in \Cref{def:nettree_cover}. Then we can construct a dynamic net tree cover $\mathcal{J}$ from $T$ such that the updates $T$ due to the insertion of a point to $S$ induce $O_{\lambda}(1)$ updates to every tree $J \in \mathcal{J}$ that are stable: they contain  $O(1)$ leaf insertions and $O(1)$ edge subdivisions. Furthermore, the updates to $\mathcal{J}$ can be identified in $O_\lambda(1)$ time.
\end{restatable}

Next, we will construct a dynamic PIT from a net tre cover $\mathcal{J}$. We call every tree $J\in \mathcal{J}$ a \EMPH{stable $(\delta,\varepsilon)$-net tree}. We basically follow the static construction in \Cref{sec:collect_pit} to construct a collection of PITs from each stable tree $J$. As we noted earlier, the construction is local, and hence, whenever a new node $(u,i)$ is inserted into $J$, we will develop a \EMPH{dynamic pairing algorithm} to examine the local neighborhood of $(u,i)$ to find nodes that can be paired up with $(u,i)$, and then update the corresponding PIT.  As the net tree $J$ is stable, we could guarantee that the dynamic PITs constructed from $J$ by our dynamic pairing algorithm are also stable. 

\begin{definition}[Stable Dynamic PIT]\label{def:dynamic-PIT} A \EMPH{stable dynamic} PIT is a PIT that is under three types of updates: adding (a null or non-null) leaf, subdividing an edge, and marking a leaf as deleted.
\end{definition}

In the theorem below, we summarize the guarantees by our dynamic pairing algorithm. The proof will be given in \Cref{sec:dynamic_pairing}. 

\begin{restatable}[Dynamic Pairing]{theorem}{ThrmDynamicPairing}\label{thrm:dynamic_pairing}
Let $\mathcal{J}$ be a  dynamic stable $(\delta, \varepsilon)$-net tree cover constructed from a  $(\delta, \varepsilon)$-net tree in \Cref{thm:nettreecover}. Then we can construct from $\mathcal{J}$ a collection of \EMPH{stable} dynamic PITs $\mathcal{T}$ such that (i) $|\mathcal{T}| = \varepsilon^{-O(\lambda)}$ and (ii) for every points $x, y \in S$ where $d_X(x, y) \in [\frac{\delta}{\varepsilon^i}, \frac{2\delta}{\varepsilon^i})$, there exists a PIT $T' \in \mathcal{T}$ such that a node at level $i$ of $T'$ is $O(\varepsilon)$-close to $(x, y)$. Furthermore, every update to a tree in $\mathcal{J}$ can be translated into $\varepsilon^{-O(\lambda)}$ updates to $\mathcal{T}$ that can be identified in $\varepsilon^{-O(\lambda)}$ time. 
\end{restatable} 

Once we have a stable dynamic PIT, we could develop a data structure to keep track of the DFS ordering of its leaves. Note that the DFS visits nodes in the PIT by order of insertion time: to break ties between children of a node, the DFS will first visit those that were inserted earlier.  Recall that the major issue in maintaining DFS leaf ordering of an unstable PIT is that when a node $u$ changes its parent to a new node, the DFS ordering, as well as the subtrees, of all ancestors of $u$ change significantly, and there could be up to $\Omega(n)$ such ancestors. 

In a stable PIT, the parent of a node could also change but only in edge subdivision. Specifically, a node $z$ could be inserted between a node $u$ and its parent $v$, so that the parent of $u$ now changes from $v$ to $z$. However, the DFS leaf ordering of ancestors of $u$ (except $z$) \EMPH{does not change} due to the subdivision by $z$. (For $z$, the DFS leaf ordering in its subtree is exactly that of $u$ since $u$ is its only child.)

On the other hand, inserting a new leaf $x$ to a PIT could still induce changes in the DFS leaf orderings of all (and up to $\Omega(n)$) ancestors of $x$. The key difference to the case of unstable PIT  is that only a \EMPH{single node} is inserted, and hence, the DFS leaf orderings and the subtrees of these ancestors only change by one, making it possible to use some lazy data structures. Here, we use a data structure by Kopelowitz and Lewenstein~\cite{KL07} to maintain a  (dynamic) centroid decomposition on top of our stable PIT. The basic idea is that in a centroid decomposition, we could identify $O(\log n)$ important ancestors of each node (which are the top endpoint of centroid paths) such that it suffices to update these ancestors only. 

A technical difficulty is that a stable PIT could have null leaves or leaves that are marked deleted.  We say that a leaf is \EMPH{active} if it is non-null and not marked deleted; otherwise, the leaf is \EMPH{inactive}. We say that a node in a PIT is \EMPH{active} if it has at least one active leaf in its subtree; otherwise, the node is \EMPH{inactive}.  In a DFS leaf ordering, we only keep track of active leaves. Here, stability also helps us in the following way: imagine that we iteratively contract an inactive node to its parent to obtain a tree of active nodes and leaves only. The edge subdivision does not really change the contracted tree by much, and hence, we could keep track of the DFS leaf ordering of the contracted tree. While the idea is rather simple, explicitly contracting nodes is expensive since inserting a new (active) leaf could turn a long chain of inactive ancestors to become active. Indeed, we only use contraction as a metaphor to develop our data structure; we do not really contract inactive nodes. All of these ideas lead to a leaf tracker data structure as defined formally below.

\begin{definition} \label{def:leaf_tracker} 
    Leaf tracker is a data structure that maintains a stable dynamic PIT $T$ and a DFS-leaf ordering $\sigma$ of \emph{active} leaves of $T$  and supports the following operations:
    \begin{itemize}
    
    \item \textsc{InsertLeaf}$(u, v, T)$: insert the node $v$ as a leaf under  node $u$.
    \item\textsc{SubdivideEdge}$(v, e, T)$: insert the node $v$ such that $v$ breaks an existing edge $e = (x, y)$ into two new edges $(x, v)$ and $(v, y)$.
    \item \textsc{DeleteLeaf}$(u, T)$: mark the leaf $u$ as deleted.
    \item\textsc{TrackLeftMostLeaf}$(u, T)$: return the left-most leaf of a node $u$.
    
    \item\textsc{TrackRightMostLeaf}$(u, T)$: return the right-most leaf of a  node $u$.
    \item\textsc{GetPredecessor$(p, T)$}: return the predecessor of an active leaf $p$ in $\sigma$.
    \item\textsc{GetSuccessor$(p, T)$}: return the successor of an active leaf $q$ in $\sigma$.
    \end{itemize}
\end{definition}

The next theorem shows how to maintain a leaf tracker data structure efficiently; the proof will be given in \Cref{sec:leaf_tracker}.

\begin{restatable}{theorem}{ThrmLeafTracker}\label{thrm:leaf_tracker}
We can construct a leaf tracker data structure for maintaining a \emph{stable dynamic PIT} $T$ with $O(\varepsilon^{-O(\lambda)} + \log{(n)})$ time per updating and tracking operation (including \textsc{InsertLeaf, SubdivideEdge, DeleteLeaf, TrackLeftMostLeaf, TrackRightMostLeaf}). Furthermore, the DFS leaf ordering of $T$ will be maintained in a doubly linked list, and hence the data structure could support  $O(1)$ time per query  (including \textsc{GetPredecessor, GetSuccessor}). 
\end{restatable}

\paragraph{About deletion.~} As we mentioned at the beginning of this section, deletions are only marked: whenever a point $p$ is deleted from $S$, we mark the leave nodes corresponding to $p$ (in dynamic net tree, net tree cover, and PITs) to be deleted. We do not explicitly delete these nodes. The standard idea to handle these is that when the number of deletions is above a certain threshold, we will rebuild the data structure. The rebuilding leads to a data structure with amortized running time, and one can de-amortize by rebuilding in the background. All of these ideas were used by Cole and Gottlieb~\cite{GC06} to handle deletions in their dynamic net tree, and we follow exactly the same strategy to handle deletions. 

Given all data structures, including net tree cover, dynamic pairing, and leaf tracker, we are now ready to prove \Cref{thrm:dynamic_collection}.

\begin{proof}[Proof of \Cref{thrm:dynamic_collection}] First, we apply \Cref{thm:nettree_ds} to maintain $O(\log \frac{1}{\varepsilon})$ dynamic $(\delta,\varepsilon)$-net trees $T$ for every $\delta \in \{1, 2, \ldots, 2^{\lceil \lg{1/\varepsilon}\rceil}\}$. Second, we maintain a dynamic net tree cover $\mathcal{J}$ for each $(\delta,\varepsilon)$-net tree $T$ by applying \Cref{thm:nettreecover}; we can choose $c = 4$. Third, we apply the dynamic pairing algorithm in \Cref{thrm:dynamic_pairing} to maintain a collection of PITs $\mathcal{T}$ from $\mathcal{J}$. Observe that the total number of PITs constructed in this process for $O(\log \frac{1}{\varepsilon})$ different net trees is $\eps^{-O(\lambda)}$.  Finally, we maintain a leaf tracker data structure for each tree in $\mathcal{T}$ to keep track of the DFS leaf ordering and querying the predecessor/successor of every active leaf. 

Now, we analyze the update time and query time. Observe that the query time is $O(1)$ by \Cref{thrm:leaf_tracker}. For the update, we note that the total update time to a net tree is $\varepsilon^{-O(\lambda)}\log n$ by \Cref{thm:nettree_ds}. This also translates to  $\varepsilon^{-O(\lambda)}\log n$ total update time to the net tree cover by \Cref{thm:nettreecover}; the same update time holds for each PIT by \Cref{thrm:dynamic_pairing}. Since the total number of PITs is  $O_{\lambda,\varepsilon}(1)$, the total update times to all PITS is $\varepsilon^{-O(\lambda)}\log n$. By \Cref{thrm:leaf_tracker}, the update time to the leaf tracker is  $O(\varepsilon^{-O(\lambda)} + \log n)$. Thus, the final update time remains $\varepsilon^{-O(\lambda)}\log n$. 
\end{proof}

In \Cref{sec:nettreecover} we construct a dynamic net tree cover to stabilize the dynamic net tree. In \Cref{sec:dynamic_pairing} we give the details of the dynamic pairing algorithm. The leaf tracker data structure is rather complicated and will be given in \Cref{sec:leaf_tracker}.

%% file: 8.nettreecover_new.tex
\subsection{Dynamic Net Tree Cover}\label{sec:nettreecover}

In this section, we describe the ideas for stabilizing a dynamic net tree using a dynamic net tree cover as claimed in \Cref{thm:nettreecover}, which we restate here for convenience.

\ThrmNetTreeCover*

Recall that the only non-trivial parent update to a node $(p,i)$ in a tree $J$ in the cover $\mathcal{J}$ is the edge subdivision: inserting a new node between $(p,i)$ and its parent in $J$. Indeed, this subdivision is jump splitting defined in \Cref{def:netttree}; that is, the parent of $(p,i)$ is $(p,j)$ associated with the same point $p$ at some level $j > i$. On the other hand, the only non-trivial parent update in a dynamic net tree $T$ is due to promotion operation. Specifically, \textsc{Promote}$(p,i-1,T)$ changes the parent of  $(p,i-1)$ from some node $(u,i)$, with $u\not=p$, at level $i$ to a newly created node $(p,i)$ associated with $p$. 

To stabilize the parent change due to \textsc{Promote}$(p,i-1,T)$, the key observation is that $d_X(p,u)$ is small; precisely, $d_X(p,u)= O(\frac{\delta}{\varepsilon^{i-1}})$. As a thought experiment, suppose that we have a version of $T$, denoted by $J$, where we only keep a subset of $Y_i$, say a $10\frac{\delta}{\eps^i}$-net\footnote{We choose a random number $10$ to make our point simpler; in reality, we construct a  $6c\frac{\delta}{\eps^i}$-net where $c$ is the constant in \Cref{def:nettree_cover}.} of $Y_i$, at level $i$ of $J$, and all other nodes in $Y_i$ are discarded. (If a node $(x,i)$ is discarded if it is too close to a node $(y,i)$ that is kept in $J$, then children of $(x,i)$ in $T$ will become children of $(y,i)$ in $J$.)  Back to \textsc{Promote}$(p,i-1,T)$, in an \EMPH{ideal situation}, if some node $(t,i)$ is kept in a version $J$ of $T$ where $t$ is close enough to both $u$ and $p$, then: (1) node $(u,i)$ will be discarded---we say that $(u,i)$ is \emph{merged by distance} to  $(t,i)$  to emphasize that $(u,i)$ is discarded due to $(t,i)$---(2) there is no need to make a parent $(p,i)$ of $(p,i-1)$ since $(p,i)$ will also be discarded, and (3) node $(p,i-1)$ is already a child of $(t,i)$ before the promotion due to the merge of $(u,i)$ to $(t,i)$, and hence no parent update is needed. Of course, the difficulty here is that there is no good way to choose a $10\frac{\delta}{\eps^i}$-net of $Y_i$ so that the ideal situation always happens. Nevertheless, this thought experiment leads us to the idea of using more than one tree: we simply partition  $Y_i$ into (a small) number of $10\frac{\delta}{\eps^i}$-nets, and for each net, construct a version $J$ of $T$ such that the ideal situation will happen in at least one of the tree. This is exactly the shifting in \Cref{def:nettree_cover}. However, shifting alone is not enough to achieve stability: as points arrive dynamically, some existing nodes could be merged to a newly inserted node, leading to parent updates of the children of the existing nodes. We then introduce the idea of \emph{merging through time} to handle this case. To expand on all ideas in this paragraph, we briefly describe how to achieve shifting and then stability.

\paragraph{Shifting.~} Observe that if two nodes $(x, i-1)$ and $(y, i-1)$ are relatively close, $d_X(x, y) = O(\frac{\delta}{\varepsilon^i})$, then their parents $(u, i)$ and $(v, i)$ are also close: $d_X(u, v) = O(\frac{\delta}{\varepsilon^{i+1}})$. As we mentioned above, the basic idea is to partition $Y_i$ into $\Delta_i$-net where $\Delta_i = 6c\frac{\delta}{\eps^i}$  using a standard coloring trick, see, e.g.,~\cite{BNF22,KLMS22}. We greedily color points in $Y_i$: when considering a new point $y$, we color $y$ by the smallest available color that is different from the colors of the nodes within distance at most $\Delta_i$ from $y$. (Our actual coloring procedure is slightly different for a technical reason described below, but the idea is largely the same.) By packing bound, the number of colors is $O_{\lambda,\eps}(1)$, and each color class will induce a $\Delta_i$-net of $Y_i$. 

Points in each color class, called \EMPH{centers}, will be at level $i+1$ of a tree $J$ in $\mathcal{J}$. The difference in one level between $T$ and $J$ is because the level $0$ of $J$ must form a bijection into $S$, but from level $0$ of $T$ we only have a subset of points. For non-center point $(v,i)$, if the distance from a center $(u,i)$ is small enough (at most $\Delta_i/2$), then it might be \EMPH{merged} to $(u,i)$ in $J$ in the sense that children of $(v,i)$ in $T$  will become children of $(u,i)$ in $J$. More formally, if $(v,i)$ is merged to $(u,i)$ in $J$, then for any child $(t,i-1)$ of $(v,i)$, $\psi_J(t,i-1)$ will be a child of $\psi_J(u,i)$. Not every non-center node will be merged to some center node; if they are far from any of the centers, then they will be left unmerged and will also appear at level $j+1$ in $J$. 

A very subtle technical problem is that when we merge a non-center node $(v, i)$ into a center node $(u, i)$, we would want any child $(w, i-1)$ of $(v, i)$ to become a child of $(u, i)$. However, it is conceivable that  $(w, i-1)$  is a non-center node at level $i-1$ and hence was merged to some other center node  $(t, i-1)$, which is not a child of either $(u, i)$ or $(v, i)$, and hence $(w, i-1)$ does not become a child of $(u, i)$. We fix this problem by using two \emph{disjoint} sets of colors for any two consecutive levels of $T$.

The key guarantee we obtained from shifting is that for any nearby pair of net points, there exists a tree $J\in \mathcal{J}$ where the corresponding nodes in $J$ of the two net points have the same parent.

\paragraph{Achieving stability.~} Observe that updates to $T$ could induce three types of parent updates in $J$: 
\begin{enumerate}
    \item A new node $(u, i)$ is inserted at level $i$ of $T$, and it is become a center at level $i+1$ of some tree $J \in \mathcal{J}$. We then have to consider whether to merge some nodes $(v, i)$ in $T$, which are currently unmerged in $J$, to $(u,i)$ in $J$. This induces parent updates of children of $\psi_J(v,i)$. (A similar but much easier case is when $(u, i)$ is inserted as a non-center node in a tree $J$, which we will discuss more below.) 
    
    \item  A new node $(x, i)$ in $T$ is inserted to split a a jump from $(x, h)$ down to $(x, l)$ where $l < i < h$.  In this case, we also split the jump from $\psi_J(x,h)$ to $\psi_J(x,l)$ in $T$ by inserting  $\psi_J(x, i)$ in $J$. Now the parent of  $\psi_J(x,l)$ changes from $\psi_J(x,h)$ to  $\psi_J(x, i)$.
    
    \item If $(u, i)$ is created by promoting $(u, i-1)$ in $T$, then $(u, i-1)$ has to change its parent from an existing node $(v,i)$ to $(u, i)$, which induces a parent update of $\psi_J(u, i-1)$.
\end{enumerate}

The only non-trivial parent update that we allow in a tree $J\in \mathcal{J}$ is in the edge subdivision. Indeed, case 2 above is an edge subdivision in $J$. A subtle point here is that the new subdividing node $(x,i)$ might possibly be merged to some center node at level $i$ of $T$. Our idea is to show that the jump isolation property forbids this case.  

For case 3, the ideal situation described above happens: by the shifting property, there exists a tree $J$ such that $(v,i)$ is a center in $J$ and $\psi_J(u,i-1)$ is a child of $\psi_J(v,i)$.  This means even if we create a new node $(u,i)$ in $T$, then one can show that $d_X(v,u)$ is small and hence $(u,i)$ will be merged to $(v,i)$. However, as the (only) child $(u,i-1)$ of $(u,i)$ is already a child of $(v,i)$ in $J$, we do not have to do anything in $J$. For every other tree $J' \in \mathcal{J}$, we still need to insert $(u,i)$ as a new leaf node (and then we insert a null leaf to be a child of $(u,i)$ in $J$), but we do not have to move $(u,i-1)$ to be a child of $(u,i)$. As a result, we will lose the hierarchical property in trees of $\mathcal{J}$: a node at level $i$ might not be associated with points from its children.  Fortunately, this property is not important for our end goal, which is to construct an LSO. 

Lastly, to handle case 1, we introduce a new rule called \EMPH{merging through time}. Specifically, we allow merging $(v, i)$ into $(u, i)$ in $J$ if $(v, i)$ is added to $T$ after $(u, i)$.  Then case 1 does not happen as $(v,i)$ was inserted before $(u,i)$. When $(u,i)$ is inserted as a non-center node in a tree $J$, we have to merge $(u,i)$ to an existing center node, say $(x,i)$, in $J$. But this is an easy case since we only insert leaves to $J$, so $(u,i)$ has no children at the time of its insertion, hence inducing no parent updates.

\subsubsection{Dynamic net tree cover construction}

First, we color nodes of the dynamic net tree $T$ when nodes are inserted to $T$. Note that we do not remove nodes out of $T$; we only mark leaves as deleted, and the same holds for trees in $\mathcal{J}$.   We say that two net points $(x, i)$ and $(y, i)$ at level $i$ are \EMPH{$r$-close} if $d_X(x, y) < r \frac{\delta}{\varepsilon^i}$. Let  $\Delta_i = \frac{6c \delta}{\varepsilon^{i}}$ be the \EMPH{coloring distance} at level $i$.   Let $k$ be the maximum number of net points $Y_i$ in a ball of diameter $\Delta_i$. By packing property of $T$ and packing property of $(X, d_X)$, $k =  O_\lambda(1)$. Basically, we will assign a color for $(u,i)$ from $[1,2k]$ depending on its levels and neighbors; the reason for this was already explained above.

\begin{mdframed}[nobreak=true]
    \textbf{\textsc{ColoringNode}$(u,i,T)$:}
    \begin{enumerate}
    \item If $i$ is odd, assign $(u,i)$ a smallest color in $[1,k]$ such that the color of $(u,i)$ is different from the colors of all level-$i$ nodes within radius $\Delta_i$ of $u$. 
    \item  If $i$ is even,  assign $(u,i)$ a smallest color in $[k+1,2k]$ such that the color of $(u,i)$ is different from the colors of all level-$i$ nodes within radius $\Delta_i$ of $u$. 
   \end{enumerate}      
   Let \EMPH{$\kappa(u,i)$} be the color of $(u,i)$. By \Cref{thm:nettree_ds}, all the nodes within distance $\Delta_i$ of $(u,i)$ can be found in $O_\lambda(1)$ time.  
\end{mdframed}

   The color of a node $(u,i)$ at level $i$ of $T$ will tell us which tree in $\mathcal{J}$ that $(u,i)$  will be a center (at level $j+1$). Therefore, the number of trees in $\mathcal{J}$  is the number of colors, which is $2k = O_{\lambda}(1)$, and each tree $J\in \mathcal{J}$ will have a \EMPH{color $\kappa(J)$} in $[1,2k]$.

   Recall in \Cref{def:netttree}, whenever a point $p$ is inserted to $S$, the dynamic net tree data structure will return a list $L_p$ of $O_{\lambda}(1)$  nodes that are either changed or inserted to $T$. We will take this list $L_p$ and call   \textsc{Insert($L_p, J$)} in \Cref{fig:update-cover-tree} to update every tree $J \in \mathcal{J}$.  By the stability property of $J$, we are only allowed to use three following operations as a black box:
 \begin{itemize}
     \item \textsc{InsertLeaf}$(x,i,J)$: inserting a leaf associated with point $x$ at level $i$ of $J$ as a leaf. One should think of $(x,i)$ as a node at level $i$ in $J$ without any children. (If $x = \textsc{Null}$, then we call the leaf a null leaf.) To keep the pseudocode clean, we do not specify the parent of the new leaf $(x,i)$. The parent will either be clear from the context or if the insertion of $(x,i)$ to $J$ is triggered by the insertion of the corresponding node $(x,i-1)$ to $T$ whose parent is $(u,i)$, then the parent in $J$ of $(x,i)$ is $\psi_J(u,i)$.
     
     \item   \textsc{Subdividing}$(u,v,z,J)$: subdividing an edge between $u$ (parent) and $v$ (child) in $J$ by creating a new node $z$ and adding $z$ between $u$ and $v$. 
 \end{itemize} 

   We will use notation $(x,i,T)$ and $(x,i,J)$ to distinguish a node $(x,i)$ at level $i$ in $T$ and a node $(x,i)$ at level $i$ in $J$, respectively. 

   Every step in \textsc{Insert($L_p, J$)} in \Cref{fig:update-cover-tree} is  self-explained, except step 3(d). Recall that $T$ and $J$ are off by one level. It is possible that for a point $x\in S$, its corresponding node $(x,0)$ in $T$ is merged to some other node at level $0$ in $T$, and hence in this case, we will have to create a node $(x,0)$ at level $0$ of $J$ to guarantee that level $0$ of $J$ contains every (non-deleted) point in $S$.

\begin{figure}[!htb]
    \centering
\begin{mdframed}[nobreak=true]
\textbf{\textsc{Insert($L_p, J$)}:}

\begin{enumerate}
    \item Sort nodes in $L_p$ by descending order of levels. Then we consider every node $(x, i, T) \in L_p$ in the sorted order and apply steps 2 and 3 below.
     \item If $(x, i, T)$ splits a jump from $(x,h)$ to $(x,l)$, call \textsc{Subdividing}$(\psi_J(x,h),\psi_J(x,l),(x,i+1),J)$. Then set $\psi_J(x, i,T) \leftarrow (x,i+1,J)$.
        \item If $(x, i, T)$ is a new-point node or a promoting node:
        \begin{enumerate}
            \item If  $\kappa(x, i) = \kappa(J)$, then $(x, i)$ is a center in $J$. We call \textsc{InsertLeaf}$(x,i+1,J)$ and set $\psi_J(x, i,T) 
 \leftarrow (x,i+1,J)$.  
            \item Otherwise, we find a center $(w, i,T)$  is $3c$-close to $(x, i, T)$ in $T$ such that $\kappa(w, i) = \kappa(J)$ and $d_X(w, x) < 3c\frac{\delta}{\varepsilon^i}$. There is at most once such node $(w, i,T)$, since we assign different colors to different nodes within the distance $\Delta_i = 6c\frac{\delta}{\varepsilon^{i}}$.
            \begin{enumerate}
                \item If $(w, i, T)$ exists, set $\psi_J(x, i,T) \leftarrow \psi_J(w, i,T)$. \textcolor{brown}{$\qquad \ll$ merge $(x,i)$ to $(w,i)$ $\gg$}
                \item Otherwise, call \textsc{InsertLeaf}$(x,i+1,J)$  and 
                set $\psi_J(x, i, T) \leftarrow (x, i+1, J)$.   
            \end{enumerate}
            \item If $(x, i, T)$ is a promoting node in $\textsc{Promote}(x, i-1, T)$, call  \textsc{InsertLeaf}$(\textsc{Null},0,J)$ to create a null leaf and assign the null leaf as a child of $\psi_J(x, i)$ in $J$. 
            \item If $(x,i,T)$ is a new-point node and $i=0$, call  \textsc{InsertLeaf}$(x,0,J)$  to create $(x, 0, J)$ as a child of $\psi_J(x, 0)$.
         \end{enumerate}
\end{enumerate}
\end{mdframed}

    \caption{Updating $J$ when a new node  is inserted to $T$.}
    \label{fig:update-cover-tree}
\end{figure}

\begin{remark}\label{rm:netcover_jump} If we update $J$ following the algorithm in \Cref{fig:update-cover-tree}, the top of a jump in $J$ can have two or more children, this occurs when we merge $(x, i, T)$ to $(w, i, T)$, and any of them is the top of a jump in $T$. In the dynamic pairing algorithm in \Cref{sec:dynamic_pairing}, it would be much easier if the top of a jump has only one child. To guarantee this, whenever we have a jump from $(x, i, T)$ down to $(x, j, T)$ (at step 2, 3(c) or 3(d) of \textsc{Insert}), we create $(x, i, J)$ as a child of $\psi_J(x, i)$ and make a jump from $(x, i, J)$ down to $\psi_J(x, j) = (x, j+1, J)$.  
    Since the jump starting at $(x, i, T)$ is $3c$-isolated, $(x, i-1, T)$ is not $3c$-close with any node at the same level. Thus, $(x, i, J)$ has only one child $\psi_J(x, j) = (x, j+1, J)$. Now, the top of any jump in $J$ has only one child, and for any node $(u, i, J)$ has two or more children, $(u, i, J)$ and all of its children are non-hidden nodes. 
\end{remark}

\subsubsection{Analysis}

In this section, we show all the properties of the dynamic net tree cover stated in \Cref{def:nettree_cover}. We start with the partial isomorphism property.
 
\begin{lemma}\label{lmm:netcover_isomorphism}
    Every $J \in \mathcal{J}$ is a tree and satisfies partial isomorphism property: for any node $(x, i+1, J)$ in $J$ where $i \geq 0$, if $(x, i, T)$ does not have parent update except by edge subdividing, then $(x, i+1, J)$ is a child of $\psi_J(u, i')$ where $(u, i', T)$ is the parent of $(x, i, T)$ in $T$.
    Furthermore, if $(x, i, T)$ is a promoted node whose parent before the promotion is $(v, i+1, T)$, then $(x, i+1, J)$ is a child of $\psi_J(v, i+1)$.
\end{lemma}

\begin{proof}   Observe that we only modify $J$ by creating leaves or subdividing edges, and hence $J$ is a tree. Inductively, assume that $J$ satisfies the partial isomorphism property before an update, and we have to show that it holds after an update, and specifically, if $(x,i,T)$ has no parent update, then the partial isomorphism holds for $(x,i,T)$.

Suppose that $(x,i,T)$ has a parent update by splitting a jump from $(x,j,T)$ to  $(x,i,T)$ for $i < j$, and in this case, a node $(x,k,T)$ is inserted between them where $i < k < j$. Now the parent of $(x,i,T)$ is $(x,k,T)$.
By induction, in $J$, there is a corresponding jump from $\psi_J(x,j)$ down to $\psi_J(x,i)$. Then in step 2 of the insert algorithm, we add $\psi_J(x,k)$ to split this jump, and hence the parent of $(x,k,T)$ is $\psi_J(x,k)$, giving the partial isomorphism property.
    
The lass claim that  $(x, i+1, J)$ is a child of $\psi_J(v, i+1)$ follows from step 3(a). 
\end{proof}

Next, we show the packing and covering properties.

\begin{lemma}\label{lmm:netcover_net}
    For any $J \in \mathcal{J}$, points at level $i+1$ is an $O(\frac{\delta}{\varepsilon^i})$-net of $S$:
    \begin{itemize}
        \item \textnormal{[Packing. ]} For any pair of nodes $(u, i+1, J)$ and $(v, i+1, J)$, $d_X(u, v) > \frac{1}{4}\frac{\delta}{\varepsilon^{i}}$.
        \item \textnormal{[Covering. ]} For any child $(x, i, J)$ of $(v, i+1, J)$, $d_X(x, v) < 4c\frac{\delta}{\varepsilon^i}$. This means, for any descendant $(y, j+1, J)$ of $(v, i+1, J)$ for $j \leq i$, $d_X(v, y) < 5c\frac{\delta}{\varepsilon^i}$.
    \end{itemize}
\end{lemma}
\begin{proof}
Since nodes at level $i+1$ in $J$ is a subset of nodes at level $i$ in $T$, for any $(u, i+1, J)$ and $(v, i+1, J)$, the packing property of $J$ follows from that of $T$ (see also \Cref{thm:nettree_ds}).

For covering property of $J$, first we observe that there exist $(x,i-1,T)$ and $(v,i,T)$ in $T$ corresponding to $(x,i,J)$ and $(v,i+1,J)$, respectively. If $(v,i,T)$  is the parent of $(x,i-1,T)$, then by \Cref{thm:nettree_ds}, $d_X(x,v)\leq \phi \frac{\delta}{\eps^i} \leq 4c\frac{\delta}{\varepsilon^i}$ as $c\geq 1$. However, it is possible that $(v,i,T)$  is not the parent of $(x,i-1,T)$ since the parent of $(x,i-1,T)$, denoted by $(u,i,T)$, is merged to $(v,i,T)$. In this case, by the construction in step 3b(i) of  \textsc{Insert}($L_p, J$), $d_X(u,v)< 3c\frac{\delta}{\eps^{i}}$. Since $(u,i)$ is the parent of $(x,i-1)$ in $T$, by the covering property of $T$, $d_X(x, u) \leq \phi\frac{\delta}{\varepsilon^{i}}$. By triangle inequality:
\begin{equation*}
    \begin{aligned}
        d_X(x, v) &\leq d_X(x, u) + d_X(u, v) \\
         &< (\phi + 3c)\frac{\delta}{\varepsilon^{i}} \leq 4c\frac{\delta}{\varepsilon^{i}} \textrm{\quad (since $\phi < 1$ and $c\geq 4$)}
    \end{aligned}
\end{equation*}

Finally, we bound the distance from $(v, i+1, J)$ to its descendants. By induction, the distance from any child $(x, i, J)$ of $(v, i+1, J)$ to one its descendants, say  $(y, j+1, J)$ is at most $5c\frac{\delta}{\varepsilon^{i-1}}$. By the covering of $J$, $d_X(v, x) < 4c\frac{\delta}{\varepsilon^{i}}$. By triangle inequality:
\begin{equation*}
    \begin{split}
        d_X(v, y) &\leq d_X(v, x) + d_X(y, v)\\
        &< \left(4c + 5c\varepsilon\right)\frac{\delta}{\varepsilon^{i}} \leq 5c\frac{\delta}{\varepsilon^{i}} \quad \text{(since $\varepsilon \leq 1/20$ in \Cref{def:nettree_cover})}
    \end{split}
\end{equation*}
as desired.
\end{proof}

Next, we show the shifting property of $\mathcal{J}$. Recall that we set the isolation parameter $b = 3c$ for the dynamic net tree $T$ so that every jump is $3c$-isolated. We will use the jump isolation property extensively to show the shifting property.  First, we claim that:

\begin{claim}\label{lmm:netcover_bclosepair}
    If $(u, i)$ and $(v, i)$ are $3c$-close, there exists a tree $J \in \mathcal{J}$ such that $\psi_J(u, i) = \psi_J(v, i)$.
    Furthermore, if we add $(u, i, T)$ to $T$ before $(v, i, T)$ then $J$ and $(u, i, T)$ have the same color. 
\end{claim}

\begin{proof} W.l.o.g, assume that we add $(u, i, T)$ to $T$, either explicitly as a node at level $i$ or as a hidden node in a jump, before $(v, i, T)$. If $(u, i, T)$ is a hidden node in a jump, since every jump is $3c$-isolated, $d_X(u,y)\geq 3c\frac{\delta}{\eps^i}$ for any node $(y,i)$ at level $i$; this contradicts that $(u, i)$ and $(v, i)$ are $3c$-close. Thus, $(u, i, T)$ is not a hidden node, and therefore, there exists a tree $J\in \mathcal{J}$ where  $\kappa(J) = \kappa(u, i)$.

When $(u,i, T)$ is added to $T$, the insert procedure will add $(u, i+1, J)$ as a node in $J$ and assign $\psi_J(u, i) = (u, i+1, J)$.  Since $d_X(u, v) < 3c\frac{\delta}{\varepsilon^i} < \Delta_i$, we have $\kappa(v, i) \neq \kappa(u, i)$. It follows that $\kappa(v, i) \neq \kappa(J)$. Thus, when $(v,i)$ was inserted to $T$,  $(v, i, T)$ and $(u, i, T)$ satisfy the step 3(b) of \textsc{Insert}($L_p, J$) with $w = u$ and $x = v$, and therefore, step i in 3(b) will set $\psi_J(v, i) \leftarrow \psi_J(u, i)$. 
\end{proof}

\begin{lemma}\label{lmm:proof_shifting}
    $\mathcal{J}$ satisfies the shifting property: For every pair of nodes $(x, i)$ and $(y, i)$ with $d_X(x, y) < c\frac{\delta}{\varepsilon^{i+1}}$, there exists a tree $J \in \mathcal{J}$, such that $\psi_J(x, i) = (x, i+1, J)$, $\psi_J(y, i) = (y, i+1, J)$ and they have the same parent in $J$.
    Furthermore, let $(u, i+1)$ and $(v, i+1)$ be the parent of $(x, i)$ and $(y, i)$ in $T$. If we add $(u, i+1)$ before $(v, i+1)$ to $T$, then $\kappa(J) = \kappa(u, i+1)$, and $\psi_J(u, i+1)$ is the parent of both $\psi_J(x, i)$ and $\psi_J(y, i)$.
\end{lemma}

\begin{proof}
    We consider a pair of net points $x, y \in Y_i$ whose $d_X(x, y) < c\frac{\delta}{\varepsilon^{i+1}}$.
    If $(x, i+1)$ (or $(y, i+1)$) is a hidden node, then the $3c$-jump isolation property is violated at $(x, i+1)$ (or $(y, i+1)$ respectively). Thus, the parents of $(x, i)$ and $(y, i)$ must be non-hidden nodes at level $i+1$ in $T$. We consider two cases:
    
   If $(x, i, T)$ and $(y, i, T)$ have the same parent in $T$, denoted by $(u, i+1, T)$. Let $J$ be a tree in $\mathcal{J}$ whose color $\kappa(J) = \kappa(u, i+1)$. Since the sets of colors used for any two consecutive levels of $J$ are disjoint, no node $(w, i)$ at level $i$ of $T$ has $\kappa(w, i) = \kappa(J)$.  Therefore, when $(x, i)$ and $(y, i)$ were inserted to $J$, step 3(b-ii) will be executed: two nodes $(x, i+1, J)$ and $(y, i+1, J)$ will be created as children of $\psi_J(u, i+1)$, and we set $\psi_J(x, i) \leftarrow (x, i+1, J)$ and $\psi_J(y, i) \leftarrow (y, i + 1, J)$.  Thus $\psi_J(x, i)$ and $\psi_J(y, i)$ have the same parent.

    Otherwise, $(x, i, T)$ and $(y, i, T)$ have different parents in $T$, denoted by $(u, i+1, T)$ and $(v, i+1, T)$ respectively. By covering property of $T$,$d_X(u, x)$ and $d_X(v, y)$ are both at most $\phi\frac{\delta}{\varepsilon^{i+1}}$. Since $d_X(x, y) < c\frac{\delta}{\varepsilon^{i+1}}$, and by triangle inequality, we obtain: 
    \begin{equation}
        \begin{aligned}
        d_X(u, v) &\leq d_X(u, x) + d_X(x, y) + d_X(y, v) \\
        &< \phi\frac{\delta}{\varepsilon^{i+1}}+ c\frac{\delta}{\varepsilon^{i+1}} + \phi\frac{\delta}{\varepsilon^{i+1}} \\
        &< 3c \frac{\delta}{\varepsilon^{i+1}} \textrm{ \quad (since $\phi < 1 \leq c$)}
        \end{aligned}
    \end{equation}
        
     Suppose w.l.o.g that we add $(u, i+1, T)$ to $T$ before $(v, i+1, T)$. Let $J$ be the tree  in $\mathcal{J}$ such that $\kappa(J) = \kappa(u, i+1)$.  By~\Cref{lmm:netcover_bclosepair}, $\psi_J(u, i+1) = \psi_J(v, i+1)$. By the same argument in the first case, since the sets of colors between two consecutive levels are disjoint, two nodes $(x, i+1, J)$ and $(y, i+1, J)$ will be created as children of $\psi_J(u, i+1)$ and $\psi_J(v, i+1)$, respectively.  Since $\psi_J(u, i+1) = \psi_J(v, i+1)$, we conclude that $\psi_J(x, i)$ and $\psi_J(y, i)$ have the same parent.
\end{proof}

Finally, we show the pairwise covering property of $\mathcal{J}$.

\begin{lemma}\label{lmm:proof_pairwisecovering_2}
    $\mathcal{J}$ satisfies pairwise covering: For every two points $x_0, y_0 \in S$ such that $d_X(x_0, y_0) \in (\frac{\delta}{\varepsilon^{i}}, \frac{2\delta}{\varepsilon^{i}}]$,  there exists a tree $J$ such that $(x_0, 0, J)$ and $(y_0, 0, J)$ have the same ancestor at level $i+1$ of $J$.
\end{lemma}

\begin{proof}
    Recall that we color nodes in $T$ by the parity of levels. 
    Let $\mathcal{J}_0\subseteq \mathcal{H}$ be the subset of trees whose centers are colored at level $i$; these colored nodes will appear in level $i+1$ of trees in $\mathcal{J}_0$.  Since all nodes at level $i-1$ in $T$ have different colors with $\kappa(J_t)$, Step 3(b-i) in  \textsc{Insert($L_p, J$)} will not be applicable to trees in $\mathcal{J}_0$ when considering nodes at level $i-1$ in $L_p$. Therefore, step 3(b-ii) will be executed, which means,  for every $J \in \mathcal{J}_0$ and every $(p, i-1, T)$, step 3(b-ii) $\psi_J(p, i-1) \leftarrow (p, i, J)$. 
    
    Observe that in different trees $J \in \mathcal{J}_0$, we may have different ancestors at level $i$ of the leaf node $(x_0, 0, J)$. 
    For each $J_t \in \mathcal{J}_0$, let $(x_t, i-1, T)$ and $ (y_t, i-1, T)$ be nodes in $T$ such that $(x_t, i, J_t)$ and $(y_t, i, J_t)$ are respectively the ancestor at level $i$ of $(x_0, 0, J_t)$ and $(y_0, 0, J_t)$. Let $R = \left\{x_1, x_2, \ldots x_{|\mathcal{J}_0|}, y_1, y_2, \ldots y_{|\mathcal{J}_0|}\right\}$; $R$ contains net points at level $i-1$ of  $T$. Now, we show that every two points in $R$ have distance at most $c\frac{\delta}{\varepsilon^i}$.
    By triangle inequality, for any $t, t' \in \{1, \ldots |\mathcal{J}_0|\}$, we have:
    \begin{equation}\label{eq:xt-ytprime}
        \begin{split}
            d_X(x_t, y_{t'}) &\leq d_X(x_t, x_0) + d_X(x_0, y_0) + d_X(y_0, y_{t'})\\
            d_X(x_t, x_{t'}) &\leq d_X(x_t, x_0) + d_X(x_0, x_{t'})\\
            d_X(y_t, y_{t'}) &\leq d_X(y_t, y_0) + d_X(y_0, y_{t'})
        \end{split}
    \end{equation}
    
    Since $(x_0, 0, J_t), (y_0, 0, J_t)$ are descendants of $(x_t, i, J_t)$ and $(y_{t'}, i, J_{t'})$,  by~\Cref{lmm:netcover_net}, $d_X(x_t, x_0)$ and $d_X(y_{t'}, y_0)$ are bounded by $5c \frac{\delta}{\varepsilon^{i-1}}$.  By the assumption of the lemma, $d_X(x_0, y_0) < \frac{2\delta}{\varepsilon^i}$. Plugging these bounds to~\Cref{eq:xt-ytprime}, we have:
    \begin{equation*}
        \begin{aligned}
            d_X(x_t, y_{t'}) &< 5c \frac{\delta}{\varepsilon^{i-1}} + \frac{2\delta}{\varepsilon^{i}} + 5c \frac{\delta}{\varepsilon^{i-1}}\\
            &= \left(2 + 10c\varepsilon \right)\frac{\delta}{\varepsilon^{i}}\\ 
            &\leq c\frac{\delta}{\varepsilon^i} \textrm{\quad (since $\varepsilon \leq \frac{1}{20}$ and $c\geq 4$ by \Cref{def:nettree_cover})}
        \end{aligned}
    \end{equation*}

    Since a jump is $3c$-isolated, every node $(x, i-1, T)$ corresponding to a point $x \in R$ has the (non-hidden) parent at level $i$.  
    
    Let $(s, i, T)$ be the parent of a node in $R$ that is added first to $T$ among all the parents of all the nodes in $R$.  Let $J\in \mathcal{J}_0$ be the tree such that $\kappa(J) = \kappa(s, i)$. By~\Cref{lmm:proof_shifting}, $\psi_J(s, i)$ is the parent of $\psi_J(x, i-1) = (x, i, J)$ for every $x \in R$. Since $J \in \mathcal{J}_0$, there exist $(x', i-1, T)$ and $(y', i-1, T)$ where $x', y' \in R$  that are ancestors of $(x_0, 0, J)$ and $(y_0, 0, J)$, respectively, such that $\psi_J(x', i-1) = (x', i, J)$ and $\psi_J(y', i-1) = (y', i, J)$.  Since $\psi_J(x', i-1)$ and $\psi_J(y', i-1)$ have the same parent, which is $\psi_J(s, i)$, the lemma holds.
\end{proof}

\begin{proof}[Proof of \Cref{thm:nettreecover}]
    By~\Cref{lmm:netcover_isomorphism} and~\Cref{lmm:netcover_net}, we show that every $J \in \mathcal{J}$ satisfies packing property, covering property, and is partial isomorphic with $T$. 
    By~\Cref{lmm:proof_shifting}, $\mathcal{J}$ satisfies shifting  property, and by \Cref{lmm:proof_pairwisecovering_2}, $\mathcal{J}$ satisfies the covering property. Since $\mathcal{J}$ has $2k$ trees where $k = O_\lambda(1)$, $|\mathcal{J}| \in O_\lambda(1)$.

    Observe that in \textsc{Insert($L_p, J$)}, we only call edge subdivisions and insert leaves to $J$ and hence every tree in $J$ is stable. Furthermore, as $|L_p| = O_{\lambda}(1)$, we only call $O_{\lambda}(1)$ update operations to $J$.  Coloring a node in $T$ takes $O(1)$ time as noted in the procedure. The most expensive step (per node in $L_p$) in  \textsc{Insert($L_p, J$)} is to find  $3c$-close nodes for $(x, i, T)$ in step 3(b), which can also be done in $O_\lambda(1)$ time  by \Cref{thm:nettree_ds}. Therefore, the total running time overhead is $O_\lambda(1)$ as claimed.
\end{proof} 

%% file: 7.dynamic_pairing.tex
In~\Cref{sec:collect_pit}, we showed the static construction for pairwise tree cover from a net tree. Here we adapt the static construction to construct a dynamic pairwise tree cover from a dynamic net tree cover as claimed in \Cref{thrm:dynamic_pairing}, which we restate below.

\ThrmDynamicPairing*

For each stable $(\delta, \varepsilon)$-net tree $J_g \in \mathcal{J}$, we will construct a collection of PITs $\mathcal{T}_g$; the final set of PITs contains all PITs $\mathcal{T} =  \mathcal{T}_1 \cup \ldots \cup \mathcal{T}_{|\mathcal{J}|}$.
We guarantee that there exists a PIT $T \in \mathcal{T}_g$ such that $T$ contains $(x, y, i)$\footnote{In the static construction, the level of a PIT and the level of a net tree differ by $1$; here the levels of a stable net tree and PITs derived from it are  the same.}  pairing up $(x, i)$ and $(y, i)$ if two following conditions hold:
\begin{itemize}
    \item[(a)] $d_X(x, y) \in R_{i}$ where $R_i = \left((1-5c\varepsilon)\frac{\delta}{\varepsilon^i}, (2 + 5c\varepsilon)\frac{\delta}{\varepsilon^i}\right]$, where $c$ is the same constant that we use in net tree cover.
    \item[(b)] $(x, i)$ and $(y, i)$ have the same parent in $J_g$.
\end{itemize}

Note that in the static construction, for net points $x, y$ at level $i-1$ of the (static) net tree, we say $(x, y)$ is a blue edge if $d_X(x, y) \in R_i$, a red edge if $d_X(x, y) \leq s_i$. Here, we do not use red edges, and instead use pairs $(x,y)$ if $(x, i)$ and $(y, i)$ have the same parent in one of the stable net trees. 
Recall that in the static construction, children of a pairwise node $(x, y, i)$ include the corresponding pairwise nodes of $(x, i-1)$'s children, $(y, i-1)$'s children, and children of some unmatched node $(z, i-1)$. Here, if $(z, i)$ is added to $J_g$ before $(x, i)$ and $(y, i)$, then we have to change the parent for the corresponding pairwise node of $(z, i)$'s children, from $(z, z, i)$ to $(x, y, i)$. As parent updates make PITs unstable, we have to avoid this case. Specifically, in the dynamic pairing algorithm, we relax the static algorithm in that the children of $(x, y, i)$ include the corresponding pairwise node of $(x, i)$'s children and $(y, i)$'s children only. Intuitively, $(x, y, i)$ is created as merging two subtrees of $J_g$ rooted at $(x, i)$ and $(y, i)$, and $(x, y, i)$ becomes an $O(\varepsilon)$-node for any pair $(x_0, y_0)$ where $x_0$ and $y_0$ are respectively a point in descendants of $(x, i)$ and $(y, i)$ in $J_g$.

In more detail, for some $J_g \in \mathcal{J}$ where $(x, i)$ and $(y, i)$ have the same parent, suppose that $(y, i)$ is added to $J_g$ after $(x, i)$. We visit all PITs in $\mathcal{T}_g$ to find a tree $T$ containing $(x, x, i)$; we will show that such tree $T$ exists as long as  $\mathcal{T}_g$ has sufficiently (but still  $\varepsilon^{-O(\lambda)}$) many trees. Then we rename the corresponding pairwise node in $T$ of $(x, i)$ from $(x, x, i)$ to $(x, y, i)$. 
By applying the pairwise covering property of $\mathcal{J}$, we guarantee that every pair of points in $S$ with a certain range of distance has an $O(\varepsilon)$-close node. 
We will show that PITs in $\mathcal{T}_g$ have the same types of updates as $J_g$, and since $J_g$ is stable, every PIT is stable.

Note that we do not use jump terminology in dynamic PITs. PITs still have a ``long'' edge between a node at level $j$ and a node at level $k$ where $j > k+1$; this long edge corresponds to some jump in the corresponding stable net tree. Therefore, splitting a jump in a net tree will correspond to subdividing a (long) edge in a PIT.

Now, we describe our dynamic pairing algorithm. 
\paragraph{Dynamic pairing algorithm.~}
If a leaf $(q, 0)$ is marked as deleted in $J_g$, in every PIT of $\mathcal{T}_g$, we mark its corresponding pairwise nodes  $(q, q, 0)$ as deleted.
When $J_g$ has a new node $(p, i)$, we create pairwise nodes for $(p, i)$ in PITs of $\mathcal{T}_g$ depending on the type of $(p, i)$: 
\begin{itemize}
    \item[(1)] If $(p, i)$ splits a jump from $(p, j)$ down to $(p, k)$ in $J_g$ where $j > i > k$: since $(p, j)$ has only one child, $(p, i)$ is not paired up with any node at the same level. Thus, for every PIT $T \in \mathcal{T}_g$, let $(p_1, p_2, j)$ and $(p, p, k)$  be the corresponding pairwise node in $T$ of $(p, j)$ and $(p, k)$, respectively---note that $(p, k)$ is the only child of $(p, j)$ before adding $(p, i)$, thus its corresponding pairwise node is single label. We create $(p, p, i)$ by subdividing the edge from $(p_1, p_2, j)$ down to $(p, p, k)$. 
    \item[(2)] If $i > 0$ and $(p, i)$ is a child of $(q, i+1)$:
    if there is some node $(x, i)$ where $d_X(p, x) \in R_i$ and $(x, i)$ is also a child of $(q, i+1)$, we will call the \EMPH{dynamic matching algorithm} described below. This algorithm will create a new pairwise node $(x, p, i)$ in a PIT $T$ of $\mathcal{T}_g$ (as well as several single-label nodes in some other trees in $\mathcal{T}_g$).
    \item[(3)] If $i = 0$, we create a leaf $(p, p, 0)$ in every PIT $T \in \mathcal{T}_g$, note that $p$ can be a null point. To find parent for $(p, p, 0)$ in a PIT $T$, let $(p', i')$ be the parent of $(p, 0)$ in $J_g$, and $(p_1, p_2, i')$ be the corresponding pairwise node in $T$ of $(p', i')$. We make $(p, p, 0)$ a child of $(p_1, p_2, i')$.
    
\end{itemize}

\paragraph{Dynamic matching algorithm.~}
This algorithm applies to the case where  a new node $(p,i)$ of a stable net tree $J_g$ has (at least one) sibling $(x,i)$ such that  $d_X(x, p) \in R_i$. Let $(q,i+1)$ be the parent of $(p,i)$ (and also $(x,i)$). Let $\mathcal{I}$ be the set of PITs in $\mathcal{T}_g$ that do not have a pairwise node of $p$; initially, $\mathcal{I} = \mathcal{T}_g$.  
For every child $(x, i)$ of $(q, i+1)$ in $J_g$, if $d_X(x, p) \in R_i$, let $T$ be a PIT in $\mathcal{I}$ where the corresponding pairwise node of $(x, i)$ is $(x, x, i)$; in the analysis below, we will show that $T$ exists. 
Then, we create the corresponding pairwise node of $(p, i)$ in $T$ by simply renaming the corresponding pairwise node of $(x, i)$ from $(x, x, i)$ to $(x, p, i)$.
Finally, for every remaining tree $T'$ in $\mathcal{I}$, as $T'$ does not have a corresponding pairwise node of $(p,i)$, we create $(p, p, i)$ as a child of the corresponding pairwise node of $(q, i+1)$ in $T'$.

\paragraph{Analysis.~} We now analyze the dynamic pairing algorithm. First, we show a bound on $|\mathcal{J}_g|$ for nice properties assumed in the dynamic pairing algorithm to exist.

\begin{lemma}\label{lmm:dynamic_pairing} It suffices to maintain $\mathcal{T}_g$ that has $|\mathcal{T}_g| = \varepsilon^{-O(\lambda)}$ trees. Furthermore, given a new node $(p, i)$ in $J_g$, the dynamic matching algorithm runs in $\varepsilon^{-O(\lambda)}$ time to update $\mathcal{T}_g$, and guarantees that: 
   for any node $(x, i)$ in $J_g$ where $(x, i)$ and $(p, i)$ have the same parent and $d_X(p, x) \in R_{i}$, there exists a PIT $T$ in $\mathcal{T}_g$ that contains $(x, p, i)$.
\end{lemma}

\begin{proof}
For every child $(x, i)$ of $(q, i+1)$ where $d_X(x, p) \in R_i$, the algorithm finds a PIT $T$ containing $(x, x, i)$ to create $(x, p, i)$. Here we show that by constructing a sufficiently large (but still $\varepsilon^{-O(\lambda)}$) number of PITs in $\mathcal{T}_g$, such a tree $T$ is guaranteed to exist.  

Let $X_i$ be the set of points labeling nodes at level $i$ of $J_g$. 
For $x \in X_i$, we define $N_b(x) = \{y \in X_{i}: d_X(x, y) \in R_{i}\}$.
Let $\delta_b$ be the maximum size of $N_b(x)$ for every $x \in X_i$. By packing property of $J_g$, it holds that $\delta_b = \varepsilon^{-O(\lambda)}$, since the minimum distance of points in $X_{i}$ is $\Theta(\frac{\delta}{\varepsilon^{i-1}})$, while $R_{i} \in \Theta(\frac{\delta}{\varepsilon^{i}})$. 
Consider a point $x \in N_b(p)$, observe that at most $\delta_b-1$ points in $N_b(x) \setminus \{p\}$ that can be paired with $x$. Therefore, there are at most $\delta_b-1$ PITs in $\mathcal{T}_g$ where $(x, i)$ has a double-label pairwise node. If we maintain $\delta_b$ trees in $\mathcal{T}_g$, there always exists a PIT $T$ to pair up $x$ and $p$. 

We now analyze the running time per update. 
By the packing and covering properties of $J_g$, $(q, i+1)$ has $\varepsilon^{-O(\lambda)}$ children, and we can check the pairwise node of a $(q, i+1)$'s child in a PIT with $O(1)$ time. For a PIT, we create a pairwise node $(p, p, i)$ or rename $(x, x, i)$ to $(x, p, i)$ in $O(1)$ time. Since there are $\varepsilon^{-O(\lambda)}$ trees, the dynamic matching algorithm totally runs in $\varepsilon^{-O(\lambda)}$ time.    
\end{proof}

\begin{lemma}\label{lmm:dynamicpit_pairing}
    For every two points $x, y \in S$ where $d_X(x, y) \in (\frac{\delta}{\varepsilon^i}, \frac{2\delta}{\varepsilon^i}]$, there exists a PIT $T \in \mathcal{T}$ such that a node at level $i$ is $O(\varepsilon)$-close to $(x, y)$.
\end{lemma}

\begin{proof}
For every two points $x, y \in S$ where $d_X(x, y) \in (\frac{\delta}{\varepsilon^i}, \frac{2\delta}{\varepsilon^i}]$, we prove that there exists a PIT that has a pairwise node containing both $x$ and $y$ as its descendant leaves.
By the pairwise covering property of the net tree cover $\mathcal{J}$, there exists a tree $J_g$ such that $(x, 0)$ and $(y, 0)$ have the same ancestor at level $i+1$ in $J_g$.
Let $(x', i)$ and $(y', i)$ be the ancestor at level $i$ in $J_g$ of $(x, 0)$ and $(y, 0)$ respectively, we know that $(x', i)$ and $(y', i)$ have the same parent $(u, i+1)$. 
By the covering property of $J_g$ (\Cref{lmm:netcover_net}), $d_X(x, x')$ and $d_X(y, y')$ are at most $5c\frac{\delta}{\varepsilon^{i-1}}$. Thus, $d_X(x', y') \in \left(\frac{\delta}{\varepsilon^i} - 5c\frac{\delta}{\varepsilon^{i-1}}, \frac{2\delta}{\varepsilon^i} + 5c\frac{\delta}{\varepsilon^{i-1}}\right] = R_i$. 
Since $d_X(x, y) > 0$, we have $x \neq y$. By~\Cref{rm:netcover_jump}, since $(u, i+1)$ has at least two children, $(x', i)$ and $(y', i)$ are non-hidden nodes in $J_g$. 
W.l.o.g, suppose that $(x', i)$ is added before $(y', i)$ to $J_g$. 
Since $(x', i)$ and $(y', i)$ have the same parent in $J_g$ and $d_X(x', y') \in R_i$, when $(y', i)$ is added to $J_g$, by \Cref{lmm:dynamic_pairing}, there exists $T \in \mathcal{T}_g$ that contains $(x', y', i)$. Observe that we pair up two nodes in $J_g$ only if they have the same parent, thus the ancestor at level $j$ in $T$ of $(x, x, 0)$ is the corresponding pairwise node of the ancestor at level $j$ in $J_g$ of $(x, 0)$.
Therefore, $(x, y, i)$ contains $x$, and similarly contains $y$ in its descendants.

Now we show that $(x', y', i)$ is $O(\varepsilon)$-close to the pair $(x, y)$. Observe that points in descendants of $(x', y', i)$ in $T$ is the union of points in descendants of $(x', i)$ and $(y', i)$ in $J_g$. Thus, for every point $t$ in descendants of $(x', y', i)$ in $T$, $(t, 0)$ must be a descendants of $(x', i)$ or $(y', i)$ in $J_g$. By the covering of $J_g$ (\Cref{lmm:netcover_net}), $d_X(t, \{x', y'\})$, $d_X(x, x')$ and $d_X(y, y')$ are at most $5c\frac{\delta}{\varepsilon^{i-1}}$. Therefore $d_X(t, \{x, y\})$ is at most $10c\frac{\delta}{\varepsilon^{i-1}} \leq 10c\varepsilon d_X(x, y)$.
\end{proof}

\begin{proof}[Proof of~\Cref{thrm:dynamic_pairing}] 
By~\Cref{lmm:dynamic_pairing}, we know that $|\mathcal{T}_g| = \varepsilon^{-O(\lambda)}$, and since $|\mathcal{J}| = O_\lambda(1)$, $\mathcal{T} =  \mathcal{T}_1 \cup \ldots \cup \mathcal{T}_{|\mathcal{J}|}$ has totally $\varepsilon^{-O(\lambda)}$ PITs, which proves item (i) of the theorem.
By~\Cref{lmm:dynamicpit_pairing}, item (ii) holds.

Since $|\mathcal{T}_g| = \varepsilon^{-O(\lambda)}$, each update of $J_g$ is translated into $\varepsilon^{-O(\lambda)}$ updates of $\mathcal{T}_g$. For every PIT in $\mathcal{T}_g$, the dynamic pairing algorithm marks a leaf as deleted with $O(1)$ time, creates $(p, p, i)$ for a new node $(p, i)$ in $J_g$ with $O(1)$ time in step 1 and step 3.
In step 2, the dynamic matching algorithm updates all PITs in $\mathcal{T}_g$ and runs in $\varepsilon^{-O(\lambda)}$ time by~\Cref{lmm:dynamic_pairing}. Therefore, the total running time is $\varepsilon^{-O(\lambda)}$.

Now we show that PITs in $\mathcal{T}_g$ are stable. Parent updates occur only in step 1 of the dynamic pairing algorithm, where we subdivide an edge.
In step 2, the dynamic matching algorithm creates single-label pairwise nodes, or renames nodes to create double-label pairwise nodes in PITs. 
These single-label pairwise nodes do not have any child and thus are leaves.
In step 3 of the dynamic pairing algorithm, we create a leaf in every PIT. Therefore, we update PITs with three types of operations: adding a leaf, subdividing an edge, and marking a leaf as deleted; which means PITs are stable.
\end{proof}

\begin{remark}\label{rm:PIT_insertions}
 In the construction of the leaf tracker data structure in~\Cref{sec:leaf_tracker}, it would be conceptually simpler (though technically not needed) if we re-arrange the order of nodes being inserted into PITs and guarantee that adding a new leaf always occurs at level $0$. Recall that in a tree $J_g$ of net tree cover $\mathcal{J}$ and PITs in $\mathcal{T}_g$, we insert nodes by: 
    \begin{enumerate}
        \item In $J_g$, $(p, i)$ splits a jump from $(p, j)$ down to $(p, k)$. 
        In every PIT of $\mathcal{T}_g$, we add $(p, p, i)$ by subdividing the edge from $(p_1, p_2, j)$ down to $(p, p, k)$. In this case, we do not add a new leaf in any PIT.  
        \item In $J_g$, we insert a leaf $(p, i)$ at level $i > 0$, after that, we insert a leaf $(p', 0)$, where $p' = p$ or $p' = \textsc{Null}$. 
        \begin{itemize}
            \item [(a)] If the corresponding pairwise node in a PIT $T$ of $(p, i)$ is $(x, p, i)$ for $x \neq q$, we add $(p', p', 0)$ as a leaf under $(x, p, i)$. 
            \item [(b)] If the corresponding pairwise node of $(p, i)$ in a PIT $T$ is $(p, p, i)$, it is a new leaf at level $i > 0$ in $T$. We need to arrange new nodes of $T$ in this case.
        \end{itemize}
        \item In $J_g$, we only insert a leaf $(p', 0)$ to the tree, where $p'$ is a new point or $p' = \textsc{Null}$ (without adding a leaf at level $i > 0$). This case already satisfies that adding a leaf occurs at level $0$.
    \end{enumerate}
    Now we show how to arrange new nodes of PITs in case 2(b). First, we run the dynamic matching algorithm in $\mathcal{T}_g$, then: If the corresponding pairwise node in a PIT $T$ of $(p, i)$ is $(p, p, i)$, let $(q, i+1)$ be the parent of $(p, i)$ in $J_g$, and $(q_1, q_2, i+1)$ be the corresponding pairwise node in $T$ of $(q, i+1)$. In $T$, we add $(p', p', 0)$ as a leaf under $(q_1, q_2, i+1)$, then add $(p, p, i)$ by subdividing the edge between $(q_1, q_2, i+1)$ and $(p', p', 0)$.
\end{remark}

%% file: 9.leaf_tracker.tex
In this section, we design the leaf tracker data structure for a dynamic PIT as claimed in \Cref{thrm:leaf_tracker} in \Cref{sec:dynamic_ptc}, which we restate below for convenience.
\ThrmLeafTracker*

In this section, all dynamic trees are stable: every update is either inserting a (null or non-null) leaf, marking a leaf deleted, or subdivision an edge. Thus, for simplicity, we use the word dynamic tree to refer to a stable dynamic tree.

By definition (\Cref{def:leaf_tracker}), a leaf tracker has to maintain a DFS ordering of only active leaves. Furthermore,    \textsc{GetPredecessor} and \textsc{GetSuccessor} operate on $\sigma$ and have to return active leaves as results. Therefore, in maintaining the DFS ordering of the leaves $T$, we have to skip over inactive leaves (which include null and mark-deleted leaves). And this is the key difficult challenge in the design of a leaf tracker data structure. 

Recall that the DFS leaf ordering $\sigma$ of $T$ is obtained by visiting the tree and writing down the leaves in the DFS order, breaking ties by insertion time. Specifically,  children of every node in $T$ are ordered linearly by their insertion time. In the DFS order,  we prioritize visiting the nodes in $T$ by their insertion time: from a node, we visit the older children first.   We will use a doubly-linked list to store $\sigma$, and hence, getting the predecessor and the successor of a point in $\sigma$ can be done in $O(1)$ time by simply following the pointers to the next and previous nodes in the list $\sigma$. (Herein, we will slightly abuse the notion by using $\sigma$ to refer to the doubly-linked list representing the DFS ordering $\sigma$.)

First, we will handle a simpler case where a dynamic tree (not necessarily a PIT) $T$ only has (non-null) leaf insertions and edge subdivisions; there are no marking leaves as deleted or inserting null leaves. We will also store the DFS ordering of leaves of $T$ in a doubly-linked list $L_T$.  Our key idea is to construct a data structure that could support querying the leftmost and rightmost \footnote{The left-right order of nodes in $T$ is determined by the insertion time; specifically, earlier inserted nodes are on the left and vice versa.} leaves of a given node $u \in T$  in $O(\log{(n)})$ time. When we add a leaf $(q, q, k)$ as a new child of a node $u$ at a level $k > 0$, we query $D$ to get the get the rightmost leaf $(x, x, 0)$ of $u$ in $O(\log(n))$ time. Assume that we are in the ideal case where $(x, x, 0)$ is active. Then we follow the pointer stored at $(x, x, 0)$ to access its position in $L_T$, and insert $q$ after $(x, x, 0)$ in $L_T$ in $O(1)$ time. The time to locate the rightmost leaf, which is $O(\log n)$, dominates the total running time to update $\sigma$. 

To search for a leftmost or rightmost leaf of a node $u$, the observation is that all the leaves in the subtree rooted at a node $u$ form a contiguous subsequence of $L_T$, where the leftmost (rightmost) leaf is the leftmost (rightmost) element of the subsequence. Then, to search for these extreme points of the subsequence, we will build a skip list on top of $L_T$ to perform some kind of binary search.  However, there seems to be no obvious way to assign keys to elements in $L_T$ to construct the skip list. Nodes in $L_T$ are not sorted in increasing orders of insertion time, and there is no natural linear order between the names of the nodes to use as keys.  To solve this problem, we introduce \emph{ancestral arrays} and a data structure for maintaining them. Roughly speaking, an \EMPH{ancestral array} of a node $u\in T$ is an \emph{array} $O(\log n)$ ``important'' ancestors stemming from a centroid decomposition of $T$ (see \Cref{def:centr-decom-anc-arry}).  We will use ancestral arrays as ``keys'' to the skip list. Though there is no linear order between the ancestral arrays to use them as keys in the traditional sense, we could use them to determine if a leaf $x$ is a descendant of a query node $u$ or not by \Cref{lm:leaf-descedant}, which turns out to be sufficient for binary search using skip lists. There are several subtleties in the implementation, which we will discuss in detail later in \Cref{sec:track_leaf}. Our ultimate result is the following data structure.

\begin{lemma}\label{lmm:search_leaf}
    Let $T$ be a dynamic rooted tree of $n$ nodes under updates by adding new leaves and subdividing edges.     Then, we can construct a data structure with $O(n)$ space that maintains the DFS leaf ordering of $T$ in a doubly-linked list with $O(\log{(n)})$ time per update and tracking operation (including \textsc{InsertLeaf, SubdivideEdge, TrackLeftMostLeaf, TrackRightMostLeaf}). 
\end{lemma}

Next, we design a data structure for a PIT which could contain inactive nodes. There are two key challenges: (a) a node $u$ might be inactive before the insertion of a new active leaf $(q,q,0)$ and hence, no descendant leaves of $u$ will appear in the DFS leaf ordering $\sigma$ since they are also inactive; (b) if $u$ is active, it is possible that most of its descendant leaves are inactive, including its leftmost and rightmost descendant leaves.  (Recall that a node in $T$ is active if it has at least one active descendant leaf, and inactive otherwise.) To resolve these issues, our basic idea is that, given a  PIT $T$, if we iteratively contract every inactive node  to its parent until there is no more inactive node, then the resulting tree only has active nodes and hence we could apply the data structure, denoted by $D$, for trees without inactive nodes.  Of course, we will not explicitly contract inactive nodes, as if we do so, when a new active leaf is inserted as a child of an inactive node, it could trigger a large number of nodes to change their status from inactive to active, resulting in  a large amount of time to undo the contractions.  Instead, we design a new data structure called active tracker (see \Cref{subsec:active-leaf-tracker}) that supports two important operations: (i) given a node $u$ in a PIT, returns an active descendant leaf of $u$, if any, and (ii) given an inactive node $u$, returns the lowest active ancestor of $u$. Operation (i) allows us to access an active leaf of a node to start the binary search on $\sigma$ using skip lists stored in $D$. Operation (ii) provides a kind of implicit contraction: if a new active leaf $(q,q,0)$ is inserted as a child of an inactive leaf $u$, we could conceptually think of $(q,q,0)$ as a new child of the lowest active ancestor $v$ of $u$ in the contracted tree, and hence we could call an update to $D$ to insert a new child to $v$. A subtle point is that the new leaf  $(q,q,0)$  might not be the rightmost leaf in the DFS ordering of descendant leaves of $v$, since the (inactive) child of $v$ that is an ancestor of $u$, denoted by $x$, might have insertion time smaller than other children of $v$. In this case, we insert $(q,q,0)$ next to the rightmost leaf of an active child $y$ of $v$ whose insertion time is largest among all children of $v$ with insertion time smaller than $x$. All these ideas lead to the following lemma, whose proof will be given in \Cref{sec:pred_succ}.

\begin{lemma}\label{lmm:maintain_ordering} Suppose that we are given a data structure in \Cref{lmm:search_leaf}, then we can construct a data structure for maintaining  the DFS leaf ordering of any given PIT in a doubly-linked list with $O(\varepsilon^{-O(\lambda)} + \log{(n)})$ time per update and tracking operation.
\end{lemma}

Observe that \Cref{lmm:search_leaf} and \Cref{lmm:maintain_ordering} together imply \Cref{thrm:leaf_tracker}.  
The rest of this section is organized as follows. \Cref{sec:track_leaf} construct a data structure for a simpler case as claimed in \Cref{lmm:search_leaf}. \Cref{sec:pred_succ} shows how to maintain $\sigma$ and proves \Cref{lmm:maintain_ordering}.

\subsection{Special Case: Trees without Inactive Nodes}\label{sec:track_leaf}

In this section, we construct a data structure for querying the leftmost and rightmost leaves of a node in a dynamic tree $T$ without inactive nodes as claimed in \Cref{lmm:search_leaf}. Note that nodes in $T$ are ordered by their insertion times. (Our data structure works for any tree with a linear order between children of every node in a tree, not just the linear order by insertion times.) Let $L_T$ be the list of leaves in $T$ obtained by visiting $T$ in the DFS order, where children of a node are visited according to their insertion times.  We observe that:

\begin{observation}\label{obs:contiguous}The descendant leaves of any node $u \in T$ form a contiguous subsequence of $L_T$.
\end{observation}

As discussed above, the key idea is to construct a data structure for querying the leftmost and rightmost descendant leaves of a node in $T$. To this end, we need a data structure to maintain an ancestral array of every node, each array holds $O(\log n)$ ancestors from a centroid decomposition. We say that a path $P$ in $T$ is \EMPH{monotone} if it is a subpath from a leaf to the root of $T$.

\begin{definition}[Centroid Decomposition~\cite{jordan69} and Ancestral Array]\label{def:centr-decom-anc-arry}   Given a rooted tree $T$, a \EMPH{centroid path} $\pi$ of $T$ is a maximal monotone path such that there exists an integer $i$ satisfying $2^i \leq |T(u)| < 2^{i+1}$ for all node $u \in \pi$, where $|T(u)|$ is the total number of nodes (size) of the subtree rooted at $u$. We say the node at the highest level of $\pi$ is the \EMPH{head} of $\pi$. A  \EMPH{centroid decomposition} of $T$ is a decomposition into a set $\mathcal{P}$ of centroid paths such that every node $u\in T$ has at most $O(\log n)$ centroid paths, each contains an ancestor of $u$, and every ancestor of $u$ (including $u$) is contained in one of these centroid paths. \\
An \EMPH{ancestral array} of a node $u$ is an array containing the heads of the centroid paths of $u$.
\end{definition}

By the definition of centroid decomposition, the ancestral array of a node has $O(\log n)$ elements, and furthermore, the first node of the array is the root of $T$.  We observe that one can extract a data structure for maintaining an ancestral array from the data structure for maintaining dynamic weighted ancestors in a rooted tree by Kopelowitz and Lewenstein~\cite{KL07}. For completeness, we will review their construction and adapt it to our notation by the end of this section.

\begin{lemma}[Kopelowitz and Lewenstein~\cite{KL07}, implicit]\label{lmm:ancestral_arrays}
    Given a rooted tree $T$ of $n$ nodes under updates by adding new leaves and subdividing edges, there is a data structure that maintains an ancestral array for every node in the tree and runs in $O(\log{(n)})$ time per update.
\end{lemma}

Let $A_v$ be the ancestral array of a node $v \in T$; note that $A_v[1]$ is the root of $T$. We now show that using ancestral arrays, one can infer if a node $x$ is a leaf descendant of a node $u$. (We assume that every node has a level such that the level of a node is smaller than the level of its parent.)

\begin{lemma}
    \label{lm:leaf-descedant} Let $u$ and $x$ be two nodes in $T$ where $x$ is a leaf. Let $h_u$ be the last element of $A_u$, and $j$ be the index of $h_u$ in $A_x$. Then $x$ is a descendant leaf of $u$ if and only if either (a) $h_u$ is the last element of $A_x$, or (b) both following conditions hold: (b.1) $A_u$ is a prefix of $A_x$ and (b.2) the parent of $A_x[j+1]$ has a level at most the level of $u$. 
\end{lemma}

\begin{proof}
    Let $\pi$ be the centroid path containing $u$. Observe that $h_u$ is the head of $\pi$. If $x\in \pi$, then (a) holds and $x$ is a descendant leaf of $u$. On the other hand, if (a) holds, then $x\in \pi$, which means it is a descendant leaf of $u$.   It remains to consider the case where $x \not\in \pi$. We will show that $x$ is a descendant leaf of $u$ if and only if both (b.1) and (b.2) hold.

    If $u$ is an ancestor of $x$, then by definition of ancestral arrays, $A_u$ must be a prefix of $A_x$; $(b.1)$ holds. Since $x\not\in \pi$, $j$ is not the last element of $A_x$, which means $A_x[j+1]$ exists. Let $p$ be the parent of $A_x[j+1]$, observe that $p \in \pi$ and $p$ is the lowest ancestor of $x$ in $\pi$. Thus, the level of $u$ must be at least the level of $p$; (b.2) holds.

    On the other hand, assuming that (b.1) and (b.2) hold. By (b.1), we know that $\pi$ is a centroid path of $x$. As $x \not\in \pi$, $A_x[j+1]$ must exist.  Let $p$ be the parent of $A_x[j+1]$. By \Cref{def:centr-decom-anc-arry}, both $u, p$ are in $\pi$. Since the level of $p$ is at most the level of $u$ by (b.2), $u$ must be an ancestor of $p$, and therefore, of $x$. 
\end{proof}

We will construct a skip list structure on top of $L_T$; nodes in $L_T$ will be referred to as leaves to be distinguished from nodes in the skip list. The skip list has $O(\log n)$ levels to ``navigate'' $L_T$; $L_T$ will be at level 0 of the skip list. As we discussed above, the ``key'' of every element in $L_T$  in the skip list is its ancestral array. Though there is no linear order between the ancestral arrays, we could use them to determine if a leaf $x$ is a descendant of a query node $u$ or not by \Cref{lm:leaf-descedant}. A node in the 
 skip, say $\Tilde{x}_i$, at a level $i$ now holds (a) the name of some leaf, say $x$, in $\sigma$ as data and (b) three pointers: a \EMPH{right pointer} which points to a node $\Tilde{y}_i$ at the same level $i$ where its corresponding leaf $y$ is to the right of $x$ in $\sigma$, and the \EMPH{down pointer} which points to $\Tilde{x}_{i-1}$ (if any), the node at level $i-1$ holding the same leaf $x$ as the data, and the \EMPH{up pointer} that points to $\Tilde{x}_{i+1}$ (if any). Note that each leaf $x$ appears as data in at most $O(\log n)$ nodes of the skip list (at different levels)

\begin{figure}[!ht]
    \begin{tikzpicture}
	 \node at (0,0){\includegraphics[width=1.0\linewidth]{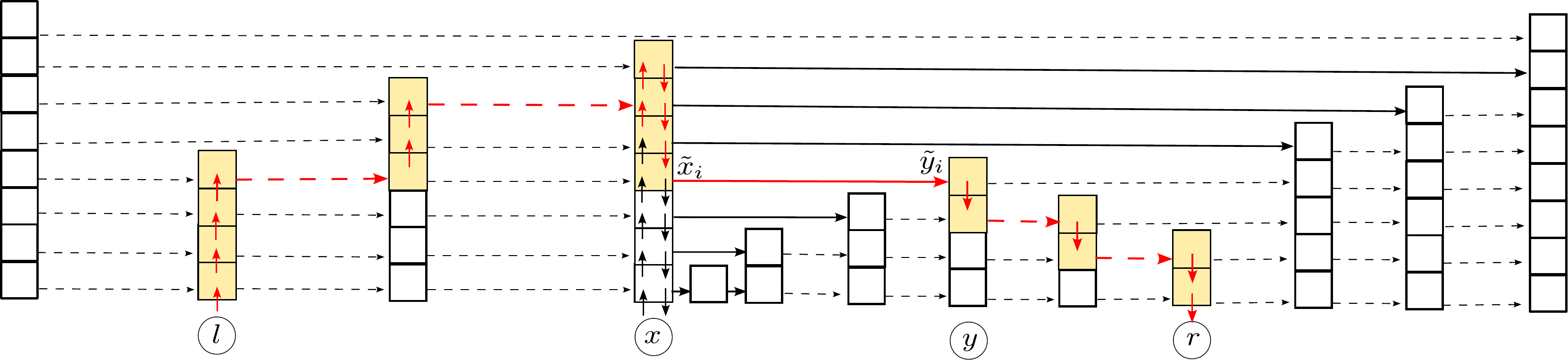}};	
    \end{tikzpicture}    

    \caption{Illustrate the skip list that we build on top of $L_T$ and the process of searching the right-most leaf $r$ of a node $u$ in the tree $T$. Four circles $l, x, y, r$ are leaves in $L_T$. A solid arrow is a direct pointer between two nodes. A dashed arrow shows that there is a path between two nodes of different leaves at the same level. We demonstrate direct pointers of nodes $\tilde{x}_i$ for one leaf $x$, other leaves have the same structure. We start searching at a leaf $l$ of $u$, then follow pointers (red arrows) of yellow nodes to find $r$. The going up stage is from $l$ to $x$ and then going down stage is from $x$ to $r$.}
    \label{fig:skiplist}
\end{figure}
 
When searching for the rightmost of a node $u\in T$, we take an arbitrary leaf $\ell$ of $u$ and start the search from node $\Tilde{\ell}_0$ corresponding to $\ell$ at level $0$  of the skip list, which is $L_T$; see the pseudocode in \Cref{fig:trackright}.  There are two stages in the search:
\begin{enumerate}
    \item \textbf{Going up stage.~} This stage starts from $\Tilde{\ell}_0$, and at an intermediate step, we have a node $\Tilde{x}_i$ at level $i$, where its corresponding leaf $x$ is always a descendant of $u$. 
We then follow the up pointers to get to the node $\Tilde{x}_{i^*}$ corresponding to the same leaf $x$ where $i^*$ is the highest level. Then we following the right pointer of $\Tilde{x}_{i^*}$  to  the next node $\Tilde{y}_{i^*}$, corresponding to a leaf $y$. If $y$ is a descendant of $u$, then we will continue this stage by jumping to $\Tilde{y}_i$. Otherwise, we follow the down pointer to  $\Tilde{x}_{i^*-1}$ and start the second stage.
  \item \textbf{Going down stage.~} At an intermediate step, the search is at some node  $\Tilde{x}_i$ at level $i$, where its corresponding leaf $x$ is always a descendant of $u$ as the first stage. We then follow the right pointer to the next node $\Tilde{y}_{i-1}$, corresponding to a leaf $y$. If $y$ is still a descendant of $u$, then we jump to  $\Tilde{y}_i$. Otherwise, we go down one step to $\Tilde{x}_{i-1}$ and continue this stage. This stage, and also the search, terminates when we reach level $0$.
\end{enumerate}

There are two subtle points in implementing the rightmost/leftmost leaf tracker algorithm. The minor point is that we will not store the ancestral arrays explicitly in the skip list since their sizes are non-constant;  instead, we only store pointers to these arrays. The major point is that we have to check whether $A_u$ is a prefix of $A_x$ in $O(1)$ time; note that their lengths are $O(\log n)$. We do so by exploiting the fact that $A_u$ and $A_x$ are stored as \emph{arrays}: simply look at the last element, say $h_u$, of $A_u$ and check if $A_u[j] = A_x[j]$ where $j$ is the length of $A_u$, which is also the index of $h_u$ is $A_u$.

\begin{figure}[!htb]
\begin{mdframed}[nobreak=true]

\textbf{\textsc{IsDescendant$(u, x, T)$}}: \textcolor{brown}{check if a leaf $x$ is a descendant of $u$}
    \begin{enumerate}
        \item Let $j$ be the last index of $A_u$, and $h_u \leftarrow A_u[j]$. 
        \item $x$ is a leaf of $u$ if either one of following conditions holds:
        \begin{itemize}
            \item[(a)] The last element of $A_x$ is $h_u$.
            \item[(b)] If the last element of $A_x$ is not $h_u$, then both (b.1) and (b.2) hold, where:
            \begin{itemize}
                \item[(b.1)] $A_x[j] = h_u$.
                \item[(b.2)] The parent of $A_x[j+1]$ is at a level at most the level of $u$.
            \end{itemize}
        \end{itemize}

    \end{enumerate}

\noindent\textbf{\textsc{TrackRightMostLeaf}}$(u, \sigma, T)$:  
\textcolor{brown}{return the rightmost leaf in $\sigma$ of a given node $u$}
    
\begin{enumerate}
    \item Let $\ell$ be an arbitrary leaf under the subtree rooted at $u$. 
    \item stage $\leftarrow \textsc{Up}$, $i\leftarrow 0$ and $\Tilde{x}_i = \Tilde{\ell}_0$.   
    \item while stage $= \textsc{Up}\qquad$  \textcolor{brown}{$\ll$ up stage $\gg$}
    \begin{itemize}
        \item $\Tilde{x}_{i^*}\leftarrow$ the node at highest level corresponding to $x$.
        \item  $\Tilde{y}_{i^*} \leftarrow
        \textsc{RightPointer}(\Tilde{x}_{i^*})$.
        \item  if $\textsc{IsDescendant}(u, y, T) = \textsc{True}$, $i\leftarrow i^*$ and $\Tilde{x}_i \leftarrow \Tilde{y}_{i^*}$.
        \item otherwise, stage $\leftarrow \textsc{Down}$, $i \leftarrow i^*-1$.
    \end{itemize}
   
    \item while $i \geq 0 \qquad$ \textcolor{brown}{$\ll$ down stage $\gg$}
    \begin{itemize}
        \item  $\Tilde{y}_{i} \leftarrow
        \textsc{RightPointer}(\Tilde{x}_{i})$.
        \item  if $\textsc{IsDescendant}(u, y, T) = \textsc{True}$,   $\Tilde{x}_i \leftarrow \Tilde{y}_{i}$.
        \item otherwise, $i \leftarrow i-1$.
    \end{itemize}
    \item Return $x$.
\end{enumerate}
\end{mdframed}
    \caption{\textsc{TrackRightMostLeaf} searches the rightmost leaf of $u$ by searching on a skip list. }
    \label{fig:trackright}
\end{figure}

\begin{proof}[Proof of~\Cref{lmm:search_leaf}]  Since all descendant leaves of $u$ form a continuous subsequence of $L_T$, we can follow pointers of $L_T$ to find the rightmost leaf of $u$ from an arbitrary leaf $\ell$. Therefore, the correctness of \textsc{TrackRightMostLeaf} follows~\Cref{lm:leaf-descedant}. 

For running time, observe that \textsc{IsDescendant$(u, x, T)$} has $O(1)$ running time. Therefore, every iteration in the while loops in steps 3 and 4 of \textsc{TrackRightMostLeaf} run in $O(1)$ time. As the height of a skip list is $O(\log n)$, the number of steps going up in the first stage is $O(\log n)$. However, we also have to bound the number of steps the search jumps to the right following the right pointers. The observation is that, in the first stage, every time we follow the right pointer in the skip list, the distance from the starting point of the search to the current node increases exponentially. Thus, the number of jumping-to-the-right steps is also $O(\log n)$,  implying that the total running time of the first stage is $O(\log n)$. The total running time of the second stage is also $O(\log n)$ for the same reason. 

Lastly, we need to keep track of $L_T$ once a new leaf is inserted into $T$.
 As we do not delete leaves from $T$, when a new leaf $x$ is inserted as children of a node $u$, it will be the new rightmost descendant leaf of $u$, and we have to insert $x$ next to its old rightmost descendant leaf, say $y$. We invoke   \textsc{TrackRightMostLeaf} to find $y$ and its position in $L_T$. Since $L_T$ is a doubly linked list, inserting $x$ next to $y$ takes only $O(1)$ time, making the total time to update $L_T$ $O(\log n)$ per insertion. 

 As $L_T$ changes, the skip list also has to change.  For this purpose, we could use the data structure of Munro, Papadakis, and Sedgewick~\cite{MPS92} that has only $O(n)$ space. Thus, the total space of our data structure is $O(n)$.
\end{proof}

\paragraph{Maintaining ancestral arrays}

Kopelowitz and Lewenstein~\cite{KL07} studied the dynamic weighted ancestor problem, where there is a dynamic weighted tree and the goal is to answer weighted ancestor queries: given a node $v$ and a value $i$, return the first node in the path from $v$ to the root whose value is less than $i$. 
Note that the weight of a node is higher than that of its parent, and the tree is updated dynamically by inserting a leaf or subdividing an edge; there are no deletions. They developed two data structures:
\begin{enumerate}
    \item  \EMPH{Ancestral representative} data structure built on top of the centroid path decomposition (\Cref{def:centr-decom-anc-arry}) of $T$. This data structure also maintains for every node $v$ a list of heads of the centroid paths of $v$, called the \EMPH{head record} of $v$.
    \item  \EMPH{Dynamic predecessor} data structure that maintains all the centroid paths in the centroid path decomposition and supports predecessor search\footnote{In the predecessor search problem, one has to design a data structure for a set of integer keys such that given an integer $x$, it has to quickly return the largest (smallest) key at most (at least, resp.) $x$, called the predecessor (successor, resp.) of $x$.} on each path.
\end{enumerate}

To query a weighted ancestor, their basic idea was to search the two data structures. First, from the head record of $v$, they determined the (head of the) centroid path $\pi$ of $v$ that contains the weighted ancestor. Then, they executed a \emph{predecessor search} supported by the dynamic predecessor data structure to search for the result\footnote{To be more precise, they either searched on $\pi$ or the centroid path of $v$ following $\pi$.} on $\pi$.

\begin{remark}\label{rm:update_query_time} 
In the work of Kopelowitz and Lewenstein~\cite{KL07}, each insertion to the tree induces a constant number of what they called \emph{predecessor updates}.  Each predecessor update could be (a) creating at most $\log{(n)}$ nodes in the head record for a new node, (b) adding a new element at the end of the head record, (c) changing the value of an element in the head record of a node, and (d) adding a new node into a centroid path. The running time of a predecessor update depends on the choice of the data structures for the head records and centroid paths. (For our purpose, we simply use an array to store each head record  and a skip list to store each central path to achieve $O(\log n)$ time per predecessor update.)

The running time per weighted ancestor query was bounded by the number of predecessor searches. One predecessor search is executed on the head record of $v$, which has size $O(\log n)$, while another predecessor search is on a centroid path of size at most $n$. Thus, the running time of querying a weighted ancestor is at most $\max\{T_1(\log{(n)}), T_2(n)\}$, where $T_1$ and $T_2$ depend on the choices of data structures for implementing predecessor search on the head records and centroid paths. 
\end{remark}

The result of Kopelowitz and Lewenstein~\cite{KL07} is summarized in the following lemma. 
\begin{lemma}[Kopelowitz and Lewenstein~\cite{KL07}, Theorem 7.1]\label{lmm:dynamic_wa}
    Given a weighted tree of $n$ nodes that can be updated by adding a new leaf or subdividing an edge, each insertion costs a constant number of predecessor updates, and each weighted ancestor query can be answered by calling a constant number of predecessor searches. 
\end{lemma}

Here, we do not need to support a weighted ancestor query. Furthermore, we only look for an $O(\log n)$ search time instead of  $O(\log^*(n))$ or faster as in the work of Kopelowitz and Lewenstein~\cite{KL07}). As a result, we could use much simpler data structures than theirs. Specifically, we will use an array to store the head record of every vertex; this is our ancestral array. We also use a skip list to store each centroid path. (Therefore, we can discard most of their dynamic predecessor data structure.) When a node $v$ is added to the tree, we update its ancestral array in $O(\log n)$ time by looking at the ancestral array of its parent and add $v$ to an appropriate centroid path also in $O(\log{(n)})$ time. As noted in \Cref{rm:update_query_time}, the insertion time is $O(\log{(n)})$.

For completeness, we now briefly zoom in on the technical ideas of Kopelowitz and Lewenstein~\cite{KL07}. This overview is not necessary to understand our work, and hence, readers could skip this part if needed. 

First, as centroid paths are determined based on their sizes, the authors~\cite{KL07} needed to maintain the sizes of the subtrees rooted at the heads of centroid paths. When a new node is added, a centroid path $\pi$ could have a new head: the current head $u$ leaves $\pi$ and joins to the end of the preceding centroid path, while the child $v$ of $u$ \EMPH{in $\pi$} becomes the new head of $\pi$. And we have to maintain the size of the subtree, denoted by $|T(v)|$, rooted at $v$. If $v$ is the only child of $u$, then $|T(v)| = |T(u)|-1$. Otherwise, other child of $u$ are also heads and hence $|T(v)| = |T(u)| - \sum_{v'\in \{\text{children of }u\}\setminus \{v\}}|T(v')| - 1$. The sum $\sum_{v'\in \{\text{children of }u\}\setminus \{v\}}|T(v')|$ can be maintained directly at $u$. Therefore, we can compute $|T(v)|$ in constant time. As inserting a node into $T$ could lead to changing at most $\log{(n)}$ heads, the running time to maintain the subtree sizes at these heads is $O(\log{(n)})$.

Maintaining the sizes of the heads is only the first step; the main challenge is to {maintain the head records} in an efficient time per insertion. 
Adding a new node could lead to changing multiple heads of the centroid paths, which induces updating the head records of many nodes.
To solve this problem, instead of updating the head records of nodes immediately, they waited for more insertions.
In the meantime, a centroid path could be ``oversized'', as more nodes are inserted into it but its head is not updated. 
They observed that using an ``outdated'' version of head records still guarantees the correctness of ancestor queries.
Furthermore, this observation allows updating the data structure in the background, then they could deamortize their construction by recursively splitting the tree into subtrees of $O(\log(n))$ nodes.
Ultimately, they achieved a worst-case constant bound on the number of predecessor updates (and predecessor searches) per insertion (weighted ancestor query, resp.).

\subsection{General Case: Maintaining DFS Ordering of a PIT}\label{sec:pred_succ}

In this section, we show how to maintain the DFS leaf ordering $\sigma$ of $T$ as claimed in \Cref{lmm:maintain_ordering}.  We will store $\sigma$
as a doubly linked list that only contains \emph{active leaves}. Our goal is to transform a structure $D$ for trees without inactive leaves, such as the data structure constructed in the previous section, to a data structure that works for PIT with inactive leaves. We refer readers to the beginning of \Cref{sec:leaf_tracker} for an overview of our ideas. A key  data structure is an \EMPH{active tracker} formally defined below.

\begin{definition}[Active Tracker]\label{def:active-tracker}
    A data structure that maintains a dynamic rooted tree under updates by inserting a leaf, subdividing an edge, and marking a leaf as deleted. It supports three following queries:
    \begin{itemize}
        \item \textsc{IsActive}$(u, T)$: check if $u$ is an active node.
        \item \textsc{GetActiveLeaf$(u, T)$}: given a node $u$, return $l_u$, where $l_u$ is an active leaf under the subtree rooted at $u$. 
        \item \textsc{GetLowestActiveAncestor$(u, T)$}: return the lowest ancestor $v$ of $u$ such that $v$ is an active node.
    \end{itemize}
\end{definition}

By the end of this section, we will design an active tracker data structure with linear space that supports fast query time, as claimed in the following lemma. Note that we only apply procedure \textsc{GetActiveLeaf$(u, T)$} on an active node $u$.

\begin{lemma}\label{lmm:active-tracker}
    There is an active tracker data structure with $O(n)$ space that has $O(\varepsilon^{-O(\lambda)} + \log{(n)})$ time per update and supports \textsc{IsActive} in $O(1)$ time, and \textsc{GetActiveLeaf} and \textsc{GetLowestActiveAncestor} in $O(\log n)$ time.
\end{lemma}

Given the active tracker data structure by \Cref{lmm:active-tracker}, we now show how to update the doubly-linked list $\sigma$ under updates. Conceptually, we could think of $\sigma$ as a DFS leaf ordering of the tree $\hat{T}$ obtained by iteratively contracting inactive nodes to their parents, and therefore, we could use the data structure in the previous section for maintaining $\sigma$. Specifically, a skip list is maintained on top of $\sigma$ for binary search, as we did with $L_T$, with ancestral arrays to be keys. However, ancestral arrays of a node in $T$ here could contain inactive nodes. The observation is that in the search of the leftmost/rightmost leaf, we only compare ancestral arrays of active nodes, and by definition, every ancestor of an active node is active. Thus, the ancestral arrays of nodes that we compare during the search only contain active nodes.

\subsubsection{Updating $\sigma$}

 When an active leaf $(q,q,0)$ corresponding to a point $q\in S$ is removed from $T$, we then follow the pointer at $(q,q,0)$ to find its corresponding node $q$ in $\sigma$. Then, we simply remove $q$ from $\sigma$, which can be done in $O(1)$ time since $\sigma$ is a doubly-linked list.  It remains to consider insertions. If a null leaf is inserted to $T$, we do nothing, so the difficult case is inserting an active leaf. 

Suppose that an active leaf $(q,q,0)$ is inserted to $T$. Let $u$ be its parent in $T$. We consider two cases:
\begin{enumerate}
    \item \textbf{$u$ is active.~} We query the active tracker data structure: $l_u \leftarrow  \textsc{GetActiveLeaf}(u, T)$.  This means $u$ is an active node, and $(q,q,0)$ will become the rightmost leaf of $u$. Thus, we simply insert $(q,q,0)$ by finding the current rightmost leaf, say $x$, of $u$ by calling  \textsc{TrackRightMostLeaf$(u, L_F)$} (in \Cref{fig:trackright}) and insert $q$ right after $x$ in $\sigma$. By \Cref{lmm:search_leaf}, the running time of this step is $O(\log n)$.
    \item \textbf{otherwise, $u$ is inactive.~~} Let $v\leftarrow\textsc{GetLowestActiveAncestor}(u, T)$. We then examine every child of $v$ to find the \emph{active child} $v_x$ that the DFS visits before $u$ and after other active children of $v$; we can afford to do so since $v$ only has $\eps^{-O(\lambda)}$ children. If $v_x$ exists, then we find the rightmost leaf, say $x$, of $v_x$ by calling \textsc{TrackRightMostLeaf$(v_x, \sigma)$} and insert $q$ after $x$ in $\sigma$. Note that in the DFS order, $x$ is followed by the active descendant leaves of $u$ in $\sigma$.  Since $(q, q, 0)$ is the only active leaf in descendants of $u$, $q$ must be the new successor of $x$ in $\sigma$. Otherwise, $v_x$ does not exist. Since $v$ is active, there exists an active child $v_y$ of $v$ such that $v_y$ is visited after $u$ and before other active children of $v$. We then find the leftmost leaf, say $y$, of $v_y$  by calling \textsc{TrackLeftMostLeaf$(v_y, L_F)$} and insert $q$ before $y$ in $\sigma$. 
    
    Since checking if a child of $v$ is active can be done in $O(1)$ time, the total time to find $v_x$ and $v_y$ is $\eps^{-O(\lambda)}$. By \Cref{lmm:search_leaf}, finding the leftmost or the rightmost leaf can be done in $O(\log n)$ time. Thus, the total running time of this step is $O(\eps^{-O(\lambda)} + \log n)$.
\end{enumerate}

By considering all  the cases, updating $\sigma$ can be done in $O(\eps^{-O(\lambda)} + \log n)$.

\subsubsection{Active tracker data structure}\label{subsec:active-leaf-tracker}

Now, we provide details of active data structures as claimed in~\Cref{lmm:active-tracker}. For each centroid path $\pi$ in the centroid decomposition of $T$, we keep track of the \EMPH{lowest active node} of $\pi$, denoted by \textsc{Lowest}$(\pi)$. Specifically, the head of $\pi$ will store a pointer that points to \textsc{Lowest}$(\pi)$, so that it takes only $O(1)$ time to find \textsc{Lowest}$(\pi)$. Initially, \textsc{Lowest}$(\pi)$ is \textsc{Null}, then it will be updated while nodes of $T$ are inserted or deleted. For now, we assume that \textsc{Lowest}$(\pi)$ is given, and we will use it to implement all other operations of the active tracker data structure; the pseudocodes are given in \Cref{fig:active-leaf-ds}. We will come back to the issue of maintaining \textsc{Lowest}$(\pi)$ for every path $\pi$ later.

\begin{figure}[!htb]
    \centering
\begin{mdframed}[nobreak=true]
    \textbf{\textsc{FindCentroidPath$(u, T)$}}: \textcolor{brown}{find the centroid path containing $u$}
    \begin{itemize}
        \item[]  Let $v$ be the last element of the ancestral array $A_u$ of $u$, and $\pi_u$ be the centroid path whose head is $v$. Then return $\pi_u$. 
    \end{itemize}

    \noindent\textbf{\textsc{IsActive}}$(u, T)$: 
    \begin{enumerate}
        \item Let $\pi_u \leftarrow \textsc{FindCentroidPath}(u, T)$.
        \item Let $t \leftarrow \textsc{Lowest}(\pi_u)$.
        \item  If $t$ is \textsc{Null} or the level of $t$ is higher than $u$, then $u$ is an inactive node, and we return \textsc{False}. Otherwise, we return \textsc{True}.
    \end{enumerate}
    
    \noindent\textbf{\textsc{GetActiveLeaf}}$(u, T)$:
    
    \textcolor{brown}{$\ll u$ is guaranteed to be active $\gg$}
    \begin{enumerate}
        \item  Let $\pi_u \leftarrow \textsc{FindCentroidPath}(u, T)$.
        \item  Let $t \leftarrow \textsc{Lowest}(\pi_u)$.
        \begin{itemize}
            \item[(a)] if $t$ is a leaf, we return $t$. 
            \item[(b)] Otherwise, we pick an active child $t'$ of $t$ (by  maintaining a pointer to an arbitrary active child at $t$) and return \textsc{GetActiveLeaf}$(t', T)$. 
        \end{itemize}
    \end{enumerate}

    \noindent\textbf{\textsc{GetLowestActiveAncestor}}$(u, T)$: \textcolor{brown}{$\ll$ see \Cref{fig:lowest_active_ancestor} $\gg$}

    \begin{enumerate}
        \item   Let $v \in A_u$ be the lowest head such that its corresponding centroid path $\pi$ has $\textsc{Lowest}(\pi) \neq \textsc{Null}$. (We find $v$ by considering every element of $A_u$.)
       
        \item  {[Case 1: $u \in \pi$.]} If the level of \textsc{Lowest}$(\pi)$ is higher than $u$, we return \textsc{Lowest}$(\pi)$. Otherwise, we return $u$.
        \item  {[Case 2: $u\not\in \pi$.]} Let $j$ be the index of $v$ in $A_u$. Let $v'$ be the parent of $A_u[j+1]$.   If $v'$ is active, return   $v'$. Otherwise, return  \textsc{Lowest}$(\pi)$.
        
    \end{enumerate}
        
\end{mdframed}
    \caption{Operations supported by the active tracker data structure.}
    \label{fig:active-leaf-ds}
\end{figure}

\begin{figure}[!ht]
    \begin{tikzpicture}
	 \node at (0,0){\includegraphics[width=1.0\linewidth]{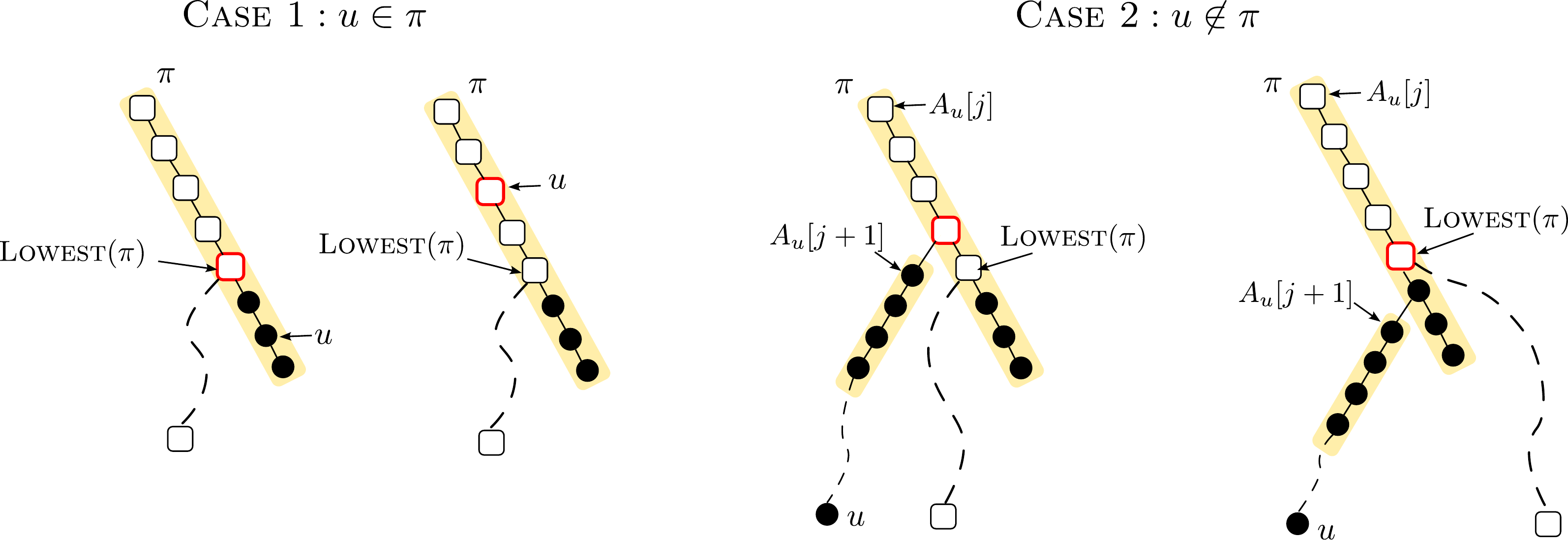}};	
    \end{tikzpicture}    

    \caption{Illustrate \textsc{GetLowestActiveAncestor} operation. Let $\pi$ be the lowest centroid path of $u$ whose head is active, and $j$ be the index of $\pi$'s head in $A_u$. Rectangular white nodes are active, round black nodes are inactive. The lowest active ancestor of $u$ is the red node. To find the lowest active ancestor, when $u \in \pi$, we consider the levels of $u$ and $\textsc{Lowest}(\pi)$, when $u \not\in \pi$, we consider the levels of $A_u[j+1]$'s parent and $\textsc{Lowest}(\pi)$.} 
    \label{fig:lowest_active_ancestor}
\end{figure}

\paragraph{Correctness.~} Since the ancestral array $A_u$ of a node $u$ contains the heads of all centroid paths of $u$ sorted by decreasing level, and $u$ belongs to some centroid path, \textsc{FindCentroidPath$(u, T)$} in \Cref{fig:active-leaf-ds} correctly returns the centroid path containing $u$ and its head.  Clearly, if $u$ is active, then by definition of $\textsc{Lowest}$, the lowest active node in $\pi_u$ in line 2 of \textsc{IsActive}$(u, T)$ has a level at most that of $u$, and hence \textsc{IsActive}$(u, T)$ correctly decides if $u$ is active or not. 

For \textsc{GetActiveLeaf}$(u, T)$, the input node $u$ is always an active node. Since $u$ is active, the node $t$ in line 2 exists and has a level at most $u$. If $t$ is a leaf, then it will be an active leaf of $u$, so the algorithm is correct. Otherwise, it recursively finds an active leaf from an active child $t'$ of $t$. By definition of an active node, $t'$ has an active leaf in its subtree, and hence the algorithm is correct.

For \textsc{GetLowestActiveAncestor}$(u, T)$, case 1 is self-explained. For case 2, by definition of $A_u$ and $v$,  $A_u[j+1]$ is the head of a centroid path, say $\pi'$, of $u$ such that every node $\pi'$ is inactive, and therefore, the lowest active ancestor of $u$ must be an ancestor of $A_u[j+1]$. Since $v'$ is the parent of $A_u[j+1]$, if it is active, then it is the lowest active ancestor of $u$, and hence the algorithm is correct. Otherwise, observe that $v'$ is in $\pi$ by the definition of centroid path decomposition. Since $v'$ is in active, every active node in $\pi$ is an ancestor of $v'$, and hence \textsc{Lowest}$(\pi)$ is the lowest active ancestor of $v'$, which is also of $u$. Therefore, the algorithm is correct.

To show \Cref{lmm:active-tracker}, it remains to bound the running time of each operation.

\begin{proof}[Proof of \Cref{lmm:active-tracker}] Observe that \textsc{FindCentroidPath}  runs in $O(1)$ time  and hence \textsc{IsActive} also runs in $O(1)$ time. 

In  \textsc{GetActiveLeaf}, observe that the algorithm recursively invokes $\textsc{Lowest}$ on centroid paths of the final active leaf, and there are only $O(\log n)$ such paths by definition of the centroid path decomposition. As we maintain an active child for every lowest active node of a path, $t'$ in step 2(b) can be found in $O(1)$ time, giving $O(\log n)$ total running time to find an active leaf. 

In \textsc{GetLowestActiveAncestor}, we can find $v$ in step 1 in $O(\log n)$ time since $|A_u| = O(\log n)$. Other steps can be implemented in $O(1)$ time, and hence, the total running time is  $O(\log n)$.
\end{proof}

\paragraph{Maintaining \textsc{Lowest}$(\pi)$ pointers.~} Cole and Gottlieb~\cite{GC06} described major ideas for maintaining \textsc{Lowest}$(\pi)$ pointers. Here, we review their ideas. We also fill in the missing detail of handling the changes to the dynamic centroid path decomposition. Recall that there are three types of simple updates to the PIT $T$: adding a leaf, marking a leaf as deleted, and subdividing an edge. Observe that subdividing an edge does not change the active/inactive status of the endpoints of the subdivided edge. Marking a leaf as deleted could turn some ancestors of the leaf from active to inactive, while adding a leaf could, on the other hand, turn its ancestors from inactive to active. 

One subtlety here is that subdividing an edge and adding a leaf could change the size of the tree $T$ and trigger updates to the centroid decomposition of $T$. These updates are handled by the underlying dynamic data structure for the centroid decomposition. While subdividing an edge does not change the status of all other nodes, adding a new leaf could cause massive changes. Therefore, when adding a new leaf, we will consider it as \emph{inactive} so that the status of all other nodes is unchanged, and hence, the underlying dynamic data structure for the centroid decomposition could proceed as usual. Note that the underlying dynamic data structure for the centroid decomposition now has to maintain an additional pointer \textsc{Lowest}$(\pi)$ for centroid path $\pi$, and with the assumption that the status of all the nodes is unchanged, this can be done by slightly augmenting the data structure by Kopelowitz and Lewenstein in the previous section. Once the dynamic updates to the structure of the centroid decomposition are done, we change the status of the newly inserted node to active and update the status of other nodes. This is exactly the same problem with marking a leaf as deleted since we simply change the status of this leaf from active to inactive. (We note that the newly inserted leaf could be null, and in this case, there is no need to turn it to active; hence, we only need to consider inserting a non-null leaf.)

Now, we will handle a leaf status change from active to inactive or vice versa.  Let $\pi$ be a centroid path, and  $t$ is the current lowest active node of $\pi$. Let $i$ be the level of $t$. If the only active leaf of $t$ changes to inactive, then $t$ becomes inactive. Then we need to find another lowest active node to update \textsc{Lowest}$(\pi)$. Observe that the head of $\pi$ is active if $\pi$ contains an active leaf or a node with an active child not on $\pi$, called an \EMPH{off-path child}. The idea of Cole and Gottlieb~\cite{GC06}  is to keep track of nodes that have an active off-path child, and this task can be efficiently done by using a balanced binary tree.  Here, we use a skip list instead of a balanced binary tree for two following reasons: (1) it was not clear how Cole and Gottlieb~\cite{GC06} updated  binary trees when their centroid paths change, (2) a list of nodes in $\pi$ with an active off-path child is a subpath of $\pi$, thus we can track these nodes in a skip list by the same way we maintain $\pi$, and therefore, we can resolve (1) effectively.

Now consider a centroid path $\pi$ and a node $t$ of $\pi$. If $t$ is an active leaf or $t$ has an active off-path child, then we say $t$ is a \EMPH{low candidate} of $\pi$. Let $B(\pi)$ be the skip list that contains all the low candidates of $\pi$, and the keys to $B(\pi)$ are the levels of the candidates. 
We claim that the lowest node of $B(\pi)$ is \textsc{Lowest}$(\pi)$. 
To see this, let $z$ be the lowest active node of $\pi$. 
If $z$ is a leaf, since $z$ is active, $z$ is a low candidate of $\pi$.
If $z$ is not a leaf, since the child of $z$ in $\pi$ is either inactive or null (in the case where $z$ is the low endpoint of $\pi$), $z$ must have an active off-path child. Thus, $z$ is also a low candidate of $\pi$. In both cases, $z$ is maintained in $B(\pi)$. Therefore, we can update \textsc{Lowest}$(\pi)$ by simply taking the lowest node in $B(\pi)$.

Let $l$ be the leaf whose status changes.  Since only ancestors of $l$ change status, we only need to update all the centroid paths of $l$ in bottom-up order, and there are only $O(\log n)$ such paths. However, we note that, since a single centroid path could have up to $\Omega(n)$ nodes, a balanced binary tree could incur $O(\log{(n)})$ time per update, potentially bringing the total update time up to $\Theta(\log^2(n))$.  Cole and Gottlieb~\cite{GC06} resolved this issue by observing that 
if a centroid path $\pi$ has at least one low candidate, then changing a leaf under $\pi$ from inactive to active does not change the status of the head $v$ of $\pi$. Similarly, if $\pi$ has more than one low candidate, then changing a leaf under $\pi$ from active to inactive does not change the status of the head.
In both cases, the parent of $v$ is still a low candidate of the \EMPH{parent path} of $\pi$, which is the path containing the parent of $\pi$'s head, and therefore, we could terminate the status update at $\pi$. For all centroid paths that are descendants of $\pi$ (which are paths whose heads are descendants of $\pi$'s head), we update their skip lists in $O(1)$ since they have at most one low candidate. Thus, in the entire process, $\pi$ is the one path that could incur in  $O(\log{(n)})$ time to update $B(\pi)$, and therefore the total running time is $O(\log{(n)})$.

In addition, recall that we maintain a pointer to an arbitrary active off-path child for every low candidate (see \textsc{GetActiveLeaf}). Let \textsc{ActiveChild}$(t)$ be an active off-path child of a low candidate $t$; \textsc{ActiveChild}$(t)$ is \textsc{Null} if $t$ does not have any.
We also count the number of active off-path children for each low candidate by \textsc{ActiveCount}$(t)$. Now we can check in $O(1)$ when $t$ in $\pi$ is no longer a low candidate to remove $t$ out of $B(\pi)$.

In \Cref{fig:status-change}, we show the pseudocode of how to change the status of a leaf from active to inactive, following the discussion above. The code for changing from inactive to active is very similar. At step 2, we add $l$ to $B(\pi)$, assign $l$ to \textsc{Lowest$(\pi)$}, and terminate if $|B(\pi)|> 1$. We decrease $\textsc{ActiveCount}(t)$ by $1$ at 3(b). We do not find an active off-path child to replace $v$ in step 3(c). We terminate if $\textsc{ActiveCount}(t) > 1$ at 3(d). At 3(e), we add $t$ to $B(\pi')$ and assign \textsc{ActiveChild$(t)=v$, ActiveCount$(t) = 1$}. At 3(f), we terminate if $|B(\pi')| \geq 2$.

\begin{figure}[!htb]
\begin{mdframed}

            
            

    \textbf{\textsc{DeActivate}}$(l, T)$: \textcolor{brown}{change the status of a leaf $l$ from active to inactive}

    \begin{enumerate}
        \item Let $\pi \leftarrow \textsc{FindCentroidPath}(l, T)$, and $v$ be the head of $\pi$.   
        \item Remove $l$ out of $B(\pi)$ and update $\textsc{Lowest}(\pi)$ by getting the lowest node in $B(\pi)$.
        If $|B(\pi) > 0|$, $v$ is still active after deleting $l$, we terminate.
        \item  Repeat the following steps as long as the head of $\pi$ is not the root of $T$:
        \begin{enumerate}
        \item  Let $\pi'$ be the parent centroid path of $\pi$, and $t$ be the parent of $v$ in $\pi'$.
        \item Decrease \textsc{ActiveCount}$(t)$ by $1$. 
        \item If $\textsc{ActiveChild}(t) = v$, we find another active off-path child to replace $v$, or assign \textsc{Null} to this pointer if $t$ does not have any.
        \item  If $\textsc{ActiveCount}(t) > 0$, $t$ is still a low candidate of $\pi$, we terminate.
        \item  Otherwise, $t$ is no longer a low candidate of $\pi'$ after deleting $l$.
        \begin{itemize}
             \item\textnormal{[Update $B(\pi')$.]} Remove $t$ out of $B(\pi')$, the skip list maintaining low candidates of $\pi'$. 
             \item\textnormal{[Update \textsc{Lowest}$(\pi')$.]} Find \textsc{Lowest$(\pi')$} by getting the lowest node in $B(\pi')$.

             \item\textnormal{[Update $t$.]} $\textsc{ActiveChild}(t) \leftarrow \textsc{Null}$ and  $\textsc{ActiveCount}(t) \leftarrow 0$
            
        \end{itemize}
  \item If $|B(\pi')| \geq 1$, terminate. In this case, the head of $\pi'$ remains active after deleting $l$, and deleting $l$ does not change the low candidates of ancestor paths of $\pi'$.
        \item Update $\pi \leftarrow \pi'$.
        \end{enumerate}
    \end{enumerate}
    
\end{mdframed}
     \caption{Changing status of a leaf $l$ from active to inactive.}
     \label{fig:status-change}
 \end{figure}

\begin{observation}
\label{obs:byinsert_rupdate}
    \textsc{Activate} runs in $O(\log{(n)})$ time, \textsc{DeActivate} runs in $O(\varepsilon^{-O(\lambda)} + \log{(n)})$ time.
\end{observation}

\begin{proof}
     Updating $t$ to $B(\pi')$ in step 3 costs the most running time, which is $O(\log{(n)})$ time if $B(\pi')$ has at least one low candidate, and $O(1)$ time if it has one candidate. There are at most $O(\log(n))$ centroid paths, and once the algorithm considers a path $\pi'$ that has more than one low candidate, it terminates. Thus, the cost of adding a node to a skip list is $O(1)$ for each descendant centroid path of $\pi'$, and $O(\log{(n)})$ only for $\pi'$.
     Therefore, \textsc{Activate} runs in $O(\log{(n)})$ time. 
     
     For \textsc{DeActivate}, it may have to update \textsc{ActiveChild} at step 3(c).
     Observe that there is at most one centroid path $\pi'$ such that $\textsc{ActiveCount}(t) > 0$ occurs, thus the running time of step 3(c) is $O(\varepsilon^{-O(\lambda)})$ when the algorithm considers $\pi'$, and $O(1)$ when the algorithm considers descendant centroid paths of $\pi'$.
     Therefore, \textsc{DeActivate} runs in $O(\varepsilon^{-O(\lambda)} + \log{(n)})$ time.
\end{proof}

%% file: 10.nettree.tex
In this section, we show how to maintain a net tree for a dynamic point set $S$ as described in \Cref{thm:nettree_ds}, which we restate below:
\ThmDynamicNetTree*

Recall that a jump from $(x, i)$ down to $(x, j)$ for $i > j+1$ intuitively hides nodes $(x, j+c)$ for $j < j+c < i$, and we call the nodes $(x, j+c)$ \EMPH{hidden nodes}.
In a net tree cover, sometimes we merge two nodes at level $i$ if their distance is at most $3c \cdot \frac{\delta}{\varepsilon^i}$ (they are $3c$-close) for some parameter $c$. Therefore, we need to guarantee that a hidden node is not $3c$-close to any existing node at the same level, which inspires the jump isolation property. 

Previously, Cole and Gottlieb~\cite{GC06} constructed a net tree for a dynamic point set with $O_{\lambda}(\log{n})$ time per update and $O(n)$ space. Specifically, they maintained the covering and jump isolation properties for the net tree by modifying only $O_\lambda(1)$ nodes per update. To this end, they introduced the concept of \emph{rings} and used $5$ rings in their construction.   Here, we simplify their construction and the analysis. We observe that Cole and Gottlieb~\cite{GC06} used one ring among the five for the search operation and four rings to consider the distance of a node to its descendants. We simplify their insert operation, which is the bulk of the technical details, and our simplified operation only requires four rings.  Furthermore, we associate each ring in our construction with specific functionality, making our overall construction simpler and more intuitive.

We remark that our close-containment property in \Cref{thm:nettree_ds} is more relaxed (which makes it easier to guarantee) than that of Cole and Gottlieb~\cite{GC06}. Specifically, in their work, for any node $(y, k)$ and any ancestor $(z, i)$ of $(y, k)$ has $d_X(z, y) < \frac{4}{5}\frac{\delta}{\varepsilon^i} - \frac{\delta}{\varepsilon^k}$, which is smaller than the upper bound in our close-containment property. Our relaxation is due to the difference in the way we set up the parameters of packing and covering properties. 

\paragraph{Overview of the dynamic net tree.~} We now sketch the structure of the dynamic net tree with the covering, jump isolation, and close-containment properties. This structure was largely developed by Cole and Gottlieb~\cite{GC06}; we simplify some parts that we will detail along the way. It is useful to think of each node $(x, i)$ in the net tree as associated with a ball $B(x, \frac{\delta}{\varepsilon^i})$; all points under the subtree rooted at $(x, i)$ must be contained in this ball by the close-containment property. (Sometimes, we use the node and ball terminologies in the net tree interchangeably.) In addition, the close-containment property implies that if $(x, i)$ is an ancestor of $(y, j)$ for $j < i$, then $B(y, \frac{\delta}{\varepsilon^j}) \subseteq B(x, \frac{\delta}{\varepsilon^i})$. We note that it is not so hard to show that the covering property implies the close-containment property when $\eps$ is sufficiently small (see \Cref{lmm:contain}). 

For the jump isolation property, recall that a jump from $(x, i)$ down to $(x, j)$ for $i > j+1$ is a long edge connecting two nodes of the same point $x$ at level $i$ and $j$. Look at two corresponding balls centered at $x$: $B(x, \frac{\delta}{\varepsilon^i})$ and $B(x, \frac{\delta}{\varepsilon^j})$.
The jump isolation means that any point at level $k < i$ outside the smaller ball, which is $B(x, \frac{\delta}{\varepsilon^j})$, must be at a distance more than $b\frac{\delta}{\varepsilon^k}$ from $x$.  Note that the dynamic net tree of Cole and Gottlieb~\cite{GC06} also has a similar structure but with a slightly different covering property and the jump isolation property.

Whenever a new point $q$ is inserted, we first need to find the parent node $(t,i)$ for $q$ at some level $i$. Once $(t,i)$ is found, we create a new node $(q, i-1)$ as a child of $(t, i)$, and finally create a jump from $(q, i-1)$ down to $(q, 0)$. Sometimes, we have to break a jump, which means adding $(t, i)$ in the middle of the edge between $(t, j)$ and $(t, k)$ for $j < i < k$. For a node $(t,i)$ to be the parent of $q$, it has to satisfy the covering property. We could find $(t,i)$ by visiting the dynamic net tree level-by-level, but doing so would result in a running time to the height of the tree, which can be up to $\Omega(n)$. Instead, we find $(t,i)$  in two steps: (i) we first find a node $(t', i')$ which is closer to the true parent $(t, i)$ of $q$, and (ii) then we find $(t, i)$ ``around the neighborhood'' of $(t', i')$. The node $(t', i')$ we found in the first step has $i'$  to be the lowest level such that $d_X(q, t')\leq \frac{\delta}{\varepsilon^{i'}}$. This means $(t', i')$ and $(q,0)$ satisfy the close-containment property, which is a relaxation of the covering property. Therefore, we could apply a binary search to find $(t', i')$ via the so-called \emph{containment search} introduced by Cole and Gottlieb~\cite{GC06}. Here, we slightly modify the search condition in the containment search to fit our purpose.  
For step (ii), we show that $(t', i')$ is very close to the parent node $(t, i)$ that we are searching for, and hence we only need to spend $O_\lambda(1)$ additional time to locate $(t, i)$. We also simplify several steps to find $(t, i)$ from $(t', i')$.

A very important subtlety in the search for the parent node $(t,i)$ of a newly inserted point $q$ is that when $(t,i)$ is found, $d_X(t,q)$ might be more than  $\phi \cdot \delta/\eps^i$ so that the covering property would be violated at level $i$ if we made $(q,i-1)$ a child of $(t,i)$. The idea to resolve this issue is to \emph{promote} $(t,i)$ to the next level (the pseudocode in \Cref{fig:add-promote}), to become $(t,i+1)$ so that at level $i+1$, $d_X(t,q)\leq \phi \cdot \delta/\eps^{i+1}$ and hence we can make a child node $(q,i)$ of $(t,i+1)$ without violating the packing property. However, promoting  $(t,i)$ to $(t,i+1)$ might lead to another violation of the covering property between   $(t,i+1)$ and its parent at level $i+2$,  which requires another promotion to resolve and consequently triggers a chain many promotions.  Cole and Gottlieb~\cite{GC06} resolved this issue with \emph{rings} and used five rings in their construction. Here, we only use four rings, from innermost to outermost, and consider which ring the distance between a node and its parent falls into to determine whether a node should be promoted. 
If the distance is in ring 4 (the outermost ring), then we will promote the node so that it belongs to the first three (inner) rings, and hence no further promotion is needed for maintaining the covering property.

We now describe the details of the dynamic net tree, which are organized as follows. 
In~\Cref{sec:rings}, we describe the idea of rings. Then in~\Cref{sec:support_nettree}, we show the containment search and operations designed by Cole and Gottlieb~\cite{GC06} that support the dynamic net tree construction. The full construction with our modifications for search and insertion is shown in~\Cref{sec:main_operations}. Finally, in~\Cref{sec:nettree_analysis}, we prove~\Cref{thm:nettree_ds}.

\subsection{Rings}\label{sec:rings}

To define rings, we will use the following constants:
\begin{equation}\label{equ:nettree_constant}
    \alpha = \frac{1}{4}, \beta = \frac{2}{4},\phi = \frac{3}{4}, \gamma = 1, \psi = \frac{5}{4}, \textrm{ and } \varepsilon \leq \frac{\alpha}{b} 
\end{equation}
where $\alpha$ is the constant in packing distance, $\phi$ is the constant in the covering distance, $\gamma$ is the constant in containment distance, $b$ is the parameter for jump isolation, and $\psi$ is a constant in a search operation (we will see later in~\Cref{sec:main_operations}). Given $b \geq 2$, we have $\varepsilon \leq \frac{1}{8}$.

The values of constants in~\Cref{equ:nettree_constant} satisfy the following inequalities:
\begin{enumerate}
    \item $\psi\varepsilon + \beta  \leq \phi$, $\varepsilon\psi  + \alpha \leq \beta$
    and $\psi\varepsilon \leq \beta$.
    These conditions are to maintain the covering property.
    \item $\phi \leq \gamma(1 - \varepsilon)$. This condition is to maintain the close-containment property.
    \item 
    $\psi\varepsilon \leq \alpha$, $\alpha + \gamma \leq \psi \leq b$,  $\varepsilon \leq \frac{\alpha}{b}$. These conditions are to maintain the jump isolation.
\end{enumerate}

Note that Cole and Gottlieb~\cite{GC06} did not parameterize the jump isolation property; their construction only gives $b = 2$. Their values of the five other constants respectively were $\alpha =\frac{1}{5}, \beta = \frac{2}{5}, \phi = \frac{3}{5}, \gamma = \frac{4}{5}$, $\psi = 1$, and $\varepsilon \leq \frac{1}{5}$. The inequalities above also hold with their values. They used $5$ rings defined by $5$ constants above, while we have $4$ rings with $4$ parameters $\alpha, \beta, \phi, \gamma$.

\begin{wrapfigure}{r}{0.4\textwidth}
\begin{tikzpicture}
    \node at (0,0){\includegraphics[width=1.0\linewidth]{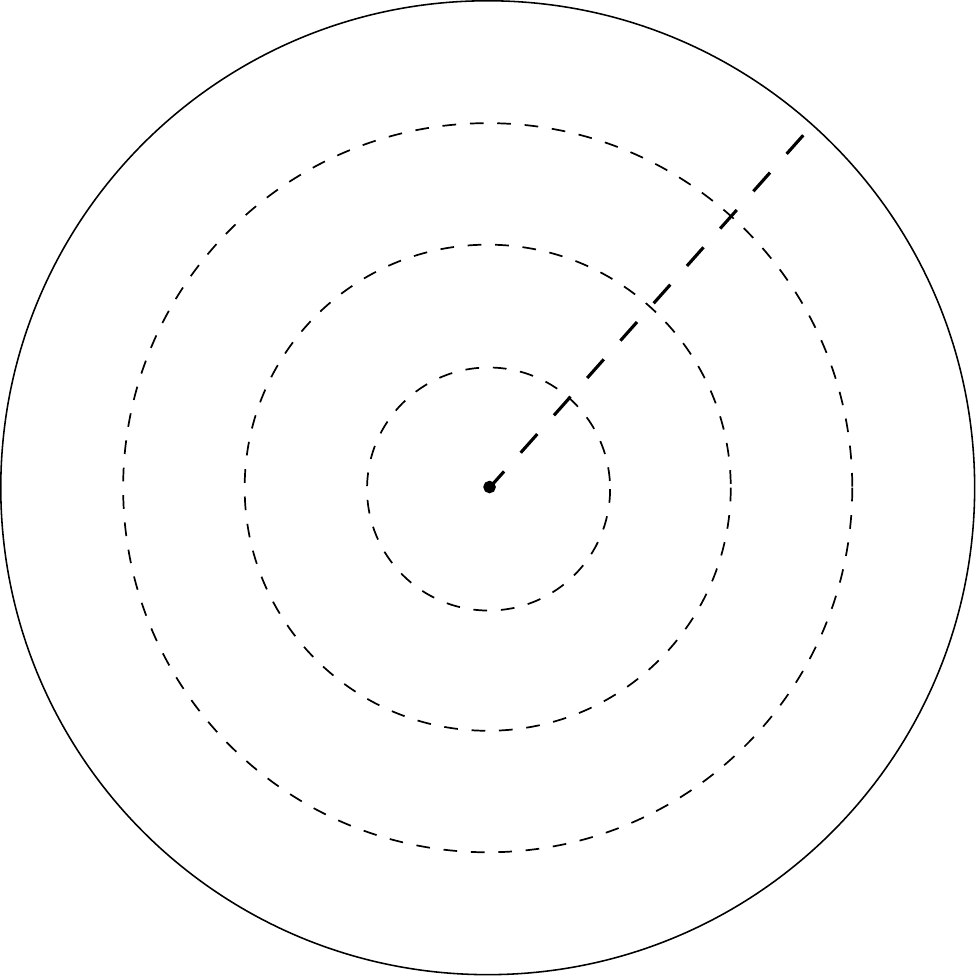} 
};		
    \node at (-0.1, -0.2){\footnotesize \textsc{$x$}};
    \node at (-0.1, 0.4){\footnotesize \textsc{$\alpha\frac{\delta}{\varepsilon^i}$}};
    \node at (0.5, 1.1){\footnotesize \textsc{$\beta\frac{\delta}{\varepsilon^i}$}};
    \node at (1.1, 1.7){\footnotesize \textsc{$\phi\frac{\delta}{\varepsilon^i}$}};
    \node at (1.7, 2.3){\footnotesize \textsc{$\gamma\frac{\delta}{\varepsilon^i}$}};
    \end{tikzpicture}  
\caption{Rings of $B(x, \frac{\delta}{\varepsilon^i})$}
\label{fig:rings}
\end{wrapfigure}

Our rings are formally defined as follows:
\begin{itemize}
    \item Ring\footnote{More precisely, ring $\alpha$ is a ball, but we use ring terminology to be consistent with other rings.} $\alpha$ of a ball $B(x, \frac{\delta}{\varepsilon^i})$ is $B(x, \alpha \frac{\delta}{\varepsilon^i})$. 
    If $p \in B(x, \alpha \frac{\delta}{\varepsilon^i})$, we say $p$ is \EMPH{in ring $\alpha$} of $(x, i)$.
    \item Ring $\beta$ of a ball $B(x, \frac{\delta}{\varepsilon^i})$ is $B(x, \beta \frac{\delta}{\varepsilon^i}) \setminus B(x, \alpha \frac{\delta}{\varepsilon^i})$.
    If $p \in B(x, \beta \frac{\delta}{\varepsilon^i}) \setminus B(x, \alpha \frac{\delta}{\varepsilon^i})$, we say $p$ is \EMPH{in ring $\beta$} of $(x, i)$.
    \item Ring $\phi$ of a ball $B(x, \frac{\delta}{\varepsilon^i})$ is $B(x, \phi \frac{\delta}{\varepsilon^i}) \setminus B(x, \beta \frac{\delta}{\varepsilon^i})$.
    If  $p \in B(x, \phi \frac{\delta}{\varepsilon^i}) \setminus B(x, \beta \frac{\delta}{\varepsilon^i})$, we say $p$ is \EMPH{in ring $\phi$} of $(x, i)$.
   \item Ring $\gamma$ of a ball $B(x, \frac{\delta}{\varepsilon^i})$ is $B(x, \gamma \frac{\delta}{\varepsilon^i}) \setminus B(x, \phi \frac{\delta}{\varepsilon^i})$.
   If $p \in B(x, \gamma \frac{\delta}{\varepsilon^i}) \setminus B(x, \phi \frac{\delta}{\varepsilon^i})$, we say $p$ is \EMPH{in ring $\gamma$} of $(x, i)$.
\end{itemize}

Given $s$ in the set $\{\alpha, \beta, \phi, \gamma\}$, we say a point $p$ is \EMPH{completely out of ring $s$} of $B(x, \frac{\delta}{\varepsilon^i})$ if $p \not\in B(x, s\frac{\delta}{\varepsilon^i})$.
Note that the notion of completely out of ring $s$ only applies to a point $p$ where $d_X(p, x) > s\frac{\delta}{\varepsilon^i}$, not to a point $q$ where it is inside $B(x, s\frac{\delta}{\varepsilon^i})$ and does not belong to ring $s$. As the rings can be linearly ordered by increasing radii, if a point is completely out of ring $\phi$, say, then it is also completely out of rings $\alpha$ and $\beta$.

We classify nodes based on their distances to parents.  Given a node $(t, j)$, we say that $(t, j)$ is a \EMPH{ring-$s$ node} if  $t$ is in ring $s$ of its parent. The following observations follow from the definition of a jump and the covering property.

\begin{observation}\label{obs:ring_nodes}
\begin{itemize}
    \item[\textnormal{(a)}] A node is either a ring-$\alpha$, ring-$\beta$, or ring-$\phi$ node.
    \item[\textnormal{(b)}] A node at the bottom of a jump or created in the middle of a jump is a ring-$\alpha$ node. 
    \item[\textnormal{(c)}] A ring-$\beta$ or ring-$\phi$ node at level $i$ has a parent at level $i+1$.
\end{itemize} 
\end{observation}

\subsection{Containment Search and Internal Operations}\label{sec:support_nettree}

In this section, for completeness, we briefly describe the dynamic net tree of Cole and Gottlieb~\cite{GC06}.
We also provide the details of operations and data structures that we reuse or modify to construct our dynamic net tree. The main goal of Cole and Gottlieb in~\cite{GC06} is to develop a dynamic data structure for solving the approximate nearest neighbor search problem with $O_\lambda(\log{(n)})$ time per update and query. To this end, they developed two data structures: \begin{inlinelist}
    \item a graph (might not be a tree) to maintain (a hierarchy of) nets with the packing, covering, and jump isolation properties and
    \item a spanning of the graph and a central path decomposition on top of the spanning to quickly implement a procedure called \emph{containment search} (which we will give more details below). 
\end{inlinelist} 

One can extract from their data structures a dynamic net tree supporting containment search as stated below.  Each node at level $i$ of the net tree is associated with a ball of radius $\delta/\eps^i$ centered at the point in the node.

\begin{definition}[Containment search data structure of Cole and Gottlieb~\cite{GC06}]\label{def:search_scheme}
There is a data structure that maintains a net tree $T$ and supports the following operations: 
\begin{itemize}
    \item \textsc{ContainmentSearch}$(q, T)$: If $q$ is an existing point, return the leaf $(q, 0)$.
    If $q$ is a new point, return the lowest ball containing $q$. That is, it returns the node $(t, i)$ such that $d_X(t, q)\leq \frac{\delta}{\varepsilon^i}$ and for all $j < i$, there is no $(z, j)$ with $d_X(z, q) \leq \frac{\delta}{\varepsilon^j}$.  This operation runs in $O_{\lambda}(\log{n})$ time.
    \item \textsc{Insert}$(q, T)$: create (several) nodes associated with $q$ in $T$. This operation invokes \textsc{Search}$(q, T)$ and executes $O_\lambda(1)$ additional basic operations; thus it runs in $O_{\lambda}(\log{n})$. 
    \item \textsc {Delete}$(q, T)$: mark $(q, 0)$ as deleted in $O_{\lambda}(\log{n})$ time.
\end{itemize} 
\end{definition}

One key technical idea in the work of Cole and Gottlieb~\cite{GC06}  was to search the parent for a new point in roughly $O(\log{n})$ time using containment search, as described in details below. When a new point $q$ is added to $S$, the insert operation will invoke the search containment to find a node that is close to the parent node. From there, in $O(1)$ additional steps, they can find the exact parent (and then modify the tree).  When a point is removed from $S$, they mark the corresponding leaf as deleted. After a predefined number of deletions, they rebuilt the data structure in the background to remove the nodes associated with deleted points completely (and also to de-amortize). A subtle issue is that the dynamic net tree might contain deleted points since deletions are only marked, while the containment search has to return non-deleted points.  Cole and Gottlieb\cite{GC06} resolved this issue by spending an extra $O(\log{(n)})$ time per deletion.

In addition to containment search, the insert operation invokes several other operations, called \emph{internal operations}, to modify the net tree. These include creating a new node, promoting a node to maintain the covering property, creating or splitting a jump, and fixing a jump to maintain the jump isolation property. We remark that the jump isolation property is important to the correctness of the containment search operation in~\Cref{def:search_scheme}.

Our simplified insert operation given in the next section reuses most of the internal operations stated in this section and only modifies the \textsc{Promote} operation. Therefore, we can still apply the containment search of Cole and Gottlieb~\cite{GC06} to construct our dynamic net tree. 

Now, we describe the containment search and the internal operations by Cole and Gottlieb~\cite{GC06}

\paragraph{Containment search.~} Recall that to maintain the covering property, whenever a point $q$ is added, we need to find a node $(t, i)$ such that $(t, i)$ and $(q, i-1)$ satisfy $d_X(q, t) \leq \phi \frac{\delta}{\varepsilon^i}$ so that we could make $(q, i-1)$ as a child of $(t, i)$ without violating the covering property. Finding $(t, i)$ directly is difficult since the distance upper bound $ \phi \frac{\delta}{\varepsilon^i}$ is very tight. Instead,  Cole and Gottlieb~\cite{GC06} relaxed this upper bound to $\frac{\delta}{\varepsilon^i}$, which is exactly the close-containment property. The relaxed upper allows them to apply binary search on (the centroid-path decomposition of) the net tree. This is because by the close-containment property, any ancestor $(z, k)$ of $(t, i)$ has $d_X(z, q) \leq \frac{\delta}{\varepsilon^k}$ and hence every node on the path from $(t, i)$ to the root satisfies the close-containment property with respect to $(q,0)$, which is ideal for binary search.
The binary search returns either $(t, i)$, which is the node that they are looking for, or a node at a level lower than $i$. In the latter case, they will spend  $O(1)$ extra steps to find $(t, i)$. In both cases, they could find the parent for $q$ in total $O_\lambda(\log{(n)})$ time.

Next, we describe the details of the binary search inside containment search. Recall that the goal is, given a point $q$, to return the lowest ball containing $q$, i.e., the node $(t, i)$ such that $d_X(t, q)\leq \frac{\delta}{\varepsilon^i}$ and for all $j < i$, there is no $(z, j)$ with $d_X(z, q) \leq \frac{\delta}{\varepsilon^j}$, in $O_{\lambda}(\log{n})$ time.  
  Cole and Gottlieb~\cite{GC06} maintained a centroid-path decomposition of the net tree, which partitions the tree into a set of paths; each path is stored as a skip list. Then, binary search is applied to each centroid path, starting from the path $\pi$ containing the root of the tree. It returns the lowest node $(s, j)$ on $\pi$ such that $B(s, \frac{\delta}{\varepsilon^j})$ contains $q$. The algorithm then examines $(s, j)$ and its nearby nodes: either some of them is the lowest ball containing $q$, or there is a child $(r, j-1)$ of a node among them such that $B(r, \frac{\delta}{\varepsilon^{j-1}})$ contains $q$. In the former case, we are done, and the containment search terminates. In the latter case, they switched to the path (in the centroid-path decomposition) containing $(r, j-1)$  to continue the search. The centroid-path decomposition guarantees that switching to a new path reduces half of the nodes in consideration, and therefore, the running time of the containment search is $O_{\lambda}(\log(n))$ in total.

\begin{remark}\label{rm:search-condition} Recall that containment search for a point $q$ returns the lowest node $(t,i)$ such that $d_X(t, q) \leq  \frac{\delta}{\varepsilon^j}$. The same idea could be applied to find the lowest $(t,i)$ such that $d_X(t, q) \leq  c \cdot\frac{\delta}{\varepsilon^j}$ as long as $c\geq 1$. Indeed, we will apply this variant of containment search in our dynamic insertion. 
\end{remark}
\vspace{-0.5cm}

\begin{figure}[!htb]
    \centering
    \begin{mdframed}[nobreak=true]
\textbf{\textsc{Add}}$(u, q, T)$:  \textcolor{brown}{add $q$ as a child of $u$}

Given a node $u = (t, i)$, we create $(q, i-1)$ as a child of $(t, i)$ then return $(q, i-1)$. 
\\

\noindent\textbf{\textsc{Promote$(t, i, T)$}}: \textcolor{brown}{promote $(t,i)$, possibly add $(t,i+1)$  }

\textcolor{brown}{$\ll$\textsc{Promote} is invoked only when $(t,i)$ is a ring-$\phi$ node.$\gg$}

Given a node $(t, i)$, let $(u, i+1)$ be the current parent of $(t, i)$; $(u, i+1)$ exists by~\Cref{obs:ring_nodes}$(c)$.
\begin{enumerate}
    \item\textnormal{\textbf{[Check the packing property.]}} We check the points in $Y_{i+1}$ that are at a distance within $2\cdot \frac{\delta}{\varepsilon^{i+1}}$ to $u$ and find $u'$ closest to $t$. If $d(t, u') \leq \alpha \frac{\delta}{\varepsilon^{i+1}}$, we change the parent of $(t, i)$ from $(u, i+1)$ to $(u', i+1)$ and terminate.
    \item \textnormal{\textbf{[Promote.]}}: If $d(t, u') > \alpha \frac{\delta}{\varepsilon^{i+1}}$ (the packing property holds), we create $(t, i+1)$ as a new parent of $(t, i)$. Next, we find a node at level $(i+2)$ to be the parent of $(t, i+1)$.
    \item \textnormal{\textbf{[Find parent for $(t, i+1)$.]}}: let $(v, i+2)$ be the parent of $(u, i+1)$, then consider nodes within $2\frac{\delta}{\varepsilon^{i+2}}$ to find $(v', i+2)$ closest to $t$. By~\Cref{obs:promote_findparent}, $(v', i+2)$ is closest to $t$ among nodes at level $i+2$.
    \begin{enumerate}
        \item If $t$ is in ring $\alpha$ or ring $\beta$ of $(v', i+2)$, we choose $(v', i+2)$ to be the parent of $(t, i+1)$.
        \item  Otherwise, we choose $(v, i+2)$ to be the parent of $(t, i+1)$.
    \end{enumerate}
    
    A corner case is when $(u, i+1)$ is the bottom of a jump starting at $(u, l)$, then $(v, i+2) = (u, i+2)$ is a hidden node. In this case, we create $(u, i+2)$ by invoking \textsc{JumpSplit($u, l, i+1, i+2, T$)}, and then proceed as above.
\end{enumerate}
\end{mdframed}
    \caption{\textsc{Add} and \textsc{Promote}.}
    \label{fig:add-promote}
\end{figure}

\begin{figure}[!ht]
    \centering
    \begin{tikzpicture}
	 \node at (0,0){\includegraphics[width=0.98\linewidth]{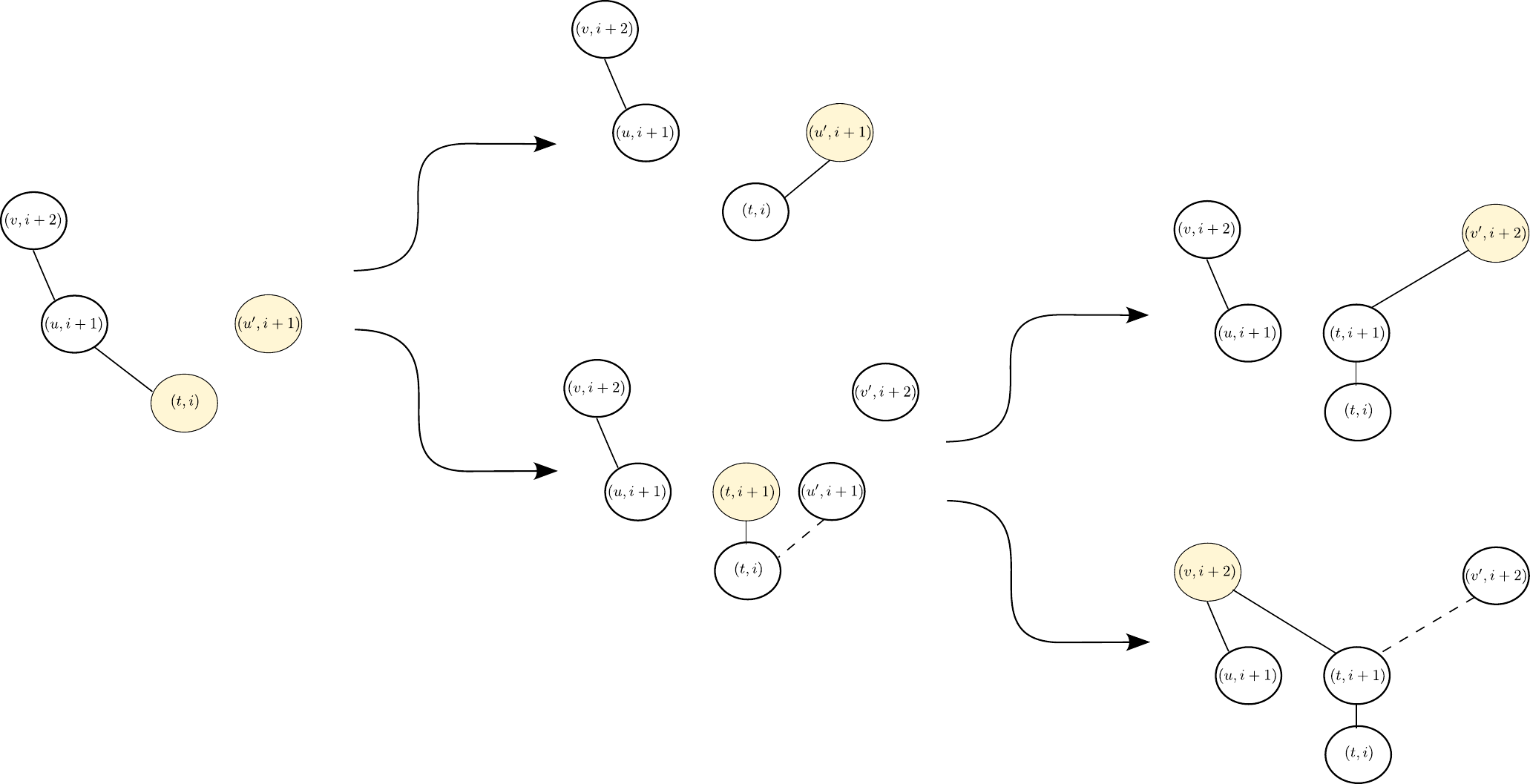}};		
			\node at (-3, 3){\footnotesize \textsc{Step 1}};
			\node at (-3, -1.25){\footnotesize \textsc{Step 2}};
            \node at (3, -0.8){\footnotesize \textsc{Step 3}};
            \node at (3.25, 1.25){\footnotesize Case 3(a)};
            \node at (3.25, -3){\footnotesize Case 3(b)};
            \node at (0.8, 2.1){\fontsize{6pt}{8} $\leq \alpha\frac{\delta}{\varepsilon^{i+1}}$};
            \node at (0.8, -1.7){\fontsize{6pt}{8} $> \alpha\frac{\delta}{\varepsilon^{i+1}}$};
            \node at (7.3, 1.){\fontsize{6pt}{8} $\leq \beta\frac{\delta}{\varepsilon^{i+1}}$};
            \node at (7.3, -2.7){\fontsize{6pt}{8} $> \beta\frac{\delta}{\varepsilon^{i+1}}$};
            \end{tikzpicture}    

    \caption{Illustrating \textsc{Promote} operation.}
    \label{fig:promote}
\end{figure}

\paragraph{Internal operations.~} These are operations that modify the current net tree to create a new node or to maintain the covering and jump isolation properties. In \Cref{fig:add-promote}, we describe \textsc{Add} and \textsc{Promote}. \textsc{Add}$(u, q, T)$ simply adds a new node $(q,i-1)$ as a child of a node $u = (t,i)$ at level $i$.  \textsc{Promote}$(t, i, T)$ promotes a node $(t,i)$ at level $i$ to $(t,i+1)$ at the next level; see \Cref{fig:promote}. Note that we only invoke \textsc{Promote$(t, i, T)$} when $(t, i)$ is a ring-$\phi$ node.

In Step 3 of \textsc{Promote}, given $(v, i+2)$, we determine $(v', i+2)$ closest to $t$ among \emph{nearby nodes} of $(v, i+2)$, where $(y, i+2)$ is a nearby node of $(v, i+2)$ if $d_X(v, y) \leq 2\cdot \frac{\delta}{\varepsilon^{i+2}}$.
This can be done in $O_\lambda(1)$ time by maintaining pointers of nearby nodes for every node in the tree as done in  Cole and Gottlieb~\cite{GC06}; by the packing bound, there are only $O_\lambda(1)$ nearby nodes. (Every time we add a new node, we will also add pointers to nearby nodes following Cole and Gottlieb~\cite{GC06}, and hence, in the pseudocodes below, we do not include this detail for a cleaner presentation.) We observe that:

\begin{observation}\label{obs:promote_findparent}
    $(v', i+2)$ is closest to $t$ among all nodes at level $i+2$.
\end{observation}

\begin{proof}
    Before \textsc{Promote}$(t, i, T)$, $(v, i+2)$ is the ancestor at level $i+2$ of $(t, i)$. By the close-containment property, $d_X(t, v) \leq \frac{\delta}{\varepsilon^{i+2}}$.
    If $(v', i+2)$ is closest to $t$ among nodes at level $i+2$, then $d_X(t, v') \leq d_X(t, v) \leq \frac{\delta}{\varepsilon^{i+2}}$, which implies $d_X(v, v') \leq 2\frac{\delta}{\varepsilon^{i+2}}$ by the triangle inequality.
\end{proof}

The idea of using \textsc{Promote}$(t,i,T)$ is to turn a ring-$\phi$ node to a ring-$\alpha$ node as in the following observation.

\begin{observation}\label{obs:Promote-goal}
    If $(t,i)$ is a ring-$\phi$ node, then after  \textsc{Promote}($t,i,T$), $(t,i)$ becomes a ring-$\alpha$ node. 
\end{observation} 
\begin{proof} In step 1, we check if $(t,i)$ is in  ring-$\alpha$ of $u'$ and in this case, it becomes a  child of $u'$ an hence a  ring-$\alpha$ node. In step 2, $(t,i)$ becomes a child of $(t,i+1)$ and hence is also a ring-$\alpha$ node.
\end{proof}

We remark that in the work of Cole and Gottlieb~\cite{GC06}, the definition of \textsc{Promote} was not precisely given; they just considered whether we could promote $(t, i)$ without violating the packing property and assigned the closest node at level $i+2$ to be the parent of $(t, i+1)$ if $(t, i)$ was promoted. 
Here, we consider more cases to find an appropriate parent for $(t, i+1)$; this is necessary to simplify the insertion 
operation.

\begin{figure}[!ht]
    \centering
\begin{mdframed}[nobreak=true]
\textbf{\textsc{JumpCreate$(u, T)$}}: \textcolor{brown}{create a leaf and a jump $u$ to the leaf}

Given a node $u = (q, i)$, we create a node $(q, 0)$ as a child of $(q, i)$.
\\

\noindent\textbf{\textsc{JumpSplit$(t, i, j, k, T)$}}: \textcolor{brown}{split the jump from $(t,i)$ to $(t,j)$ by inserting $(t,k)$ and possibly $(t,k-1)$}

Given a jump starting from $(t, i)$ down to $(t, j)$, a level $k$ where $j < k < i$, we create a node $(t, k)$ as a child of $(t, i)$, create $(t, k-1)$ as a child of $(t, k)$ if $(t, k-1)$ does not exist, and change the parent of $(t, j)$ to $(t, k-1)$.
\\

\noindent\textbf{\textsc{JumpFix$(t, i, T)$}}: \textcolor{brown}{fixing jump isolation property at $(t,i)$}

Given a jump starting from $(t, i)$ down to $(t, j)$, if the jump isolation at $(t, i)$ is violated, we create $(t, i-1)$ as the only child of $(t, i)$ and make it the parent of $(t, j)$. 
\\

\noindent\textbf{\textsc{MaintainJumpIsolation$(t, i, T)$}}: \textcolor{brown}{check and fix jump isolation property of jumps nearby $(t,i)$}

Given a node $(t, i)$, we find $(v, i)$ where $d_X(v, t) \leq (\alpha + \gamma)\frac{\delta}{\varepsilon^i}$.
If $(v, i)$ exists,  then:
\begin{enumerate}
    \item If $(v, i)$ is the top of a jump, which means the jump isolation at $(v, i)$ is violated, we invoke \textsc{JumpFix($v, i, T$)} to create $(v, i-1)$.
    \item If $(t, i)$ is the top of a jump, which means the jump isolation at $(t, i)$ is violated, we invoke \textsc{JumpFix$(t, i, T)$} to create $(t, i-1)$.
\end{enumerate}
\end{mdframed}
    \caption{Operations to create jumps and maintain jump isolation.}
    \label{fig:jump-isolation}
\end{figure}

\Cref{fig:jump-isolation} describes the operations to create a jump or maintain jump isolation. The operation \textsc{MaintainJumpIsolation} checks if a jump is isolated based on the definition of jump isolation below.

\begin{definition}[Jump isolation~\cite{GC06}, invariant 1]\label{def:olddef_jumpisolation}
    A jump from $(x, i)$ down to $(x, j)$ is \EMPH{isolated} if for any $(y, k)$ where $k \leq i$ and $(y, k)$ is not a descendant of $(x, i)$, $d_X(x, y) > \alpha\frac{\delta}{\varepsilon^i}+ \gamma\frac{\delta}{\varepsilon^k}$.
\end{definition}

If jump isolation is violated, it will call \textsc{JumpFix} to fix the jump isolation property. Basically,  \textsc{JumpFix}$(t,i,T)$ will create a new child $(t,i-1)$ of the node $(t,i)$, which is the top a jump from $(t,i)$ to $(t,j)$. By adding  $(t,i-1)$, $(t,i)$ is no longer the top of a jump---instead, $(t,i-1)$ now becomes the top--- and hence the jump isolation property does not apply $(t,i)$ by definition. We will show in~\Cref{lmm:maintain_bisolated} that jump isolation in \Cref{def:olddef_jumpisolation} implies our $b$-jump isolation property in \Cref{thm:nettree_ds} when $\varepsilon$ is sufficiently small, and therefore, we could reuse all operations in \Cref{fig:jump-isolation} to maintain our $b$-jump isolation property.

Operations \textsc{Add, Promote, JumpSplit, JumpFix} run in $O_\lambda(1)$ time; each of them will add nodes associated with a new point and modify the current net tree. Accordingly, we will have different types of nodes created or modified by one insertion. These are the types of nodes mentioned in \Cref{def:dynamic_nettree} given in~\Cref{sec:nettreecover}. 

\paragraph{Find nearby nodes.~} Given a node $(x, i)$ that is not the bottom or hidden node of a jump, we want to find nodes $(y, i)$ such that $d_X(x, y) \leq g\frac{\delta}{\varepsilon^i}$ for a constant $g$. First, we apply the same idea of Cole and Gottlieb~\cite{GC06}: maintain pointers for every $(q, i)$ to nodes $(p, i)$ if $d_X(p, q) \leq \frac{2\delta}{\varepsilon^i}$, and $(q, i)$ is not the bottom or hidden node of a jump. After that, we follow pointers of $(x, i)$ to visit nodes $(y_0, i)$ within distance $2\frac{\delta}{\varepsilon^i}$ from $(x, i)$, then follow pointers of $(y_0, i)$ to visit nodes $(y_1, i)$ within distance $4\frac{\delta}{\varepsilon^i}$ from $(x, i)$, and so on. After $O(g)$ steps, we reach nodes $(y, i)$ whose $d_X(x, y) \leq g\frac{\delta}{\varepsilon^i}$.  

To maintain pointers for every node $(q, i)$ to any $(p, i)$ where $d_X(q, p) \leq \frac{2\delta}{\varepsilon^i}$, we find $(p, i)$ when $(q, i)$ is added to the tree.
Specifically, after \textsc{Insert}, $(q, i)$ is added to the tree as a child of $(t, i')$ for $i' > i$. 
If $i' > i+1$, we do nothing.
If $i' = i+1$, we follow pointers of $(t, i+1)$ to find $(v', i+1)$ whose $d_X(t, v') \leq 2\frac{\delta}{\varepsilon^{i+1}}$.
Then for every child $(p', i)$ of $(v', i+1)$, if $d_X(q, p') \leq \frac{2\delta}{\varepsilon^i}$, we create pointers from $(p', i)$ to $(q, i)$ and $(q, i)$ to $(p', i)$. 

To see the correctness of this idea, we consider the parents of $(q, i)$ and $(p, i)$. If $i' > i+1$, then $(t, i')$ is the top of a jump, and $(q, i)$ is the bottom node or in the middle of that jump (in this case, $(q, i)$ is created by splitting a jump). By the $b$-jump isolation property, any node $(y, i)$ has $d_X(q, y) > b\frac{\delta}{\varepsilon^i} \geq 2 \frac{\delta}{\varepsilon^i}$, thus $(q, i)$ has no pointers to nearby nodes.
When $i' = i+1$, observe that $(p, i)$ is also not the bottom node or in the middle of a jump. Consider the parent $(v, i+1)$ of $(p, i)$, we have $d_X(t, v) \leq d_X(q, p) + d_X(q, t) + d_X(v, p)$.
By the covering property, $d_X(q, t)$ and $d_X(v, p)$ are at most $\phi\frac{\delta}{\varepsilon^{i+1}}$, thus 
$d_X(t, v) \leq 2\frac{\delta}{\varepsilon^i} + 2\phi\frac{\delta}{\varepsilon^{i+1}} \leq 2\frac{\delta}{\varepsilon^{i+1}}$ since $\phi = \frac{3}{4}$ and $\varepsilon \leq \frac{1}{8}$.
Therefore, from the parent $(t, i+1)$ of $(q, i)$, it suffices to consider children of $(v, i+1)$ whose $d_X(v, t) \leq 2\frac{\delta}{\varepsilon^i}$ to find $(p, i)$ nearby $(q, i)$.

Back to finding nodes $(y, i)$ whose $d_X(x, y) \leq g\frac{\delta}{\varepsilon^i}$, the process runs in $O(g)$ steps. 
By the packing property, each node has $O_\lambda(1)$ pointers. Therefore:
\begin{claim}\label{claim:find_nearbynodes}
    Given $(x, i)$ that is not the bottom or a hidden node of a jump, we can find $(y, i)$ where $d_X(x, y) \leq g\frac{\delta}{\varepsilon^i}$ from $(x, i)$ with $O(g)^\lambda$ time.
\end{claim}

\subsection{Dynamic Net Tree Operations}\label{sec:main_operations}

We are now ready to construct our dynamic net tree. We use the same containment search of Cole and Gottlieb~\cite{GC06} to search for a slightly different variant of the lowest ball containing $q$:

\begin{mdframed}[nobreak=true]
\textbf{\textsc{ContainmentSearch$(q, T)$}}  

Given a new point $q$, we apply the containment search of Cole and Gotlieb~\cite{GC06}, to find the lowest node $(t, i)$ in $T$ such that:
\begin{equation}~\label{eq:search_condition}
\begin{aligned}
    d_X(q, t) \leq \psi\frac{\delta}{\varepsilon^i} 
\end{aligned}
\end{equation}
where $\psi = \frac{5}{4}$ (in \Cref{equ:nettree_constant}); see \Cref{rm:search-condition}. If more than one node at the same level satisfies~\Cref{eq:search_condition}, we return the node closest to $p$.
\end{mdframed}

Note that the containment search might return a node $(t, i)$ such that $(t, i)$ and $(q, i-1)$ do not satisfy the covering property. We will handle this case later in the insert operation described below.

\paragraph{Insertion.~}
Given a new point $q$, our goal is to find a parent for $q$. First, we will invoke  \textsc{ContainmentSearch}$(q,T)$ to find the lowest node $(t,i)$ satisfying \Cref{eq:search_condition}. If $d_X(q, t) \leq \phi \frac{\delta}{\varepsilon^i}$, we can make $(q,i-1)$ a child of $(t,i)$. In the complementary case, making $(q,i-1)$ a child of $(t,i)$ will violate the covering property, and hence we have to find another node to be the parent of $q$. To quickly find the parent for $q$, we use the idea of \emph{chains} and \emph{obligations} introduced by Cole and Gottlieb~\cite{GC06}. Chains and obligations also have another very important role, specifically in avoiding a cascading sequence of promotions described in the overview.

\begin{definition}[Chain]\label{def:chain} Given a node $(t,i)$, let $(u,j)$ be the lowest ring-$\alpha$ node that is an ancestor of  $(t,i)$; it is possible that $(u,j)=(t,i)$. We define the \EMPH{chain} of $(t, i)$ to be the sequence of nodes that starts at $(u,j)$ and ends at $(t, i)$. 
\end{definition}

Observe by~\Cref{def:chain}, if $(t, i)$ is a ring-$\alpha$ node, then the chain of $(t, i)$ has only one node, which is $(t, i)$. A chain of a ring-$\beta$ or ring-$\phi$ node contains the chain of its parent as a subsequence. We say a chain is \EMPH{safe} if it has at most one ring-$\phi$ node. We will maintain the safe invariant for the net tree. 

\begin{invariant}[Safe invariant]\label{inv:safe}
The chain of every node in the net tree is safe. 
\end{invariant}

If a chain of $(t, i)$ is safe and contains (exactly one) ring-$\phi$ node, we will keep track of this node at $(t,i)$ as its obligation.

\begin{definition}[Obligation]\label{def:obligation}
Given a node $(t, i)$, the \EMPH{obligation} of $(t, i)$, denoted by \textsc{Obligate}$(t,i)$, is the lowest ring-$\phi$ node in its (safe) chain. If there is no ring-$\phi$ node in the chain, then \textsc{Obligate}$(t, i) = \text{null}$. 
\end{definition}

\begin{figure}[!ht]
    \centering
    \begin{mdframed}[nobreak=true]
\textbf{\textsc{Insert$(q, T)$}:}

\begin{enumerate}
    \item Let $u \leftarrow{ \textsc{ContainmentSearch}(q, T)}$ $\qquad \ll \text{$u$ is a candidate parent of $q$} \gg$
    \item  Node $u$ could be the parent of $q$, unless in the following two cases:
    \begin{enumerate}
        \item If $u$ is the top of a jump and $q$ is in ring $\alpha, \beta$, or $\phi$ of $u$, then the parent of $q$ is a (possibly hidden) node in the jump starting at $u$. We invoke $u \leftarrow{ \textsc{FindParentInJump}(q, u, T)}$ to find $q$'s parent.
        \item   If $q$ is completely out of ring $\phi$ of $u$ (and $u$ does not have to be the top of a jump), then we need to find another node to be the parent of $p$, by invoking $u \leftarrow{\textsc{FindParent}(u, T)}$.    
    \end{enumerate}
    \item Let $u = (t', i')$. We create a node $(q, i'-1)$ as a child of $(t', i')$ by invoking $\textsc{Add}(u, q, T)$. Then we invoke $\textsc{JumpCreate}$ to create a jump from $(q, i'-1)$ down to $(q, 0)$.
    
    \item If $q$ is in ring $\phi$ of $u$ and $\textsc{Obligate}(u) \not = \text{null}$, we invoke $\textsc{ChainFix}(u, T)$. This step modifies the tree to guarantee that all chains are safe after we add $(q, i'-1)$ as a child of $u$ in Step 3.

    \item For every new node $(x, j)$ that we create from Step 1 to Step 4, we check (and fix) the jump isolation of jumps nearby $(x, j)$ by invoking \textsc{MaintainJumpIsolation}$(x, j, T)$. (There are only $O(1)$ such nodes.)
\end{enumerate}

\end{mdframed}
    \caption{The insert procedure.}
    \label{fig:insert-code}
\end{figure}

\Cref{fig:insert-code} describes the pseudo-code of the insertion operation. It might call two helper procedures \textsc{FindParentInJump}$(q, u, T)$ and \textsc{FindParent}$(q, u, T)$ described in \Cref{fig:find-parent}. The former finds a parent for $q$ in the jump starting at a node $u$, while the latter finds a parent for $q$ by checking nearby nodes of $u$. The following observation is immediate from the construction.

\begin{observation}\label{obs:parent_q}
    Given a new point $q$, let $u = (t, i)$ be the result of \textsc{ContainmentSearch$(q, T)$}.
    In the end of \textsc{Insert}$(q,T)$, we create $(q, i'-1)$ as a child of an existing node $(t', i')$ in the tree where $i' \leq i+1$.
\end{observation}

\begin{figure}[!htb]
    \centering
    
\begin{mdframed}[nobreak=true]
\textbf{\textsc{FindParentInJump}$(q, u, T)$:}

Let $u = (t, i)$. We only invoke this procedure when $u$ is the top of a jump, and $q$ is in ring $\alpha, \beta$ or $\phi$ of $u$. Let $(t, j)$ be the bottom of the jump at $u$.
\begin{enumerate}
    \item Let $k \in (j, i]$ be such that $ \alpha \frac{\delta}{\varepsilon^{k-1}} < d_X(t, q) \leq \alpha \frac{\delta}{\varepsilon^{k}}$.
    \item Invoke \textsc{JumpSplit($t, i, j, k, T$)} to create $(t, k)$ and (possibly) $(t, k-1)$ in the middle of the jump. 
    \item Return $(t, k)$.
\end{enumerate}

\noindent\textbf{\textsc{FindParent}$(q, u, T)$:} \textcolor{brown}{find a node at level $i+1$ to be the parent of $q$}

Let $u = (t,i)$. We only invoke this procedure when $q$ is completely out of ring $\phi$ of $u$. To maintain the covering property, we guarantee that $t$ is not completely out of ring $\phi$ of its parent. 

\begin{enumerate}
    \item Consider the nodes within the distance $2\frac{\delta}{\varepsilon^{i+1}}$ to the parent of $(t, i)$; the parent
 has pointers to all these nodes, and there are only $O_{\lambda}(1)$ of them.  Let $(v, i+1)$ be the node closest to $q$ among them.
    By~\Cref{obs:find_closestnode}, $(v, i+1)$ is closest to $t$ among nodes at level $i+1$.
    
    A corner case is when $(t, i)$ is the bottom of a jump, and hence, $(t, i+1)$ is a hidden node. Any node $(y, i+1)$ has $d_X(y, t) > 2\frac{\delta}{\varepsilon^{i+1}}$. Thus, $(t, i+1)$ is closest to $q$, we create $(t, i+1)$ by using \textsc{JumpSplit}.

\item Now we consider $q$ and rings of $(v, i+1)$: 
\begin{itemize}
        \item[\textnormal{(a)}] If $q$ is in ring $\alpha$, $\beta$ or $\phi$ of $(v, i+1)$: return $(v, i+1)$.
        \item[\textnormal{(b)}] If $q$ is comletely out of ring $\phi$ of $(v, i+1)$: We call \textsc{Promote}$(t, i,T)$ to promote $(t, i)$ and return the parent of $(t, i)$. We will show in \Cref{claim:covering_ringphi_node}
 that $(t, i)$ is a ring-$\phi$ node.
    \end{itemize}

\end{enumerate}

\noindent\textbf{\textsc{ChainFix}$(u, T)$}
\label{sssc:ins_3}

Let $u = (t, i)$, if $q$ satisfies $d_X(t, q) \in (\beta \frac{\delta}{\varepsilon^i}, \phi \frac{\delta}{\varepsilon^i}]$, we promote $\textsc{Obligate}(u)$ if it is not null. 

\end{mdframed}
    \caption{Find parent for a new point $q$ given access to a node $u$, and fix a chain at a node $u$.}
    \label{fig:find-parent}
\end{figure}

\begin{claim} \label{obs:find_closestnode}
    In step 1 of \textsc{FindParent}$(q,u,T)$, we consider nodes within $2\frac{\delta}{\varepsilon^{i+1}}$ to the parent of $u=(t, i)$ and choose $(v, i+1)$ closest to $q$. Then $(v, i+1)$ is closest to $q$ among nodes at level $i+1$.
\end{claim}
\begin{proof}
    Let $(t', i+1)$ be the parent of $(t, i)$; if $(t, i)$ is the bottom of a jump, then $(t', i+1) = (t, i+1)$.
    By the covering property, $d_X(t, t') \leq \phi\frac{\delta}{\varepsilon^{i+1}}$. Recall that $u = (t, i)$ is the result of \textsc{ContainmentSearch}$(q, T)$, thus $d_X(q, t) \leq \psi \frac{\delta}{\varepsilon^i}$.
    By triangle inequality, we have:
    \begin{equation*}
        \begin{aligned}
            d_X(t', q) &\leq d_X(t', t) + d_X(q, t) \leq \phi\frac{\delta}{\varepsilon^{i+1}} + \psi \frac{\delta}{\varepsilon^i} \\
            &= (\phi + \varepsilon\psi) \frac{\delta}{\varepsilon^{i+1}} \leq \frac{\delta}{\varepsilon^{i+1}} \textrm{\quad (since $\psi = \frac{5}{4}, \phi = \frac{3}{4}$ and $\varepsilon \leq \frac{1}{8}$)}
        \end{aligned}
    \end{equation*}
    If $(v, i+1)$ is closest to $q$ among nodes at level $i+1$, then $d_X(v, q) \leq d_X(t', q) = \frac{\delta}{\varepsilon^{i+1}}$. This implies $d_X(v, t') \leq d_X(v, q) + d_X(q, t') \leq 2\frac{\delta}{\varepsilon^{i+1}}$. 
    Thus considering nodes within a distance  $2\frac{\delta}{\varepsilon^{i+1}}$ to $(t', i+1)$ suffices to find $(v, i+1)$.
\end{proof}

In step 2(b) of  \textsc{FindParent}$(q,u,T)$, we promote $(t,i,T)$ when $q$ is completely of ring $\phi$ of $(v,i+1)$. As promote could only be applied to ring-$\phi$ node, we show below that $(t,i)$ is a ring-$\phi$ node. 

\begin{claim}\label{claim:covering_ringphi_node}
Node $(t, i)$ in step 2(b) of  \textsc{FindParent}$(q,u,T)$ is a ring-$\phi$ node.
\end{claim}
\begin{proof}
For contradiction, suppose that $(t, i)$ is not a ring-$\phi$ node. By the covering property, it must be either a ring-$\alpha$ or ring-$\beta$ node. Let $(t', i+1)$ be the parent of $(t, i)$; if $(t, i)$ is the bottom of a jump, then let $(t', i+1) = (t, i+1)$. We have $d_X(t, t') \leq \beta \frac{\delta}{\varepsilon^{i+1}}$.
Since $u = (t, i)$ is the result of \textsc{ContainmentSearch}$(q,T)$, by~\Cref{eq:search_condition}, we have $d_X(t, q) \leq \psi \frac{\delta}{\varepsilon^i}$. Recall that in step 2 of  \textsc{FindParent}, $(v, i+1)$ is closest to $q$ among nodes at level $i+1$ and hence, $d_X(v, q) \leq d_X(t', q)$. We have:
\begin{equation*}
\begin{aligned}\label{equ:proof_covering}
    d_X(v, q) &\leq d_X(t', q)
    \leq d_X(t, q) + d_X(t, t') \\
    &\leq \psi \frac{\delta}{\varepsilon^{i}} + \beta \frac{\delta}{\varepsilon^{i+1}}  \leq \phi \frac{\delta}{\varepsilon^{i+1}} \qquad  \text{(since $\varepsilon\psi + \beta \leq \phi$ in~\Cref{equ:nettree_constant})}, 
\end{aligned}
\end{equation*}
implying that $q$ is in ring $\alpha$, $\beta$ or $\phi$ of $(v, i+1)$, and hence step 2(b) will not be invoked, a contradiction.  
\end{proof}

We remarked earlier that chains and obligations help avoid a cascading sequence of promotions. This is because when we promote a ring-$\phi$ node $(t,i)$, it becomes a ring-$\alpha$ node (\Cref{obs:Promote-goal}). However, its parent, which is $(t,i+1)$, might still be ring-$\phi$. By the chain safe invariant (\Cref{inv:safe}), the parent of $(t,i+1)$ is not a ring-$\phi$ node, and hence further promotion is needed (to guarantee the covering property).

\paragraph{About maintaining obligations.} For simplicity of the presentation, we do not explicitly include obligation maintenance in the above pseudocodes. We use the (simple yet clever) idea of Cole and Gottlieb~\cite{GC06}, which we now briefly describe.  Every node has a pointer to its obligation: a ring-$\alpha$ points to null, a ring-$\phi$ node points to itself, and a ring-$\beta$ node points to the same value to which its parent points.  (Specifically, if the parent of a ring-$\beta$ node is another ring-$\beta$ node, then they point to the same obligation, which could be a ring-$\phi$ node or null.)  So, the focus is on maintaining the obligation of a ring-$\beta$ node.

Think about a ring-$\beta$ node $(t,i)$ that points to an ancestor ring-$\phi$ node $(x,j)$ for some level $j > i$.  Then all the nodes between $(t,i)$ and $(x,j)$---except $(x,j)$---are ring-$\beta$ nodes. The key observation is that when a new point $q$ is inserted, it will never be inserted as an intermediate node between $(t,i)$ and $(x,j)$, so the pointers of these nodes do not change. But it is possible that $(x,j)$ will be promoted due to \textsc{ChainFix} and hence  $(x,j)$ is no longer a ring-$\phi$ node. However, by \Cref{obs:Promote-goal}, after the promotion,   $(x,j)$ becomes a ring-$\alpha$ node, and hence its obligation is null. By setting $\textsc{Obligate}(x,j)$ to null, the obligation pointers of all the nodes from $(x,j)$ down to $(t,i)$ are also automatically set to null since they all point to $\textsc{Obligate}(x,j)$. Therefore, maintaining obligations only adds $O(1)$ overhead.

\subsection{Analysis}\label{sec:nettree_analysis}

\paragraph{Space and time.~} We use the containment search data structure of Cole and Gottlieb~\cite{GC06} and modify \textsc{Insert} with $O(1)$ steps, thus our data structure takes $O(n)$ space and the running time is $O_{\lambda}(\log(n))$ for each search, insertion, deletion, as claimed in \Cref{thm:nettree_ds}. 

By~\Cref{claim:find_nearbynodes}, given $(x, i)$ is not a bottom or a hidden node of a jump, we can find $(y, i)$ where $d_X(x, y) \leq g\cdot\frac{\delta}{\varepsilon^i}$ in $O(g)^\lambda$ time.

Next,  we focus on showing packing, covering, close-containment, and $b$-jump isolation properties by induction. Specifically, we assume that these properties hold before an update to the net tree, and we will show them after the update.  Deleting a point is simply marking the corresponding leaf of that point as deleted, and hence, none of the properties will be violated after a deletion. The difficult case is insertions, which involve creating new nodes and updating the parents of existing nodes in the tree, potentially violating the net tree properties. 

\paragraph{Packing property.~} Before showing the packing property, we give some simple observations.

\begin{observation}\label{obs:distance_hiddennode}
    Given a jump from $(t, i)$ down to $(t, j)$, if $(t, k)$ is a hidden node or $(t, k)$ is created by operations \textsc{JumpSplit} or \textsc{JumpFix} for some level  $k \in (i,j)$, then $d_X(t, v) > b\frac{\delta}{\varepsilon^k}$ for every node $(v, k)$ at the same level $k$.
\end{observation}
\begin{proof}
    By the definition of a jump, any node $(v, k)$ for $v \neq t$ and $k \in (j, i)$ is not a descendant of $(t, i)$. By $b$-jump isolation property, $d_X(t, v) > b\frac{\delta}{\varepsilon^k}$. 
\end{proof}

\begin{observation}\label{obs:packing_promote}
 \textsc{Promote} maintains the packing property.
\end{observation}
\begin{proof}
 \textsc{Promote} creates at most one new node, and it checks the packing property in Step 1:
before create $(t, i+1)$ as a new parent of $(t, i)$, if there exists a node $(v, i+1)$ such that $d_X(v, t) \leq \alpha \frac{\delta}{\varepsilon^i}$, then it changes the parent of $(t, i)$ to $(v, i+1)$ and terminates.
Thus, the packing property is maintained.
\end{proof}

\begin{lemma}[Packing property]\label{lmm:packing}
    For any pair of nodes $(x, i)$ and $(y, i)$ at level $i$, $d_X(x, y) > \alpha \frac{\delta}{\varepsilon^i}$ for $\alpha = 1/4$.
\end{lemma}

\begin{proof}
    It suffices to show that when a new point is added, creating new nodes does not violate the packing property.
    There are three cases where new nodes are created by an insertion:
    \begin{inlinelist} 
    \item nodes created by \textsc{Promote} (called in Step 2 case (b) and Step 4 of \textsc{Insert})
    \item nodes created by \textsc{JumpSplit} (called in Step 2 cases (a) and (b) of \textsc{Insert}), or \textsc{JumpFix} (possibly invoked in Step 5 of \textsc{Insert})
    \item node $(q, i'-1)$ for a new point $q$ created by Step 3 of \textsc{Insert} for $i' > 0$
    \end{inlinelist}.

    \Cref{obs:packing_promote} takes care of case (i).
    For (ii), by~\Cref{obs:distance_hiddennode}, a node at level $k$ created by \textsc{JumpFix} or \textsc{JumpSplit} is at distance $b\frac{\delta}{\varepsilon^k}$ to any existing node at level $k$, which implies the packing property since $b > \alpha$ ($b \geq \frac{5}{4}, \alpha = \frac{1}{4}$). Both cases (i) and (ii) imply that Steps 2, 4, and 5 of \textsc{Insert} maintain the packing property.

     It remains to consider case (iii). Suppose that there exists a node $(x, i'-1)$ at the same level with $(q,i'-1)$ such that $d_X(x, q) \leq \alpha\frac{\delta}{\varepsilon^{i'-1}}$. Let $u $ be the result of \textsc{ContainmentSearch}$(q, T)$. We claim that:
     \begin{equation}\label{eq:claim-same-level-u}
         \text{$u= (t, i)$ has the same level as $(q,i'-1)$; that is, $i = i'-1$.}
     \end{equation}
     
      To see \eqref{eq:claim-same-level-u}, recall that the containment search finds the lowest node $u=(t, i)$ where $d_X(t, q) \leq \psi \frac{\delta}{\varepsilon^{i}}$, for $\psi = \frac{5}{4}$.
    Since $\alpha = \frac{1}{4} < \psi$, $d_X(x, q) < \psi \frac{\delta}{\varepsilon^{i'-1}}$, implying that $i \leq i'-1$.  By~\Cref{obs:parent_q}, the parent of $q$ is a node at a level at most $i+1$, giving $i' \leq i+1$ and hence $i\geq i'-1$. We conclude that $i' = i+1$ as claimed in \eqref{eq:claim-same-level-u}. 

    Since $u$ has the same level as $(q,i'-1)$, it cannot be $q$'s parent or ancestor. Observe that only in Step 2 of \textsc{Insert$(q, T)$} we might find a parent for $q$ different from $u$. We consider two cases:
    \begin{itemize}
        \item \textbf{Step 2(a):} In this case, $u$ is the top of a jump and $q$ is in ring $\alpha, \beta$ or $\phi$ of $u$. Then, the parent of $q$ is a descendant of $u$, contradicting \eqref{eq:claim-same-level-u}. 
        \item \textbf{Step 2(b):} In this case, $q$ is completely out of ring $\phi$ of $u$ and hence $d_X(t, q) > \phi\frac{\delta}{\varepsilon^{i'-1}}$. By the definition of containment search, $u$ is the node closest to $q$ at level $i'-1$. Thus, $d_X(x',q)\geq d_X(t,q) > \phi\frac{\delta}{\varepsilon^{i'-1}}$, contradicting the assumption that $d_X(x, q) \leq \alpha\frac{\delta}{\varepsilon^{i'-1}}$. 
    \end{itemize}
 Both cases above imply the packing property. 
\end{proof}

\paragraph{Covering property.} The proof is similar to the packing property. We observe that:

\begin{observation}\label{obs:promote_parent} 
    Given a node ring-$\phi$ node $(t, i)$, if the safe invariant (\Cref{inv:safe}) and the covering property are maintained, then the parent of $(t, i)$ is either a ring-$\alpha$ or ring-$\beta$ node.  
\end{observation}

\begin{proof}
    By~\Cref{obs:ring_nodes} item (c), the parent of $(t, i)$ is a node at level $i+1$, denoted by $(u,i+1)$.    Consider the safe chain of $(t, i)$, starting from the lowest ring-$\alpha$ ancestor of $(t, i)$ and ending at $(t, i)$.  Since the chain is safe, it has at most one ring-$\phi$ node, and in this case, must be $(t, i)$. Therefore, no other node in the chain, which includes $(u,i+1)$, is a ring-$\phi$ node. By the covering property, $(u, i+1)$ must either be a ring-$\alpha$ or ring-$\beta$ node.    
\end{proof}

\begin{claim}\label{lmm:cover_promote}
    \textsc{Promote} maintains the covering property.  
\end{claim}
  
\begin{proof}  Recall that \textsc{Promote}$(t, i)$ updates the parent of $(t, i)$ and possibly creates one new node $(t, i+1)$ as the parent of $(t, i)$. By \Cref{obs:Promote-goal}, after \textsc{Promote}$(t, i, T)$, $(t, i)$ becomes a ring-$\alpha$ node, which implies the covering property for $(t, i)$.

    It remains to consider the case where \textsc{Promote} creates $(t, i+1)$, and we have to show the covering property for $(t, i+1)$. Let $(u, i+1)$ be the parent of $(t, i)$ before \textsc{Promote}$(t, i, T)$,  $(v, i+2)$ be the parent of $(u, i+1)$, and $(v', i+2)$ be the node at level $i+2$ that is closest to $t$. By definition of $(v',i+2)$, $d_X(t, v') \leq d_X(t, v)$. In step 3 of \textsc{Promote}, we choose either $(v, i+2)$ or $(v', i+2)$ to be the parent of $(t, i+1)$.
    Recall that $(t, i)$ must be a ring-$\phi$ node as this is the condition to invoke \textsc{Promote}, giving $d_X(t, u) \leq \phi \frac{\delta}{\varepsilon^{i+1}}$.
    In addition, by~\Cref{obs:promote_parent}, $(u, i+1)$ is a ring-$\alpha$ or ring-$\beta$ node, giving $d_X(u, v) \leq \beta \frac{\delta}{\varepsilon^{i+2}}$.  
    By triangle inequality, we obtain:
    \begin{equation*}
        \begin{aligned}
            d_X(t, v') &\leq d_X(t, v) \leq d_X(t, u) + d_X(u, v) \leq \phi \frac{\delta}{\varepsilon^{i+1}} + \beta\frac{\delta}{\varepsilon^{i+2}} \\
            &\leq \phi\frac{\delta}{\varepsilon^{i+2}} \textrm{\quad(since $\beta = \frac{2}{4}, \phi = \frac{3}{4},  \varepsilon \leq \frac{1}{8}$)}
        \end{aligned}
    \end{equation*}
    Therefore, the covering property holds for $(t, i+1)$.
\end{proof}

Now, we are ready to show the covering property for all nodes in the tree.

\begin{lemma}[Covering property]\label{lmm:cover}
    If $(y, i)$ has a child $(x, i')$ for $i' < i$ then $d_X(x, y) \leq \phi \frac{\delta}{\varepsilon^i}$ with $\phi = \frac{3}{4}$.
\end{lemma}

\begin{proof} It suffices to focus on the edges of the net tree, which are changed or added by \textsc{Insert} since deletions are only marked. Two types of update:
\begin{inlinelist}
    \item modifying existing points in the tree, involving changing parents, or creating new nodes and edges for existing points.
    \item finding the parent for a newly inserted point.
\end{inlinelist}

In case (i), the tree is modified by one of the following internal operations: (1) \textsc{Promote}, or (2) \textsc{JumpFix} or \textsc{JumpSplit}.  \Cref{lmm:cover_promote} takes care of (1). 
For (2), by the definition of a jump, the new node in the middle of a jump is a ring-$\alpha$ node, thus the covering property is maintained.

For case (ii), let $u = (t, i)$ be the result of \textsc{ContainmentSearch}$(q, T)$. If step 2 of  \textsc{Insert} is not applied, then  $q$ is in ring $\alpha, \beta$ or $\phi$ of $u$ and $u$ is not the top of a jump. The covering property follows from the fact that $(q, i-1)$ as a child of $u = (t, i)$. Thus, it remains to consider step 2.

In step 2(a), $u = (t,i)$ is the top of a jump and $q$ is in ring $\alpha, \beta$ or $\phi$ of $u$. The parent of $q$ is a node $(t,k)$ in the jump from $(t, i)$ down to $(t, j)$ for some $k \in (j,i]$. If $k = i$, since $q$ is in ring $\alpha, \beta$ or $\phi$ of $u = (t, i)$, the covering property holds. Otherwise, $k < i$. By the definition of $k$, $d_X(t, q) \leq \alpha \frac{\delta}{\varepsilon^k}$ and since $\alpha < \phi$ ($\alpha = \frac{1}{4}, \phi = \frac{3}{4}$), the covering property holds.

In step 2(b), we invoke \textsc{FindParent} to find a node at level $i+1$ to be the parent of $q$. In step 2(a) of \textsc{FindParent}, the parent $(v,i+1)$ of $q$ has the property that $q$ is in its ring $\alpha$, $\beta$ or $\phi$, and hence the covering property holds. In step 2(b) of \textsc{FindParent}, $q$ is completely out of ring $\phi$ of $(v, i+1)$, we promote $(t, i)$ and choose the parent of $(t, i)$ to be the parent of $q$. By \Cref{claim:covering_ringphi_node}, $(t,i)$ is a ring-$\phi$ node.  Let $(t'', i+1)$ be the new parent of $(t, i)$ after $\textsc{Promote}(t, i)$; $(t'',i+1)$ is also the parent of $q$ by construction in step 2(b). By \Cref{obs:Promote-goal}, $(t, i)$ becomes a ring-$\alpha$ node, which means $d_X(t, t'') \leq \alpha \frac{\delta}{\varepsilon^{i+1}}$. We have:
\begin{equation*}
    \begin{aligned}
        d_X(t'', q) &\leq d_X(t'', t) + d_X(t, q) \leq \alpha \frac{\delta}{\varepsilon^{i+1}} + \psi\frac{\delta}{\varepsilon^i} \\
        &\leq \beta\frac{\delta}{\varepsilon^{i+1}} \textrm{\quad (since $\alpha + \varepsilon\psi \leq \beta$ in~\Cref{equ:nettree_constant})}
        \\
        &< \phi \frac{\delta}{\varepsilon^{i+1}},
    \end{aligned}
\end{equation*}
the covering property holds.
\end{proof}

 \paragraph{Close-containment property.~} Close-containment property follows directly from the covering property.

\begin{lemma}[Close-containment]\label{lmm:contain}
    If $(x, i)$ has a descendant $(y, k)$ then $d_X(x, y) \leq \gamma \frac{\delta}{\varepsilon^i}-\gamma\frac{\delta}{\varepsilon^k}$. 
\end{lemma}

\begin{proof} 
    Let $(y, k)$ be a descendant of $(x, i)$ for $k < i$.
    By~\Cref{lmm:cover}, we have $d_X(x, y) \leq \phi \frac{\delta}{\varepsilon^i} + \phi \frac{\delta}{\varepsilon^{i-1}} + \ldots + \phi \frac{\delta}{\varepsilon^{k+1}}$. 
    If $k = i-1$, then:
    \begin{equation*}
        \begin{aligned}
            d_X(x, y) \leq \phi \frac{\delta}{\varepsilon^i} \leq \gamma \frac{\delta}{\varepsilon^i} - \gamma \frac{\delta}{\varepsilon^{i-1}}~.
        \end{aligned}
    \end{equation*}
    The last inequality holds since $\phi \leq \gamma (1 - \varepsilon)$ for $\phi = \frac{3}{4}, \gamma = 1, \varepsilon \leq \frac{1}{8}$.  By induction:
    \begin{equation*}
    \begin{aligned}
        d_X(x, y) &\leq \phi \frac{\delta}{\varepsilon^i} +  \ldots + \phi \frac{\delta}{\varepsilon^{k+1}} + \phi \frac{\delta}{\varepsilon^{k+1}} \\
        &\leq  \left(\gamma \frac{\delta}{\varepsilon^i} - \gamma\frac{\delta}{\varepsilon^{i-1}}\right) + \left(\gamma\frac{\delta}{\varepsilon^{i-1}} - \gamma\frac{\delta}{\varepsilon^{i-2}}\right) \ldots \left(\gamma\frac{\delta}{\varepsilon^{k+1}} - \gamma\frac{\delta}{\varepsilon^{k}}\right)\\
        &=  \gamma \frac{\delta}{\varepsilon^i} - \gamma\frac{\delta}{\varepsilon^{k}}~,
    \end{aligned}
    \end{equation*}
as desired.
\end{proof}

\paragraph{Safe chain invariant.~} Recall that the chain of a node $(t,i)$ is the sequence of nodes starting from the lowest ring-$\alpha$ ancestor of $(t, i)$, denoted by $(t^*,i^*)$, to $(t,i)$. It is safe if it contains at most one ring-$\phi$ node. Recall that the obligation of $(t,i)$ is the ring-$\phi$ node in its (safe) chain. Directly from the definition:

\begin{observation}\label{obs:ring_parent_promotednode}
    \begin{itemize}
    \item[\textnormal{(a)}] The chain of a ring-$\alpha$ node is safe.
    \item[\textnormal{(b)}]  The chain of a ring-$\beta$ node is safe if the chain of its parent is safe.
    The obligation of a ring-$\beta$ node is the obligation of its parent.
    \item [\textnormal{(c)}] The chain of a ring-$\phi$ node is safe if its parent has a null obligation.
    \end{itemize} 
\end{observation}

We now show the safe invariant.

\begin{lemma}[Safe invariant]\label{lmm:nettree_safechain}
The chain of every node is always safe.
\end{lemma}

\begin{proof} Assume that all chains are safe before an update to the net tree; we show that they remain safe after an update.  After step 2 of \textsc{Insert}, we found a node $u = (t, i)$ to be a parent of $q$ such that $q$ is not completely out of ring $\phi$ of $u$. (We use the notation $(t,i)$ for $u$ instead of $(t',i')$ as in step 2 to avoid clutter.) If (a) $q$ is in ring $\alpha$ or $\beta$ of $(t, i)$ or (b) $q$ is in ring $\phi$ of $(t, i)$ and the chain of $(t, i)$ has no obligation, then adding $(q, i-1)$ as a child of $(t, i)$ does not violate the safe invariant. The remaining case is when $q$ is in ring $\phi$ of $(t, i)$ and the chain of $(t, i)$ has a ring-$\phi$ node.  In this case, step 4 of \textsc{Insert} promotes the obligation of $(t, i)$. Let $(t_j, j) = \textsc{Obligate}(t, i)$. We have to show two things:
\begin{itemize}
    \item[\textnormal{(i)}] after promoting $(t_j, j)$, we have to show that the chain of $(t, i)$ has no ring-$\phi$ node, thus we can add $(q, i-1)$ as a ring-$\phi$ child of $(t, i)$.
    \item[\textnormal{(ii)}] Since promoting $(t_j, j)$ changes the parent of $(t_j, j)$ and possibly creates a new node $(t_j, j+1)$, we also have to show that the chains of $(t_j, j)$ and $(t_j, j+1)$ are safe. 
\end{itemize}
 We first focus on (i). Before $\textsc{Promote}(t_j, j, T)$, the chain of $(t, i)$ starts from the lowest ring-$\alpha$ ancestor of $(t, i)$, say $(t^*, i^*)$. 
At this point, the sequence of nodes from $(t^*, i^*)$ to $(t, i)$ has only one ring-$\alpha$ node, which is $(t^*, i^*)$, and only one ring-$\phi$ node, which is $(t_j, j)$.
After $\textsc{Promote}(t_j, j, T)$, $(t_j, j)$ becomes a ring-$\alpha$ node by~\Cref{obs:Promote-goal} and therefore, the lowest ring-$\alpha$ ancestor of $(t, i)$. Thus, the chain of $(t, i)$ starts from $(t_j, j)$, and this chain has no ring-$\phi$ node, as desired. 

For (ii), as  $(t_j, j)$ becomes a ring-$\alpha$ node after after $\textsc{Promote}(t_j, j, T)$, its chain is safe by \Cref{obs:ring_parent_promotednode}.  We now focus on the chain of $(t_j, j+1)$. If $(t_j, j+1)$ is a ring-$\alpha$ or ring-$\beta$ node, then its chain is safe by induction. The remaining case is when $(t_j, j+1)$ is a ring-$\phi$ node. We first claim that:

\begin{claim}\label{clm:ancster-tj}
Before $\textsc{Promote}(t_j, j, T)$, let $(t^*, i^*)$ be the lowest ring-$\alpha$ ancestor of $(t, i)$.
After $\textsc{Promote}(t_j, j, T)$,  if $(t_j, j+1)$ is a ring-$\phi$ node, then the chain of $(t_j, j+1)$ starts at $(t^*, i^*)$.
\end{claim}
To see the claim, let $(x,j+1)$ be the parent of $(t_j, j)$ before $\textsc{Promote}(t_j, j, T)$. Since $(t_j, j+1)$ is created,   step 3 of \textsc{Promote} will find a parent for $(t_j, j+1)$. Since  $(t_j, j+1)$ is a  ring-$\phi$, by the construction of step 3, the parent of $(x,j+1)$, denoted by $(v,j+2)$, will be chosen as the parent of $(t_j,j+1)$. Furthermore, $(x,j+1)$ cannot be a ring-$\alpha$ node since otherwise, by triangle inequality:
\begin{equation*}
\begin{split}
    d_X(v,t_j) &\leq d_X(v,x) + d_X(x,t_j) \\
     &\leq \alpha \frac{\delta}{\varepsilon^{j+2}} + \phi \frac{\delta}{\varepsilon^{j+1}} \qquad \text{(since $(x,j+1)$ is a ring-$\alpha$ node)}\\
     &\leq \beta \frac{\delta}{\varepsilon^{j+2}} \qquad \text{(since $\alpha + \phi\varepsilon \leq \beta$ by~\Cref{equ:nettree_constant})}~~,
\end{split}    
\end{equation*}
contradicting that $(t_j,j+1)$  is a ring-$\phi$ node.  As $(x,j+1)$ and $(t_j, j)$ are not ring-$\alpha$ node, $(t^*, i^*)$ is also the lowest ring-$\alpha$ ancestor of $(v, j+2)$; it could be that $(t^*, i^*) = (v,j+2)$. Since $(t_j,j+1)$ is not a ring-$\alpha$ node and has  $(v, j+2)$ as the parent,  $(t^*, i^*)$ is also the lowest ring-$\alpha$ ancestor of $(t_j,j+1)$, implying \Cref{clm:ancster-tj}.

Observe that before the promotion of $(t_j, j)$, the path from $(t^*, i^*)$ to $(t_j, j)$ has only one ring-$\phi$ node, which is $(t_j, j)$. As $(t^*, i^*)$ is an ancestor of $(t_j, j+1)$ after the promotion of $(t_j, j)$ by \Cref{clm:ancster-tj}, there is no ring-$\phi$ node from $(t^*, i^*)$ to the parent of $(t_j, j+1)$. Thus, even if $(t_j, j+1)$ becomes a ring-$\phi$ node, the chain of $(t_j, j+1)$ is still safe.
\end{proof}

\paragraph{Jump isolation property.~} This is the last property that we have to show to complete the proof of \Cref{thm:nettree_ds}.  In \textsc{Insert}, after we add  $O(1)$ new nodes and modify existing nodes from step 1 to step 4, we check and fix the jump isolation property in step 5 by invoking \textsc{MaintainJumpIsolation}$(y, i, T)$ for every new node $(y, i)$. While the jump isolation property for a jump from $(x, i)$ down to $(x, j)$ is defined w.r.t \emph{every node} at level $k\leq i$, the checking procedure \textsc{MaintainJumpIsolation} only looks at top of the jump (and hence the checking and fixing can be done in $O_{\lambda}(1)$ time), which is justified by the following lemma.

\begin{lemma}\label{lmm:maintain_bisolated}
    Given $\varepsilon \leq \frac{\alpha}{b}$ and a jump $(x_0, i_0)$ down to $(x_0, j_0)$, if $d_X(x_0, v_0) > \alpha \frac{\delta}{\varepsilon^{i_0}} + \gamma \frac{\delta}{\varepsilon^{i_0}}$ for every other node $(v_0, i_0)$ at level $i_0$, then the jump starting at $(x_0, i_0)$ is $b$-isolated:
    any non-descendant node $(z, m)$ of $(x_0, i_0)$ for $m < i_0$ has $d_X(x_0, z) > b \frac{\delta}{\varepsilon^m}$. (This includes the case where $(z,m)$ is a newly created node.)
\end{lemma}

\begin{proof}
First, we consider when $(z, m)$ does not have an ancestor at level $i_0$.
In this case, there must be a jump $J_t$ from a node $(t, i')$ to $(t, j')$ where $j' < i_0 < i'$ such that $(t,j')$ is an ancestor of $(z, m)$.  Observe that $(x_0, i_0)$ is not a descendant of $(t, i')$. By induction, $J_t$ is $b$-isolated,  implying that $d_X(x_0, t) > b\frac{\delta}{\varepsilon^{i_0}}$. 
Since $(z, m)$ is a descendant of $(t, j')$, by the close-containment property,  we have $d_X(z, t) \leq \gamma\frac{\delta}{\varepsilon^{j'}} - \gamma\frac{\delta}{\varepsilon^m} \leq \gamma\frac{\delta}{\varepsilon^{i_0-1}} - \gamma\frac{\delta}{\varepsilon^m}$.
By triangle inequality:
\begin{equation*}
    \begin{aligned}
        d_X(x_0, z) &\geq d_X(x_0, t) - d_X(t, z)\\
        &> b\frac{\delta}{\varepsilon^{i_0}} - (\gamma\frac{\delta}{\varepsilon^{i_0-1}} - \gamma\frac{\delta}{\varepsilon^m}) \\
        &\geq (b-\gamma\varepsilon)\frac{\delta}{\varepsilon^{i_0}}  = \frac{(b-\gamma\varepsilon)}{\varepsilon}\frac{\delta}{\varepsilon^{i_0-1}} \geq b\frac{\delta}{\varepsilon^m}
    \end{aligned}
\end{equation*}
where the last inequality holds since $m \leq i_0-1$, $\alpha = \frac{1}{4}, \gamma = 1, b \geq \frac{5}{4}$, and $\varepsilon \leq \frac{\alpha}{b} \leq \frac{b}{b+\gamma}$.

Now, we consider the case where $(z, m)$ has an ancestor at level $i_0$, let this node be $(v_0, i_0)$.
For contradiction, suppose that $d_X(z, m) \leq b\frac{\delta}{\varepsilon^m}$.
By the close-containment property, we have $d_X(z, v_0)~\leq~\gamma\frac{\delta}{\varepsilon^{i_0}}~-~\gamma\frac{\delta}{\varepsilon^m}$. Thus:
\begin{equation*}
    \begin{aligned}
        d_X(x_0, v_0) &\leq d_X(x_0, z) + d_X(z, v_0) \\
        &\leq b\frac{\delta}{\varepsilon^m} + (\gamma\frac{\delta}{\varepsilon^{i_0}} - \gamma\frac{\delta}{\varepsilon^m}) \leq \alpha\frac{\delta}{\varepsilon^{i_0}} + \gamma\frac{\delta}{\varepsilon^{i_0}}
    \end{aligned}
\end{equation*}
where $b\frac{\delta}{\varepsilon^m} \leq \alpha\frac{\delta}{\varepsilon^{i_0}}$ holds since $m \leq i_0-1$ and $\varepsilon \leq \frac{\alpha}{b}$. Thus if $d_X(x_0, v_0) > \alpha\frac{\delta}{\varepsilon^{i_0}} + \gamma\frac{\delta}{\varepsilon^{i_0}}$, then $d_X(x_0, z) > b\frac{\delta}{\varepsilon^m}$.
\end{proof}
 We are now ready to show the $b$-jump isolation property.

\begin{lemma}[$b$-Jump Isolation]\label{lmm:isolation}
    Every jump is $b$-isolated.    
\end{lemma}

\begin{proof}
    Recall that in Step 5 of \textsc{Insert}, we invoke \textsc{MaintainJumpIsolation$(y, i, T)$} for any new node $(y, i)$.
    In this operation, we find $(x, i)$ where $d_X(x, y) \leq (\alpha + \gamma)\frac{\delta}{\varepsilon^i}$.

    If $(x, i)$ does not exist, \textsc{MaintainJumpIsolation} does nothing. If $(y, i)$ is the top of a jump $J_y$, then $J_y$ is $b$-isolated by~\Cref{lmm:maintain_bisolated}. On the other hand,  the $b$-jump isolation property is maintained for every existing jump $J_u$ starting at $(u, i')$. Specifically, if $i' > i$, adding a node at level $i$ does not violate the $b$-isolation property of $J_u$ by applying~\Cref{lmm:maintain_bisolated} with $i_0 = i'$ (and $(z,m) = (y,i)$).  If $i' = i$, by induction$J_u$ is $b$-isolated with respect to existing nodes before adding $(y, i)$.   When adding $(y, i)$, since $d_X(u, y) > (\alpha + \gamma)\frac{\delta}{\varepsilon^i}$,  by~\Cref{lmm:maintain_bisolated}, the $b$-jump isolation property at $J_u$ is maintained. When $i' < i$, adding a node at level $i$ does not change anything at level $i'$, and hence ~\Cref{lmm:maintain_bisolated} also applies here. In all cases, the construction maintains the $b$-jump isolation property.

    We now consider the complementary case where there exists a node $(x, i)$ such that $d_X(x, y) \leq (\alpha + \gamma) \frac{\delta}{\varepsilon^{i-1}}$. If $(x, i)$ (or $(y, i)$) is the top of a jump $J_x$ (or $J_y$), in \textsc{MaintainJumpIsolation}$(y, i, T)$, we invoke \textsc{JumpFix} to create $(x, i-1)$ (or $(y, i-1)$).  Node $(x, i)$ and $(y, i)$ are no longer the top of their jumps, and hence   $J_x$ and $J_y$ are effectively replaced by two new jumps, denoted by $J_{x}'$ and $J_{y}'$, starting at $(x, i-1)$ and $(y, i-1)$, respectively. And we need to argue that after adding these new jumps, the jump isolation property is fixed.

    Let us consider $J_{y}'$ first. We claim that:
    \begin{equation}\label{eq:claim-y-newpoint}
        \text{for $J_{y}'$ to exist, $y$ must be a newly inserted point.}
    \end{equation}
    For contradiction, suppose that $y$ is an existing point. Then $(y, i)$ is a new node created by \textsc{Promote} or \textsc{JumpSplit} in step 2 or step 4 of \textsc{Insert$(q, t)$} for some point $q \neq y$. If $(y, i)$ is created by promoting $(y, i-1)$, then $(y, i-1)$ exists and there is no jump starting from $(y, i)$ to fix. In the other case, $(y, i)$ is created by splitting a jump from  $(y, i'')$ down to $(y, j'')$  at level $i$ where  $i'' > i$.  Since the jump starting at $(y, i'')$ is $b$-isolated, by definition, $d_X(y, t) > b\frac{\delta}{\varepsilon^k}$ for any node $(t, k)$ where $j'' < k < i''$. It follows that the jump starting at $(y, i)$, which is $J_y$, is also $b$-isolated, and hence \textsc{JumpFix} is not called on $(y,i)$. Therefore, \eqref{eq:claim-y-newpoint} holds. 

    Since $y$ is a newly inserted point, observe that
    any existing node $(v, i-1)$ must satisfy $d_X(v, y) > \psi\frac{\delta}{\varepsilon^{i-1}}$ since otherwise, $\textsc{ContainmentSearch}(y, T)$ will return a node at a level at most $i-1$, and hence after \textsc{Insert}, the parent of $y$ is a node at a level at most $i$ by~\Cref{obs:parent_q}.  Furthermore, since $\psi \geq \alpha + \gamma$ by~\Cref{equ:nettree_constant}, we have $d_X(v, y) > (\alpha + \gamma)\frac{\delta}{\varepsilon^{i-1}}$, implying the $b$-jump isolation of $J_y'$ by~\Cref{lmm:maintain_bisolated}.

    Finally, we consider the jump $J_{x}'$.  Before adding $(y, i)$, $J_x$ satisfies the  $b$-jump isolation. Adding a new node $(y, i)$ does not change the distance from $x$ to other nodes, and hence, the only possible jump violation to  $J_{x}'$ is due to $(y, i-1)$.   Since $(x,i)$ and $(y,i)$ are two nodes at level $i$, by packing property, $d_X(x, y) > \alpha \frac{\delta}{\varepsilon^i}$, giving $d_X(x, y) > \psi \frac{\delta}{\varepsilon^{i-1}} \geq (\alpha  + \gamma) \frac{\delta}{\varepsilon^{i-1}}$ since $\alpha \geq \psi\varepsilon \geq (\alpha+\gamma)\varepsilon$ is given by~\Cref{equ:nettree_constant}. By~\Cref{lmm:maintain_bisolated}, $J_{x}'$ is $b$-isolated with respected to new node $(y, i-1)$, and hence $J_x'$ is $b$-isolated overall.     
\end{proof}

%% file: 4.application.tex
\subsection{Dynamic VFT Spanners}

\VFTS*

\begin{proof} Our algorithm follows that of Chan, Har-Peled and Jones~\cite{CHJ20}. Statically, given a $(\tau,\eps)$-LSO $\Sigma$, we construct a $k$-VFTS $H$ as follows:

\begin{quote}
    Initially, $H = (S,\emptyset)$. For each ordering $\sigma \in \Sigma$ and each point $q\in \sigma$, we add  $2(k+1)$ edges incident to $q$ to $H$  where $k+1$ edges are from $q$ to its $k+1$ nearest predecessors in $\sigma$ and the other $k+1$ edges are from $q$ to its nearest successors in $\sigma$.  (If $q$ is close to the endpoints of $\sigma$, then we might add less than $2(k+1)$ edges.) 
\end{quote}

The claim (which we will prove later) is that:

\begin{claim}\label{clm:VFTS-spanner} $H$ is a $k$-VFTS where every vertex has degree at most $\tau \cdot 2(k+1) = O(\tau k)$.
\end{claim}

To maintain $H$ dynamically, whenever a point $q$ is added to $S$, we invoke \textsc{Insert}$(q, \Sigma)$. Then, given $\sigma$ is the ordering $i^{th}$ of $\Sigma$, we iteratively find $k+1$ nearest predecessors $p_1 \prec_{\sigma} p_2 \prec_{\sigma} \ldots \prec_{\sigma} p_{k+1}$ by $p_j = \textsc{GetPredecessor}(p_{j+1}, i, \Sigma)$ where $p_{k+2}=q$.
Similarly, using \textsc{GetSuccessor}, we find $k+1$ nearest successors of $q$: $s_1\prec_\sigma s_2 \prec_\sigma \ldots s_{k+1}$.
For every $j \in [1, k+1]$, we add to $H$ edges $(p_j, q)$ and $(q, s_j)$, then remove from $H$ the edge $(p_j, s_j)$.

When a point $q$ is deleted from $S$, first we reconnect neighbors of $q$ in $H$ as follows. For each ordering $\sigma \in \Sigma$, get $k+1$ nearest predecessors and $k+1$ nearest successors of $q$ in $\sigma$ as described above. Let $p_1 \prec_\sigma \ldots \prec_\sigma p_{k+1}$ be $k+1$ those predecessors, and $s_1 \prec_\sigma \ldots \prec_\sigma s_{k+1}$ be $k+1$ those successors.
Add the edge $(p_j, s_j)$ to $H$ for all $j \in [1, k+1]$.
Finally, remove all edges of $q$ out of $H$ and invoke \textsc{Delete}$(q, \Sigma)$. 

By \Cref{thrm:dynamic_lso}, \textsc{Insert} and \textsc{Delete} of $\Sigma$ take $O(\log(1/\eps)(\log n + \eps^{-O(\lambda)}))$ time, while \textsc{GetPredecessor} and \textsc{GetSuccessor} run in $O(1)$ per operation.
Hence, the total time to add and remove edges 
regarding an insertion or deletion
is $O(\tau k) = k\eps^{-O(\lambda)}$. In summary, the insertion and deletion time 
is in $(\log n + k)\eps^{-O(\lambda)}$ as claimed in the theorem.

To complete the proof of \Cref{thm:vfts}, we prove \Cref{clm:VFTS-spanner}. 
By~\Cref{thrm:dynamic_lso}, $\Sigma$ is stable, thus it suffices to get predecessors and successors at the point that is updated (inserted or deleted); all other edges remain in $H$.
By the construction, every vertex has a degree at most $\tau\cdot 2(k+1)$.
Now we show $H$ is a $k$-VFTS of $S$.
Let $F$ be the subset of $S$ with size at most $k$.
Consider two points $s, t$, there is an ordering $\sigma \in \Sigma$ such that all points $p$ between $s, t$ have $d_X(p, s) \leq \varepsilon d_X(s, t)$ or $d_X(p, t) \leq \varepsilon d_X(s, t)$. 
Let $\sigma'$ be the ordering obtained from $\sigma$ by removing points in $F$.
Observe that among adjacent pairs in $\sigma'$, there are $s', t'$ such that (i) $s \preceq_{\sigma'} s' \prec_{\sigma'} t$ and $ s \prec_{\sigma'} t' \preceq_{\sigma'} t$, (ii) $d_X(s, s') \leq \varepsilon d_X(s, t)$ and $d_X(t, t') \leq \varepsilon d_X(s, t)$. 
Since we add edges of a point with its $k+1$ predecessors and $k+1$ successors to $H$, if $u$ and $v$ are adjacent in $\sigma'$, there is an edge $(u, v) $ in $H \setminus F$ and $d_{H\setminus F}(u, v) = d_X(u, v)$.
It follows that $d_{H\setminus F}(s', t') = d_X(s', t')$.
Now we prove the claim by induction, and suppose that $d_{H\setminus F}(s, s') \leq (1 + c\varepsilon) d_X(s, s')$ and $d_{H\setminus F}(t, t') \leq (1+ c\varepsilon) d_X(t, t')$.
By triangle inequality, 
we have:
\begin{equation*}
\begin{split}
    d_{H\setminus F}(s, t) &\leq d_{H\setminus F}(s, s') + d_{H\setminus F}(s', t') + d_{H\setminus F}(t', t) \\
    &\leq (1+c\varepsilon)d_X(s, s') +
    d_X(s', t') + (1+c\varepsilon)d_X(t, t') \\
    &\leq (1+c\varepsilon)(d_X(s, s') + d_X(t, t')) + (d_X(s', s) + d_X(s, t) + d_X(t, t')) \\
    &\leq (1+c\varepsilon)2\varepsilon d_X(s, t) + ( d_X(s, t) +2\varepsilon d_X(s, t)) \\
    &= (1 + 4\varepsilon + 2c\varepsilon^2)d_X(s, t)
    \\
    &\leq (1 + c\varepsilon) d_X(s, t)  
\end{split}
\end{equation*}

Setting $c = 8$, the inequality holds
when $\varepsilon \leq \frac{1}{4}$. 
By scaling $\varepsilon$ with the constant factor $c=8$, \Cref{clm:VFTS-spanner} holds. 
\end{proof}

\subsection{Dynamic Tree Cover}
 \subfile{11.treecover.tex}

\subsection{Dynamic Closest Pair}

\begin{theorem}\label{thm:ClosestPair}
    Given a dynamic point set $S$ in a doubling metric of dimension $\lambda$, there is a data structure that maintains the closest pair for $S$ in $O_\lambda(\log(n))$ time per update.
\end{theorem}

\begin{proof}
    
The data structure consists of a $(\varepsilon^{-O(\lambda)}, \varepsilon)$-LSO $\Sigma$ for $\eps = 1/2$, and a min-heap $H$, where
$H$ maintains all pairs $(u, v)$ such that $u$ and $v$ are adjacent in an ordering of $\Sigma$, and is keyed by the distances between the pair.  The closest pair is determined by the pair with the minimum distance in $H$.

If $q$ is inserted into $S$, we invoke \textsc{Insert}$(q, \Sigma)$, then for every ordering $\sigma_i \in \Sigma$, we find the successor $s_i$ and the predecessor $p_i$ of $q$ in $\sigma_i$ by calling \textsc{GetSuccessor}$(q, i, \Sigma)$ and \textsc{GetPredecessor}$(q, i, \Sigma)$. Next, we insert two pairs $(p_i, q)$ and $(q, s_i)$ to $H$, and remove the pair $(p_i, s_i)$ from $H$. 

If $q$ is deleted from $S$, for every ordering $\sigma_i \in \Sigma$, we find the successor $s_i$ and the predecessor $p_i$ of $q$ in $\sigma_i$.  We remove $(p_i, q)$ and $(q, s_i)$ from $H$, and add $(p_i, s_i)$ to $H$.

First, we analyze the running time.  By \Cref{thrm:dynamic_lso}, the $(\varepsilon^{-O(\lambda)}, \varepsilon)$-LSO $\Sigma$ runs in $O( \varepsilon^{-O(\lambda)}\log{n})$ per update.
Getting the successor and the predecessor of a point in an ordering takes $O(1)$, in all orderings takes $O(|\Sigma|) = \varepsilon^{-O(\lambda)}$. When we insert or delete a point in $\Sigma$, there are $O(|\Sigma|) = \varepsilon^{-O(\lambda)}$ pairs updated (inserted or deleted) in $H$. $H$ maintains $O(n|\Sigma|)$ pairs, thus its running time is $O(\log{|\Sigma|} + \log{n})$ per insertion or deletion of a pair.  Since  $\varepsilon=1/2$, the total running time per update of the data structure maintaining the closest pair for $S$ is $2^{O(\lambda)}(\log{n})$, as claimed.

Next, we show the correctness. It suffices to get the predecessor and the successor of $q$ only since $\Sigma$ is stable by~\Cref{thrm:dynamic_lso}. Observe that if $(a,b)$ is the closest pair, then by the definition of LSO, there exists an ordering in $\Sigma$ where $a$ and $b$ are adjacent, which means $(a, b)$ is maintained in $H$. Suppose otherwise, there exists an ordering $\sigma \in \Sigma$ and a point $u$ where $a \prec_\sigma u \prec_\sigma b$ where $\min\{d_X(a, u),d_X(b, u)\} \leq \varepsilon\cdot d_X(a, b) <  d_X(a,b)$  as $\eps = 1/2$. Then either  $u$ is closer to $a$ than $b$ or $u$ is closer to $b$ than $a$; both cases contradict the fact that $(a,b)$ is the closest pair. 
\end{proof}

\subsection{Approximate Bichromatic Closest Pair}

\BCP*

\begin{proof} We can find $(1+\varepsilon)$-approximation for the bichromatic closest pair under insertions and deletions of $B$ and $R$ by using a min-heap $H$, and a $(\varepsilon^{-O(\lambda)}, \varepsilon)$-LSO $\Sigma$. The key idea is we apply the LSO to find adjacent pairs $(r, b)$ where $r \in R, b\in B$ in every ordering $\sigma \in \Sigma$, and then use $H$ to maintain these pairs sorted by descending order of $d_X(r, b)$.

To find the adjacent pair from a new point, suppose that $r$ is newly added to $R$. First, we invoke \textsc{Insert$(r, \Sigma)$}.
After that, we invoke \textsc{GetSuccessor$(r, i, \Sigma)$} to find the successor $s$ of $r$, and   \textsc{GetPredecessor$(r, i, \Sigma)$} to find the predecessor $p$ of $r$ in $\sigma$. Now we obtain $p \prec_\sigma r \prec_\sigma s$.
If $s \in B$, we add $(r, s)$ to ${H}$. 
If  $p \in R$ and $s \in B$, we remove the pair $(p, s)$ from ${H}$.
For a new point $b \in B$, we invoke \textsc{Insert$(b, \Sigma)$}, then we follow a similar way to find $p \prec_\sigma b \prec_\sigma s$. If $p \in R$, we update $(p, b)$ to ${H}$. If $p \in R$ and $s \in B$, we remove $(p, s)$ from ${H}$.
    
With deletion, when a point $r \in R$ is deleted, 
we retrieve from $r$ the predecessor $p$ and the successor $s$ in $\sigma$ to obtain $p \prec_\sigma r \prec_\sigma s$. 
If $s \in B$, then we remove $(r, s)$ out of ${H}$.
If $p \in R$ and $s \in B$, we add $(p, s)$ to ${H}$. After that, we invoke \textsc{Delete}$(r, \Sigma)$ to remove $r$ out of all orderings in $\Sigma$. Similarly when a point $b \in B$ is deleted, we retrieve its predecessor $p$ and successor $s$ in $\sigma$: $p \prec_\sigma b \prec_\sigma s$. If $p \in R$, we remove $(p, b)$ out of $H$. If $p \in R$ and $s \in B$, we add back to $H$ the pair $(p, s)$. Finally, we invoke \textsc{Delete$(b, \Sigma)$}.

 Since each point is adjacent to at most $2$ other points with different colors, and we have $\tau = \varepsilon^{-O(\lambda)}$ orderings, ${H}$ maintains at most $\varepsilon^{-O(\lambda)}O(n)$ pairs. Thus, operations of insertion and deletion in ${H}$ run in $O(\lambda\log\frac{1}{\varepsilon} + \log{n})$ time per update. By~\Cref{thrm:dynamic_lso}, $(\varepsilon^{-O(\lambda)}, \varepsilon)$-LSO $\Sigma$ runs in $O(\varepsilon^{-O(\lambda)}\log{n})$ time per update and $O(1)$ per predecessor/successor query. 
Therefore, the running time totally is $O( \varepsilon^{-O(\lambda)}\log{n})$ per update as claimed.

To prove the correctness, we consider the closest pair $(r, b)$. By~\Cref{thrm:dynamic_lso}, $\Sigma$ is stable, thus it suffices to query the predecessor and the successor of a new point or deleted point only. By the definition of LSO, there is an ordering $\sigma$ such that: for every point $t$ where $r \prec_\sigma t \prec_\sigma b$, $d_X(r, t) \leq \varepsilon d_X(r, b)$ or $d_X(t, b) \leq \varepsilon d_X(r, b)$. 
Observe that if $d_X(r, t) \leq \varepsilon d_X(r, b)$, $t$ must be a point in $R$, otherwise $(r, b)$ is not the closest bichromatic pair. Similarly, if $d_X(t, b) \leq \varepsilon d_X(r, b)$, $t$ must be a point in $B$.
Thus in $\sigma$ from $r$ to $b$, there is an adjacent pair $(r', b')$ such that $r' \in R, b' \in B$, and both $d_X(r, r')$, $d_X(b', b)$ are at most $\varepsilon d_X(r, b)$. 
By triangle inequality, we have:

\begin{equation}
    \begin{aligned}
        d_X(r', b') &\leq d_X(r, b) + d_X(r, r')  + d_X(b', b) \\
        &\leq d_X(r, b) + \varepsilon d_X(r, b) + \varepsilon d_X(r, b) \\
        &= (1 + 2\varepsilon) d_X(r, b).
    \end{aligned}
\end{equation}
Since $(r', b')$ is maintained in ${H}$, we correctly find a pair with distance at most $(1 + 2\varepsilon) d_X(r, b)$. Adjusting $\varepsilon$ by a constant factor, the theorem follows.
\end{proof}

\subsection{Dynamic Approximate Nearest Neighbors}

\ANN*

\begin{proof}
    We can directly use an $(\varepsilon^{-O(\lambda)}, \varepsilon)$-LSO $\Sigma$ to find approximate nearest neighbours. 
    When we add a new point $q$ to $S$, we invoke \textsc{Insert}$(q, \Sigma)$; when we delete an existing point $q$, we invoke \textsc{Delete}$(q, \Sigma)$. 
    To find an approximate nearest neighbour of $x$, we follow 3 steps: (1) insert $x$ to $\Sigma$ by \textsc{Insert$(x, \Sigma)$}, (2) for each ordering $\sigma_i \in \Sigma$, find the predecessor $p_i$ and the successor $s_i$ of $x$, then return the point who is closest to $x$ among $\{p_1, s_1, \ldots p_{|\Sigma|}, s_{|\Sigma|}\}$, (3) remove $x$ out of $\Sigma$ by \textsc{Delete$(x, \Sigma)$}. Note that if $x$ is a point that we already add into $S$, we run only step (2).

    By~\Cref{thrm:dynamic_lso}, $\Sigma$ has $O(\varepsilon^{-O(\lambda)}\log{(n)})$ time per update, and we invoke $2\varepsilon^{-O(\lambda)}$ predecessor and successor queries, each takes $O(1)$ time. Thus we obtain the running time per update and per query as claimed. 

    Now to show the correctness. Given a query point $x$, consider the closest point $y$ of $x$. By the definition of LSO, there is an ordering $\sigma$ such that any point $t$ between $x$ and $y$ has
    $d_X(t, x) \leq \varepsilon d_X(x, y)$ or $d_X(t, y) \leq \varepsilon d_X(x, y)$. 
    Without loss of generality, suppose that $x \prec_\sigma y$.
    Since $y$ is the point closest to $x$, thus \begin{inlinelist}
    \item $x$ and $y$ must be adjacent in $\sigma$
    \item or any point $t$ between $x$ and $y$ in $\sigma$ must have $d_X(t, x) \geq d_X(x, y)$ and $d_X(t, y) \leq \varepsilon d_X(x, y)$
    \end{inlinelist}.
    For (i), we are done. 
    For (ii), consider the successor $s$ of $x$ in $\sigma$. 
    By triangle inequality, we obtain:
    \begin{equation}
        \begin{aligned}
            d_X(x, s) &\leq d_X(x, y) + d_X(y, s) \\
            &\leq d_X(x, y) + \varepsilon d_X(x, y) 
            \\
            &= (1 + \varepsilon) d_X(x, y)
        \end{aligned}
    \end{equation}
    Similarly, when $y \prec_\sigma x$, we consider the predecessor $p$ of $x$ in $\sigma$ and obtain:
    \begin{equation}
        \begin{aligned}
            d_X(x, p) &\leq (1 + \varepsilon) d_X(x, y)
        \end{aligned}
    \end{equation}
    
    We return a point $x'$ adjacent with $x$ in an ordering of $\Sigma$ such that $d_X(x, x') \leq d_X(x, s)$ and $d_X(x, x') \leq d_X(x, p)$. Thus $x'$ is $(1+\varepsilon)$-approximate nearest neighbour of $x$. 
\end{proof}

%% file: 11.treecover.tex
\Treecover*

 We can construct a tree cover $\mathcal{J}$ from a pairwise tree cover $\mathcal{T}$ by adding weights to edges: 
\begin{itemize}
    \item{[\textbf{Step 1.}]} For every PIT $T \in \mathcal{T}$, we
    create a tree $J \in \mathcal{J}$ such that $J$ and $T$ has the same set of nodes and edges.
    \item{[\textbf{Step 2.}]} For every edge connecting two nodes $a = (x, y, i)$ and $b = (u, v, j)$ in $J$, we assign a weight to the edge $(a, b)$ of $J$ as $w_J(a, b) = d_X(\{x, y\}, \{u, v\})$. (We use $d_X(A,B) = \min_{x\in A, y\in B} d_X(a,b)$ to denote the distance between two sets of points $A$ and $B$.)
    \item{[\textbf{Step 3.}]}  We now update the weights of the edges from a node  $a = (x, y, i)$ to its children. (Note that not all edges from $a$ to its children get their weights updated.) Let $c = (s, t, i')$ be the parent of $a$. 
    \begin{itemize}
        \item If $d_X(x, \{s, t\}) \leq d_X(y, \{s, t\})$, for every child $b = (u, v, j)$ of $(x, y, i)$ such that $d_X(y, \{u, v\}) \leq d_X(x, \{u, v\})$ 
        we add $d_X(x, y)$ to $w(a, b)$.
        \item If $d_X(x, \{s, t\}) > d_X(y, \{s, t\})$, for every child $b = (u, v, j)$ of $(x, y, i)$ such that 
        $d_X(x, \{u, v\}) < d_X(y, \{u, v\})$,
        we add $d_X(x, y)$ to $w(a, b)$.
    \end{itemize}
    
\end{itemize}

The intuition of the Step 3 is as follows.  Suppose that $b = (u, v, j)$ is a child of a node $a = (x, y, i)$, we temporarily assign the weight of edge $(a,b)$ to be the closest distance, namely $d_X(\{x, y\}, \{u, v\})$, between its labeled points in Step 2. One can think of this as using the closest pair to ``represent'' the edge $(a,b)$.  Next, consider the parent $c = (s, t, i')$ of $a$; in the same way, Step 2  also uses the closest pair in the labels of $c$ and $a$ to represent $(a,c)$. But this means the path from $b$ to $c$ (passing through $a$) might ``miss'' the edge $(x,y)$, and therefore, if $\{u, v\}$ is closer to $x$ and $\{s, t\}$ is closer to $y$, we need to add $d_X(x, y)$ to the weight of edge $(a,b)$, as in Step 3. 

For a dynamic point set $S$, whenever we update $\mathcal{T}$, $\mathcal{J}$ have the same updates as $\mathcal{T}$, and we assign weights to edges of trees in $\mathcal{J}$ as Step 2 and Step 3 above. 
Note that $\mathcal{J}$ and $\mathcal{T}$ share many properties: the number of trees, update time, and the covering property. Since $\mathcal{T}$ has $\varepsilon^{-O(\lambda)}$ PITs, the tree cover $\mathcal{J}$ also has $\varepsilon^{-O(\lambda)}$  trees. The running time per update to $S$ of $\mathcal{T}$ is $O(\varepsilon^{-O(\lambda)}\log{n})$, thus $\mathcal{J}$ has the same update time. 

It remains to show the stretch of $\mathcal{J}$. We rely on the pairwise covering property of $\mathcal{T}$ for this: for any pair of points $x_0, y_0 \in S$ whose distance in $[\frac{\delta}{\varepsilon^i}, \frac{2\delta}{\varepsilon^{i}})$ for some $\delta\in \{1,2^{1},2^{2},\ldots, 2^{\lceil \log(1/\eps) \rceil}\}$, there exists a $(\delta,\eps)$-PIT $T\in \mathcal{T}$ such that a node $(x, y, i)$ at level $i$ of $T$ is $O(\eps)$-close to pair $(x_0,y_0)$.
Recall that $(x, y, i)$ is $O(\varepsilon)$-close to $(x_0, y_0)$ means $x_0, y_0 \in C_i(x, y)$, and for any point $t \in C_i(x, y)$, $d_X(t, x)$ or $d_X(t, y)$ are at most $O(\varepsilon)d_X(x_0, y_0)$.

Also recall from~\Cref{def:pit} that $T$ satisfies the covering property: \begin{itemize}
    \item \textnormal[Children covering.] If $(u, v, j)$ is a child of $(x, y, i)$ for $j < i$, then $d_X(u, \{x, y\})$ and $d_X(v, \{x, y\})$ are $O(\frac{\delta}{\varepsilon^{i-1}})$.
    \item \textnormal[Bounded diameter.]  The cluster $C_i(x, y)$ of $(x,y,i)$ is the union of all leaf labels in the subtree rooted as $(x, y, i)$, and furthermore, the diameter of $C_i(x, y)$ is $O(\frac{\delta}{\varepsilon^i})$.
\end{itemize}

\Cref{lmm:treecover_stretch} below concludes the stretch analysis, implying \Cref{thm:treecover}.

\begin{lemma}\label{lmm:treecover_stretch}
    For every pair $(x_0, y_0)$ whose distance $d_X(x_0, y_0) \in [\frac{\delta}{\varepsilon^i}, \frac{2\delta}{\varepsilon})$, there exists a tree $J \in \mathcal{J}$ containing a path $\pi$ from the leaf of $x_0$ to the leaf of $y_0$ such that the total weight of the edges along $\pi$ is at most $(1+O(\varepsilon))d_X(x_0, y_0)$.
\end{lemma}

\begin{proof}
    For any pair $x_0, y_0 \in S$ whose $d_X(x_0, y_0) \in [\frac{\delta}{\varepsilon^i}, \frac{2\delta}{\varepsilon^i})$, by the pairwise covering property of $\mathcal{T}$, there exists a tree $J\in \mathcal{J}$ that covers $(x_0, y_0)$. That is, $J$ has a node $a = (x, y, i)$ where $(x, y, i)$ is $O(\varepsilon)$-close to $(x_0, y_0)$.

    Suppose that $d_X(x_0, \{x, y\})$ and $d_X(y_0, \{x, y\})$ are at most $c\varepsilon d_X(x_0, y_0)$ for some constant $c$.
    First, we argue that $d_X(x, y) = \Theta(\frac{\delta}{\varepsilon^i})$.
    By triangle inequality, we obtain:
    \begin{equation}\label{equ:treecover_1}
    \begin{aligned}
            d_X(x_0, y_0)-2c\varepsilon d_X(x_0, y_0) &\leq d_X(x, y) \leq  d_X(x_0, y_0)+2c\varepsilon d_X(x_0, y_0) \\
          \Leftrightarrow  (1-2c\varepsilon) \frac{\delta}{\varepsilon^i}&\leq d_X(x, y) \leq(2+2c\varepsilon) \frac{\delta}{\varepsilon^i}
        \end{aligned}
    \end{equation}

    The path $\pi$ from $x_0$ to $y_0$ in $J$ travels from the leaf of $x_0$ to $(x, y, i)$, then from $(x, y, i)$ down to the leaf of $y_0$. 
    Let $l_J(x_0), l_J(y_0)$ be the leaf of $x_0$, $y_0$ in $J$.
    Since $d_X(x, y) \leq (1 + O(\varepsilon)) d_X(x_0, y_0)$, if $w(l_J(x_0), l_J(y_0)) \leq O(\varepsilon) d_X(x, y) + d_X(x, y)$, the stretch follows.

    To compute $w(l_J(x_0), l_J(y_0))$, we bound the total weight of edges in $J$ from a leaf to a child of $a = (x, y, i)$.
    Consider a child $b = (u, v, j)$ of $a = (x, y, i)$ where $j < i$, and a point $t \in C_j(u, v)$.

    \begin{claim}\label{clm:dj-bt}  If $b$ is the ancestor at level $j$ of $l_J(t)$, then    $w(l_J(t), b) = O(\frac{\delta}{\varepsilon^{j}})$
    \end{claim}
    \textbf{Proof:}  Consider the base case where $b$ is the parent of $l_J(t)$. By the construction of $J$, we have two cases of $w(l_J(t), b)$ when  $d_X(u, t) \leq d_X(v, t)$:
    \begin{itemize}
        \item If $d_X(u, \{x, y\}) \leq d_X(v, \{x, y\})$, then $w(l_J(t), b) = d_X(t, u)$.
        \item If $d_X(u, \{x, y\}) > d_X(v, \{x, y\})$, then $w(l_J(t), b) = d_X(t, u) + d_X(u, v)$.
    \end{itemize}
    By the children covering of $J$,     
    $d_X(t, u) = O(\frac{\delta}{\varepsilon^{j-1}})$, and by the bounded diameter  property, $d_X(t, v) = O(\frac{\delta}{\varepsilon^{j}})$ and $d_X(u, v) = O(\frac{\delta}{\varepsilon^{j}})$. Thus $w(l_J(t), b) =  O(\frac{\delta}{\varepsilon^j})$. 
    
    Similarly, if $b$ is the parent of $l_J(t)$  and $d_X(u, t) > d_X(v, t)$, by the construction of $J$, we have two cases:
    \begin{itemize}
        \item If $d_X(u, \{x, y\}) \leq d_X(v, \{x, y\})$, then $w(l_J(t), b) = d_X(t, v) + d_X(u, v)$.
        \item If $d_X(u, \{x, y\}) > d_X(v, \{x, y\})$, then $w(l_J(t), b) = d_X(t, v)$.
    \end{itemize}
    By the children covering and the bounded diameter properties, we again obtain  $w(l_J(t), b) = O(\frac{\delta}{\varepsilon^j})$.
    
    Now, for the inductive case where $b$ is an ancestor at level $j$ of $l_J(t)$. By induction, let $b' = (u', v', j')$ be a child of $b$ that is the ancestor at level $j' < j$ of $l_J(t)$, and that $w(l_J(t), b') = O(\frac{\delta}{\varepsilon^{j'}})$.
    By the weight update in Step 3,  $w(b, b') \leq  d_X(\{u, v\}, \{u', v'\}) + d_X(u, v)$. By the children covering property of $J$,     
    $d_X(\{u, v\}, \{u', v'\}) = O(\frac{\delta}{\varepsilon^{j-1}})$, and by the bounded diameter property, $d_X(u, v) = O(\frac{\delta}{\varepsilon^{j}})$. We obtain:
    \begin{equation}
        \begin{aligned}
            w(l_J(t), b) &= w(l_J(t), b') + w(b', b) \\
            &\leq O(\frac{\delta}{\varepsilon^{j'}}) + 
            d_X(\{u, v\}, \{u', v'\}) + d_X(u, v)  \\
            &=  O(\frac{\delta}{\varepsilon^{j-1}}) + O(\frac{\delta}{\varepsilon^{j-1}}) + O(\frac{\delta}{\varepsilon^{j}}) \\
            &= O(\frac{\delta}{\varepsilon^{j}})
        \end{aligned}
    \end{equation}
    as desired.  \qed
    \linebreak

    Back to proving the stretch, observe that $x_0$ (or $y_0$) can not be within the distance $O(\frac{\delta}{\varepsilon^{i-1}})$ to both $x$ and $y$, since $d_X(x_0, y_0)$ and $d_X(x, y)$ are $\Theta(\frac{\delta}{\varepsilon^i})$.
    Without loss of generality, suppose that $d_X(x, x_0) = O(\frac{\delta}{\varepsilon^{i-1}})$ and $d_X(y, y_0) = O(\frac{\delta}{\varepsilon^{i-1}})$.
    Let $b_x = (u_x, v_x, j_x)$ be a child of $a = (x, y, i)$ and the ancestor of $l_J(x_0)$.
    By the bounded diameter property, $d_X(x_0, u_x)$ and $d_X(x_0, v_x)$ are $O(\frac{\delta}{\varepsilon^{j_x}}) = O(\frac{\delta}{\varepsilon^{i-1}})$ since $j_x < i$. 
    By triangle inequality, $d_X(u_x, x)$ and $d_X(v_x, x)$ are at most $O(\frac{\delta}{\varepsilon^{i-1}})$.
    Similarly, let $b_y = (u_y, v_y, j_y)$ be a child of $a = (x, y, i)$ and the ancestor of $l_J(y_0)$, we have $d_X(u_y, y)$ and $d_X(v_y, y)$ are at most $O(\frac{\delta}{\varepsilon^{i-1}})$.
    We obtain that labeled points of $b_x$ are close to $x$, and labeled points of $b_y$ are close to $y$. Thus, there is only one node between $b_x$ and $b_y$ such that we add $d_X(x, y)$ to $w(b_x, a)$ or $w(b_y, a)$ in Step 3. We obtain:
    \begin{equation}\label{equ:bx_by}
    \begin{aligned}
        w(b_x, a) + w(a, b_y) &= d_X(\{u_x, v_x\}, \{x, y\}) + d_X(\{x, y\}, \{u_y, v_y\}) + d_X(x, y) 
    \end{aligned}
    \end{equation}

    By the children covering property of $J$,     
    $d_X(\{x, y\}, \{u_x, v_x\}) = O(\frac{\delta}{\varepsilon^{i-1}})$ and $d_X(\{x, y\}, \{u_y, v_y\}) = O(\frac{\delta}{\varepsilon^{i-1}})$.
    Thus, $w(b_x, a) + w(a, b_y) =  O(\frac{\delta}{\varepsilon^{i-1}}) + d_X(x, y)$.
    Besides that, by~\Cref{clm:dj-bt}, $w(l_J(x_0), b_x) = O(\frac{\delta}{\varepsilon^{i-1}})$ and $w(l_J(y_0), b_y) = O(\frac{\delta}{\varepsilon^{i-1}})$,  since $j_x, j_y \leq i-1$.
    The total weight of edges in $\pi$ is bounded as follows:
    \begin{equation}
        \begin{aligned}            
        w(l_J(x_0), l_J(y_0)) &= w(l_J(x_0), b_x) + \left(w(b_x, a) + w(a, b_y)\right) + w(b_y, l_J(y_0)) \\
            &= O(\frac{\delta}{\varepsilon^{i-1}}) + \left(O(\frac{\delta}{\varepsilon^{i-1}}) + d_X(x, y)\right)  + O(\frac{\delta}{\varepsilon^{i-1}})
            \\&\leq (1 + O(\varepsilon)) d_X(x_0, y_0) 
        \end{aligned}
    \end{equation}
    where the last inequality holds since $d_X(x, y) \leq (1 + O(\varepsilon))d_X(x_0, y_0)$ by~\Cref{equ:treecover_1} and $d_X(x_0, y_0) = \Theta(\frac{\delta}{\varepsilon^i})$. 
\end{proof}